\documentclass{LMCS}

\def\doi{8(4:13)2012}
\lmcsheading%
{\doi}
{1--71}
{}
{}
{Nov.~\phantom08, 2011}
{Nov.~19, 2012}
{}   

\usepackage{enumerate,hyperref}

 \newcommand{\nm}[1]{\ensuremath{\mathsf{#1}}}
 \newcommand{\nmu}[1]{\ensuremath{\mathtt{#1}}}
 \newcommand{\midef}{\ensuremath{\stackrel{\mathsf{def}}{=}}}

\newcommand{\new}[1]{#1}

\usepackage{japerez}
\usepackage{xspace}
\usepackage{davidemacro} 
\usepackage{evolvablemacro}
\usepackage{myproof}
\usepackage{xytree}
\usepackage{mathpartir}
\usepackage{amssymb}
\usepackage{stmaryrd}
\usepackage{subfigure}

\theoremstyle{plain}

\begin{document}

 \title[Adaptable Processes]{Adaptable Processes}

\author[M.~Bravetti]{Mario Bravetti\rsuper a}	
\address{{\lsuper{a,d}}Laboratory Focus (University of Bologna/INRIA),
  Italy}	
\email{bravetti@cs.unibo.it, zavattar@cs.unibo.it}

\author[C.~Di Giusto]{Cinzia Di Giusto\rsuper b}	
\address{{\lsuper b}CEA, List, France}	
\email{cinzia.digiusto@gmail.com}

\author[J.~A.~P\'erez]{Jorge A. P\'erez\rsuper c}	
\address{{\lsuper c}CITI and Department of Computer Science, FCT New University
  of Lisbon, Portugal}
\email{japerezp@gmail.com}

\author[G.~Zavattaro]{Gianluigi Zavattaro\rsuper d}	
\address{\vskip-6 pt}





\begin{abstract}
We propose the concept of \emph{adaptable processes} as a way of overcoming the limitations that process calculi have for describing patterns of dynamic \emph{process evolution}.
Such patterns rely on direct ways of controlling the behavior and location of \emph{running} processes, and so 
they are at the heart of the \emph{adaptation} capabilities present in many modern concurrent systems.
Adaptable processes have a location and are sensible to actions of \emph{dynamic update} at runtime; 
this allows to express a wide range of evolvability patterns for concurrent processes.
We introduce a core calculus of adaptable processes and propose two verification problems
for them: \emph{bounded} and \emph{eventual adaptation}.
While the former ensures that the number of consecutive erroneous
states that can be traversed during a computation is bound by some 
given number $k$,
the latter ensures that if the system 
enters into a state with errors then a state without errors  will be eventually reached. 
We study the (un)decidability of these two problems in several variants of the calculus, which  
result from considering dynamic and static topologies of adaptable processes as well as different evolvability patterns.
Rather than a specification language, our calculus intends to be a basis for 
investigating the fundamental properties of evolvable processes and for 
developing richer languages with evolvability capabilities.
\end{abstract}

\maketitle

\section{Introduction}\label{S:one}
Process calculi aim at describing formally the behavior of concurrent systems. 
A leading motivation in the development of process calculi has been 
properly capturing the \emph{dynamic character} of concurrent behavior.
In fact, much of the success of the $\pi$-calculus~\cite{MilnerPW92a} can be fairly attributed to 
the way it departs from CCS~\cite{Milner89} so as to describe 
mobile systems in which  communication topologies can change dynamically.
Subsequent developments can be explained similarly. 
For instance, the Ambient calculus~\cite{CardelliG00} 
builds on $\pi$-calculus mobility 
to describe the dynamics of interaction within 
boundaries and hierarchies, as
required in distributed systems.
A commonality in these calculi is that 
the dynamic behavior of a system is realized through a number of \emph{local changes}, 
usually formalized by reduction steps.
Indeed, while in the $\pi$-calculus mobility is enforced by the reconfiguration of individual linkages in the communication topology, 
in the Ambient calculus spatial mobility is obtained by individual modifications to the containment relations within the ambient hierarchy.
This way, the combined effect of a series of changes at a local level (links, containment relations) 
suffices to explain dynamic behavior at the global (system) level.


There are, however, interesting forms of dynamic behavior that cannot be satisfactorily 
described as a combination of local changes, in the above sense.
These are behavioral patterns which concern change at the \emph{process} level (i.e., the process as a whole), 
and describe \emph{process evolution} along time.
In general, forms of process evolvability are 
characterized by an enhanced control/awareness over 
the current behavior and location of running processes. 
Crucially, this increased control is central to the 
\emph{adaptation} capabilities by which processes modify 
their behavior in response to exceptional 
circumstances in their environment.
As a simple example, 
consider  a scheduler in an operating system which 
manages the execution of 
a  set of processes.
To specify the behavior of the scheduler, the processes, and their evolution, 
we would need mechanisms for \emph{direct} process manipulation, 
which appear hard  to represent 
in calculi enforcing 
local changes only. 
More precisely, 
it is not clear at all how to represent 
the \emph{intermittent evolution} of a process under the scheduler's control: that is, precise ways of describing that 
its behavior 
``disappears'' (when the scheduler suspends the process)
and
``appears'' (when the scheduler resumes the process). 
Emerging applications and programming paradigms 
provide challenging examples of
evolvable processes. 
In workflow applications, we would like to be able to replace  a running activity, 
suspend the execution of a set of activities, or even suspend and relocate the whole workflow.
Similarly, in component-based systems we would like to
reconfigure parts of a component, a whole component, or groups of components.
Also, 
we would like to specify the 
context-aware policies that dynamically adapt the
computational power 
of cloud computing 
applications.
At the heart of these applications we find 
forms of process evolution and adaptation which appear very difficult (if not impossible) to express
in existing process calculi. 

\subsection*{A Core Calculus of Adaptable Processes}
In an attempt to address these shortcomings, this paper introduces the concept of \emph{adaptable processes}. 
Adaptable processes have a location and 
are sensible to actions of \emph{dynamic update} at runtime.
While locations are useful to designate and structure processes into hierarchies,
dynamic update actions 
implement a sort of built-in adaptation mechanism.
We illustrate this novel concept by 
 introducing \evol{}, a core process calculus of adaptable processes. 
The \evol{} calculus arises as
 a variant of CCS without restriction and relabeling, and 
extended with primitive notions of \emph{location} and \emph{dynamic update}.
In $\mathcal{E}$,  
$\component{a}{P}$ denotes the adaptable process $P$ located at $a$.
Name $a$ acts as a \emph{transparent} locality: $P$ can 
evolve on its own but also  
interact freely with its environment.
Localities can be nested, and are sensible to interactions with \emph{update prefixes}.
An update prefix $\update{a}{U}$ decrees the update of the adaptable process at $a$
with the behavior defined by 
$U$, a \emph{context} with zero or more holes,  denoted by $\bullet$.
The \emph{evolution} of $\component{a}{P}$  is realized by 
its interaction  
with the update prefix 
$\update{a}{U}$, which leads to 
the process obtained by replacing every hole $\bullet$ in $U$ by $P$, denoted 
\fillcon{U}{P}.

We consider several variants of \evol{},  obtained via  
two orthogonal characterizations.
The first one is 
\emph{structural}, and 
defines 
\emph{static} and \emph{dynamic} topologies of adaptable processes.
In a static topology, the number of adaptable processes does not vary along the evolution of the system:
they cannot be destroyed nor new ones can appear. 
In contrast, in the more general dynamic topology this restriction is lifted. 
We will use the subscripts $s$ and $d$ to denote the 
variants of \evol{} with static and dynamic topologies,
respectively.
The second characterization is 
\emph{behavioral}, and  
concerns 
\emph{update patterns}---the 
context 
$U$ in an update prefix $\update{a}{U}$.
As hinted at above, update patterns determine the 
 behavior of running processes after an update action.
In order to account for different evolvability patterns, 
we consider three kinds of update patterns, which determine three 
families of \evol{} calculi---denoted by the superscripts 
1, 2, and 3, respectively.
The first update pattern 
admits all kinds of contexts, and so it represents  
the most expressive form of update. 
In particular, holes $\bullet$ can appear behind prefixes.
The second update pattern forbids such guarded holes in contexts. 
In the third update pattern we further require contexts to have exactly one hole, 
thus preserving the current behavior (and possibly adding new behaviors): 
this is the most restrictive form of update. 


In our view, 
these variants
capture a fairly ample spectrum of scenarios that arise in the 
joint
analysis of correctness and adaptation concerns in evolvable systems.
They borrow inspiration from 
existing programming languages,
development frameworks, and 
component\break models.
The structural characterization follows the premise that 
while it is appealing to define the runtime evolution of
 the structures underlying 
aggregations of behaviors, in some scenarios 
it is also sensible to specify evolution and adaptation preserving such structures.
For instance, 
we would like software updates to preserve the main architecture of our operating system; 
conversely, an operating system could be designed to disallow runtime updates that alter its basic organization in dangerous ways.  
A static topology is also consistent with settings in which 
adaptable processes represent \emph{located resources}, 
whose creation 
is disallowed or 
comes with a cost (as in cloud computing scenarios).
The behavioral characterization
is inspired in the  
(restricted) forms 
of reconfiguration and/or update  available in component models 
in which evolvability is specified in terms of 
patterns, such as SOFA 2~\cite{HnetynkaP06}.
Update patterns are also related to functionalities present in 
programming languages such as Erlang~\cite{erlang,Armstrong03} and in
development frameworks such as the Windows Workflow Foundation (WWF)~\cite{wwf}.
In fact, 
forms of dynamic update behavior for workflow services in the WWF
include the possibility of replacing and removing service contracts 
(analogous to our first and second update patterns)
and also the addition of new service contracts and operations 
(as in our third update pattern, which preserves existing behavior and operations).

\subsection*{Verification of Adaptable Processes}
Rather than a specification language, the \evol{} calculus intends to be a basis for 
investigating the fundamental properties of evolvable processes.
In this presentation, we study  two  \emph{verification problems} associated to $\mathcal{E}$ processes and their (un)decidability.
They are defined in terms of standard observability predicates (\emph{barbs}), which indicate the presence of a designated error signal.
We thus distinguish between \emph{correct states} (i.e., states in which no error barbs are observable) and \emph{error states} (i.e., states exhibiting error barbs).
The first verification problem, 
\emph{bounded adaptation} 
(abbreviated \OG)
ensures that,
given a finite $k$, 
at most $k$ consecutive error states can arise  in computations of the system---including those reachable as a result of
dynamic updates.
The second one, 
\emph{eventual adaptation} (abbreviated \LG), 
is similar but weaker: 
it ensures that if the system enters into an error state then it will eventually reach a correct 
state. 
We believe that 
\OG and \LG
fit well in 
the kind of correctness analysis that is required in a number of emerging applications.  
For instance, 
on the provider side of a cloud computing application, 
these properties allow to check
whether a 
client is able to assemble faulty systems
via the aggregation of the provided services and the possible subsequent updates.
On the client side, it is possible to 
carry out forms of 
\emph{traceability analysis}, so as to 
prove that if 
the system exhibits an incorrect behavior, then 
it follows from a bug in the provider's infrastructure
and not from the initial aggregation and dynamic updates 
provided by the client.

In addition to  
error occurrence, 
the correctness of
adaptable processes 
must consider the fact 
that the number of 
modifications (i.e. update actions)
that can be 
applied to the system is typically \emph{unknown}. 
For this reason, we 
consider \OG and \LG in conjunction with 
the notion of \emph{cluster} of adaptable processes.
Given a system $P$ and a set $M$ of possible updates that
can be applied to it at runtime, 
its associated cluster considers $P$ together 
with an arbitrary number of instances of the updates in $M$.
This way, 
a cluster  formalizes adaptation and correctness properties 
of an initial system configuration (represented by an aggregation of adaptable processes)
in the presence of arbitrarily many sources of update actions.
For instance, in a cloud computing scenario 
the notion of cluster captures 
the cloud application as initially deployed by the client along with 
the options offered by the provider for 
its evolution at runtime.


\subsection*{Contributions}
The main technical results of the paper are summarized in Table \ref{t:results}.
The calculus \evol{1} is shown to be Turing complete, 
and both \OG and \LG are shown to be 
\emph{undecidable} for \evol{1} processes. 
The Turing completeness of \evol{1} says much on the expressive power of update actions.
In fact, it is known that fragments of CCS without restriction
can be translated into finite Petri nets (see, e.g.,  the discussion in \cite{Bravetti09}), and so they are not Turing complete.
Update actions in \evol{} thus allow to ``jump''  from finite Petri nets to a Turing complete model.
We then show that in \evol{2} 
\OG is \emph{decidable}, 
while \LG remains \emph{undecidable}. 
Interestingly,  \LG is already undecidable in \evold{3}, while it is \emph{decidable} in \evols{3}.

\begin{table}[t]
\begin{center}
\begin{tabular}{c|c|c}
		& \evold{} -- Dynamic Topology & \evols{} -- Static Topology \\
\hline \hline
\evol{1}	&~ \OG undec ~/~\LG undec ~& ~\OG undec~/~\LG undec~\\
\hline
\evol{2}	& ~ \OG dec~/~\LG undec ~ & ~ \OG dec~/~\LG undec \\
\hline
\evol{3}	& ~ \OG dec~/~\LG undec ~ & \OG~dec~/~\LG dec 
\end{tabular}
\end{center}
\caption{\label{t:results} Summary of  (un)decidability results for dialects of \evol{}.}
\end{table}

We now comment on the proof techniques.
The decidability of 
\LG in \evols{3} 
is proved by resorting to 
Petri nets: 
\LG is reduced to a problem on Petri nets that we call \emph{infinite visits} which, in turn, can 
be reduced to \emph{place boundedness}---a decidable problem  for Petri nets. 
For the decidability of 
\OG we appeal to the theory of 
\emph{well-structured transition systems}  \cite{FinkelS01,AbdullaCJT00}
and its associated results. 
In our case,  such a theory 
must be coupled with Kruskal's theorem~\cite{kruskal60} 
(which allows to deal with terms whose syntactical tree structure has an unbounded depth), and
with the calculation of the predecessors of target terms in the context of trees with unbounded depth 
(which is necessary in
order to deal with arbitrary aggregations
and dynamic updates that may generate new adaptable processes). 
This combination of techniques 
proved to be very challenging.
In particular, 
the technique is more complex than the one given in~\cite{AB09}, which relies on a bound on the depth of trees, or the one 
in~\cite{WZH10}, where only topologies with bounded paths are taken into account.
Kruskal's theorem is also used in \cite{Bravetti09} for studying the decidability properties of calculi with 
exceptions and compensations. The calculi considered in \cite{Bravetti09}
are \emph{first-order}; in contrast, 
\evol{} can be considered as a \emph{higher-order} process calculus (see Section~\ref{s:rw}).

The undecidability results are obtained via encodings of Minsky machines \cite{Minsky67}, 
a well-known Turing complete model. 
In particular, the encodings that we provide 
for showing undecidability of  \LG in \evol{2} and \evold{3} 
do not reproduce 
faithfully the corresponding machine, but only finitely many steps
are wrongly simulated. 
Similar techniques have been used to prove the undecidability
of repeated coverability in reset Petri nets~\cite{E03}, 
but in our case
their application revealed much more complex;
this is particularly true for the case of \evold{3} 
where there is 
no native mechanism for \emph{removing} an arbitrary amount of processes.  
Moreover, as in a cluster there is no a-priori knowledge of the 
number of modifications that will be applied to the system, we need
to perform a parametric analysis. Parametric verification has been
studied, e.g., in the context of broadcast protocols
in both fully connected~\cite{EsparzaFM99} and ad-hoc 
networks~\cite{DelzannoSZ10}.
 Differently from \cite{EsparzaFM99,DelzannoSZ10}, 
in which 
the number of nodes (or the topology) of the network 
is unknown,
we consider systems in which there is a known part  (the initial
system $P$), and there is another part composed of an unknown
number of process instances  (taken from 
$M$, the set of possible modifications). \\

\noindent Summing up, in the present paper we make the following contributions:
\begin{enumerate}[(1)]

\item We introduce $\evol{}$, a core calculus of adaptable processes.
\evol{} allows to express a wide range of patterns of process evolution and runtime adaptation.
By means of structural and behavioral characterizations, we 
identify different meaningful variants of the language.  
We are not aware of other process calculi 
tailored to the joint representation of 
evolution and adaptation concerns 
in concurrent systems.

\item 
We introduce bounded and eventual adaptation---two correctness properties for adaptable processes---and study their (un)decidability 
in each of the  variants of \evol{}.
We do so by considering systems as part of clusters which define their evolvability along time.
To the best of our knowledge, ours is the first study of the (un)decidability of 
adaptation properties for dynamically evolvable processes.

\end{enumerate}

\bigskip

\noindent This paper is an extended, revised version of the conference paper~\cite{BGPZFMOODS}. 
In addition to provide full details of the technical results, 
here we 
thoroughly develop the structural and behavioral characterizations of adaptable processes.
This way, we present a unified treatment of the 
 distinction between the static and dynamic topologies of adaptable processes, as well as of the three different update patterns.
These ideas were treated only partially  in~\cite{BGPZFMOODS}, 
where the family \evol{2} was called \evol{-}.
In particular, new results not presented in~\cite{BGPZFMOODS}  include 
the relationship between static and dynamic topologies (Section~\ref{ss:stdyn}) and 
the decidability of \LG in \evols{3} by resorting to Petri nets (Section~\ref{sec:petri}).
Moreover, several 
examples and extended discussions are included.  
Section~\ref{ss:facs} is based on the short paper~\cite{BGPZFACS}.

\subsection*{Structure of the document}
The rest of this paper is structured as follows.
The \evol{} calculus, its different variants, and several associated results  are presented in Section~\ref{s:calculi}. 
The two verification problems are defined in Section~\ref{s:prop}.
Section~\ref{s:examp} presents extended examples of modeling in \evol{}, in several emerging applications.
Section~\ref{s:prelim} collects basic definitions and results on Minsky machines, well-structured transition systems, and Petri nets.
(Un)decidability results for \evol{1}, \evol{2}, and \evol{3} are detailed in Sections \ref{s:ev1}, \ref{s:ev2}, and \ref{s:ev3}, respectively.
Section~\ref{s:rw} presents some additional discussions, and reviews some related works.
Section~\ref{s:conc} concludes. While proofs of the main results are included in the main text, technical details for some other results 
 (most notably, correctness proofs for the encodings) 
 are collected in the Appendix.


\section{A Calculus of Adaptable Processes}\label{s:calculi}


We begin by presenting the $\evol{}$ calculus and its different variants.
Then, we introduce the operational semantics of the calculus, 
and establish the relationship
between static and dynamic topologies of adaptable processes.

\subsection{Syntax} 

The $\mathcal{E}$ calculus 
is a variant of CCS \cite{Milner89} without restriction and relabeling, 
and extended with constructs for evolvability. 
As in CCS, in $\mathcal{E}$ 
processes can perform actions or synchronize on them.  
We presuppose a countable
set $\mathcal{N}$ of names, ranged over by $a,b$, possibly decorated as 
$\overline{a},  \overline{b} \ldots $ and $\til{a}, \til{b} \ldots $. 
As customary, we use $a$ and $\outC{a}$ to denote atomic input and output actions, respectively.
The syntax of $\mathcal{E}$ processes 
extends that of CCS with 
 primitive notions of \emph{adaptable processes} $\component{a}{P}$
 and \emph{update prefixes} $\update{a}{U}$:
 \[
P        ::=  \component{a}{P} \sepr 
          P \parallel P  \sepr \sum_{i \in I} \pi_i.P_i  \sepr ! \pi.P 
          \quad \quad \quad 
\pi   ::=  a \sepr \outC{a} \sepr \update{a}{U}
\]
Above, 
the $U$ in the update prefix
$\update{a}{U}$ is an \emph{update pattern}: it represents   
a context, i.e., a process with zero or more \emph{holes}   (see Definition \ref{d:contexts} below).
The intention is that when an update prefix is able to interact, 
the current state of  an adaptable process named $a$ 
is used to fill the holes in the update pattern $U$. 
Given a process $P$, 
process $\component{a}{P}$ denotes 
the adaptable process $P$ \emph{located at} $a$.
Notice that $a$ acts as a \emph{transparent} locality: process $P$ can evolve on its own, and  interact freely with external processes.
Localities can be nested, so as to form suitable hierarchies of adaptable processes.
The rest of the syntax follows standard lines.
A process $\pi.P$ performs prefix $\pi$ and then behaves as $P$. 
Parallel composition $P \parallel Q$ decrees the concurrent execution of $P$ and $Q$.
We abbreviate $P_{1} \parallel \cdots \parallel P_{n}$ as $\prod_{i=1}^{n} P _{i}$, and
use $\prod^{k} P$ to denote the parallel composition of $k$ instances of process $P$.
Given an index set $I = \{1,..,n\}$, the guarded sum $\sum_{i \in I} \pi_{i}.P_{i}$ represents an exclusive choice
over $\pi_{1}.P_{1}, \ldots, \pi_{n}.P_{n}$.
As usual, we write $\pi_{1}.P_{1} + \pi_{2}.P_{2}$ if ${|}I{|}=2$, and $\nil$ if $I$ is empty. 
Process $!\, \pi.P$ defines guarded replication, i.e., 
infinitely many occurrences of $P$ in parallel, which are triggered by prefix $\pi$.
 
We now define  a general way 
of extending the grammar of process languages with holes, so as to define update patterns.
Intuitively, we extend rule productions with a hole (denoted~$\bullet$), distinguishing between rule productions
for process expressions (so-called \emph{process categories}) from the rest.
In particular, we would like to avoid adding holes to rule productions for prefixes (i.e., productions for $\pi$ in the syntax).

\begin{defi}\label{d:contexts}
%
Given a process category $E$, we denote with $E_{\bullet}$ the process category with rule productions obtained from those of $E$ by: 
\begin{enumerate}[(1)]
\item adding a new rule ``$E_{\bullet} ::= \bullet$''; 
\item replacing every rule ``$E ::= term$'' of $E$
with a rule ``$E_{\bullet} ::= term_{\bullet}$'', 
where ``$term_{\bullet}$'' is obtained from ``$term$'' by syntactically replacing all process categories $F$ occurring in ``$term$'' by $F_{\bullet}$. 
\end{enumerate}\smallskip
\end{defi}
 
\noindent  Given an update pattern $U$ and a process $Q$, we define $\fillcon{U}{Q}$
 as the process obtained by filling in those holes in $U$ not occurring inside update prefixes with $Q$.

\begin{defi}\label{d:fillit}
The effect of replacing the holes in 
an update pattern $U$ with a process $Q$, 
denoted \fillcon{U}{Q}, 
 is defined inductively on $U$ as follows:
\begin{align*}
\fillcon{\bullet\,}{Q} &= Q & 
\fillcon{(U_{1} \parallel U_{2})}{Q} & =  \fillcon{U_{1}\,}{Q} \parallel \fillcon{U_{2}\,}{Q} \\
\fillcon{\component{a}{U}}{Q} &= \component{a}{\fillcon{U}{Q}} &
\fillcon{\bigg(\sum_{i \in I}\pi_{i}.U_{i}\bigg)}{Q} &= \sum_{i \in I}\pi_{i}.\fillcon{U_{i}\,}{Q} \\
\fillcon{(! \pi.U)}{Q} &= !\pi.(\fillcon{U}{Q})
\end{align*}
\end{defi}\smallskip

\noindent This way, $\{ \cdot \}$ can be intuitively seen as a scope delimiter for holes $\bullet$  in $\update{a}{U}$.
Indeed, it is worth observing that Definition \ref{d:fillit} does not replace holes inside prefixes; 
this ensures a consistent treatment of nested update actions.

We now move on to consider different variants of this basic syntax by means of two different characterizations.

\subsubsection{A Structural Characterization of Update}
As anticipated in the Introduction, 
our structural characterization of update in
$\mathcal{E}$ 
defines
two families of languages,
 namely
$\mathcal{E}$ \emph{with dynamic topology} (denoted \evold{}) 
and $\mathcal{E}$ \emph{with static topology} (denoted \evols{}).
Here, ``dynamic'' refers to the ability of
creating and deleting new adaptable processes, something allowed in languages in \evold{} 
but not in those in \evols{}.
The definition of  $\evold{}$ and $\evols{}$ is
parametric on update patterns
$U_{}$.






\begin{defi}[Dynamic $\mathcal{E}$ -- \evold{}]\label{d:finiteccs}
The class of $\mathcal{E}$ processes with dynamic topology (\evold{}) is described by the following
grammar: 
$$
P        ::= \component{a}{P} \sepr
          P \parallel P  \sepr ! \pi.P \sepr \sum_{i \in I} \pi_i.P_i  \qquad \quad  \pi   ::=  a \sepr \outC{a} \sepr \update{a}{U} 
$$
where $U::=P_{\bullet}$, as in Definition \ref{d:contexts}.
\end{defi}

The definition of 
\evols{}
makes
use of two distinct process categories: $P$ and $A$.
Intuitively, $P$ correspond to processes defining the (static) topology of adaptable processes; these are populated by
terms $A$, which do not include subprocesses of the kind $\component{a}{Q}$.

\begin{defi}[Static $\mathcal{E}$ -- \evols{}]\label{d:eccsstatic}
The class of $\mathcal{E}$  processes with static topology (\evols{}) is described by the following
grammar: 
$$
\begin{array}{lcl}
P       & ::=& \component{a}{P} \sepr
          P \parallel P \sepr A \\ 
A       & ::=&  A \parallel A \sepr
          ! \pi.A \sepr \sum_{i \in I} \pi_i.A_i \qquad
\pi  ::=  a \sepr \outC{a} \sepr \update{a}{\component{a}{U} \parallel A}
\end{array}
$$ 
where the syntax $U::=P_{\bullet}$, as in Definition \ref{d:contexts}, considering both $P$ and $A$ as process categories. 
\end{defi}

Definition \ref{d:eccsstatic} relies on syntactic restrictions  to ensure that 
the nesting structure of adaptable processes in \evols{}  remains invariant.
The first restriction (i.e., no adaptable process is removed) is manifest in update prefixes, which are always of the form $\component{a}{U} \parallel A$; 
this forces the recreation of the adaptable process $a$ after every update, thus maintaining  the static structure of adaptable processes 
invariant.
For the same reason, 
holes can only occur inside the recreated adaptable process: this way, 
processes cannot be relocated outside $\component{a}{U}$.
The second restriction (i.e., no adaptable process is created) appears in the definition of $A$, 
which decrees that no new adaptable processes occur behind a prefix.
As we will discuss below, the
operational semantics 
ensures that these syntactic restrictions  
are preserved along process execution.

\begin{rem}\label{rem:stdyn}
Observe that 
every \evols{} process is, from a syntactic point of view,
also an \evold{} process.
In fact, the 
update pattern 
$U = a[U'] \parallel A$
is a particular case of the possible update patterns  for \evold{} processes.
The correspondence between processes in \evols{}
and \evold{} from the point of view of their operational
semantics will be made more precise by Lemma \ref{lem:statvsdyn}.
\end{rem}

\subsubsection{A Behavioral Characterization of Update}

We now move on to consider 
three concrete instances of 
update patterns $U$
and their associated 
variants of $\evold{}$ and $\evols{}$.

%

\begin{defi}[Update Patterns]\label{def:varianti}
We shall consider the following three instances of update patterns for \evols{} and \evold{}:
\begin{enumerate}[(1)]
\item {\bf Full $\mathcal{E}$ (\evold{1} and \evols{1}).}
The first 
update pattern
admits all kinds of contexts for update prefixes, i.e., $U ::= P_{\bullet}$.
These variants, corresponding to the above \evold{} and \evols{},
are denoted also with \evold{1} and \evols{1}, respectively. 

%
%
%

\item {\bf Unguarded $\mathcal{E}$ (\evold{2} and \evols{2}).}
In the second
update pattern, 
holes cannot occur in the scope of prefixes in $U$:
$$
U  ::=  P \sepr \component{a}{U}  \sepr    U \parallel U  \sepr \bullet  
$$
The variants of \evold{} and \evols{} that adopt this update pattern are denoted
\evold{2} and \evols{2}, respectively. 

\item {\bf Preserving $\mathcal{E}$ (\evold{3} and \evols{3}).}
In the third update pattern, the 
current state of the adaptable process is always preserved.
Hence,  
it is only possible to add new adaptable processes and/or behaviors in parallel or to relocate it:
$$
U  ::=  \component{a}{U}  \sepr U \parallel P \sepr \bullet
$$
The variants of \evold{} and \evols{} that adopt this update pattern are denoted 
\evold{3} and \evols{3}, respectively.

\end{enumerate}

\end{defi}

\subsection{Semantics} 

The semantics of \evol{} processes is given in terms of a 
Labeled Transition System (LTS). 
We  introduce some auxiliary definitions first.

\begin{defi}\label{d:srtcong}
\emph{Structural congruence} is  the smallest
congruence relation generated by the following laws: 
$P \parallel Q \equiv Q \parallel P$; \quad $P \parallel (Q \parallel R) \equiv (P \parallel Q) \parallel R$.
\end{defi}

\begin{defi}[Normal Form]\label{d:nform}
An \evol{} process $P$ is said to be in \emph{normal form} iff
$$P = \prod_{i=1}^{m} P_i \parallel \prod_{j=1}^{n} \component{a_j}{P'_{j}} $$
  where, for $i \in \{1,\ldots,m\}$, $P_i$ is not in the form $Q \parallel Q'$ or $a[Q]$, 
  and, for all $j \in \{1,\ldots,n\}$,  
${P'_{j}}$ is in normal form.
Note that if $m = 0$ then 
the normal form is simply 
$P = \prod_{j=1}^{n} \component{a_j}{P'_{j}}$; similarly, if $n=0$ then the normal form is $P = \prod_{i=1}^{m} P_i$.

\end{defi}


 \begin{lem}\label{lem:normalform}
 Every \evol{} process
 is structurally congruent
  to a process in normal form. \qed
 \end{lem}\nobreak%
We now define the \emph{containment structure denotation} of a process.
Intuitively, it captures the tree-like structure induced by the nesting of adaptable processes.


  \vspace{-4 pt}
 \begin{defi}[Containment Structure]\label{def:cstr}
  Let $P = \prod_{i=1}^{m} P_i \parallel \prod_{j=1}^{n} \component{a_j}{P'_{j}}$ be an \evol{} process in normal form.
The \emph{containment structure denotation} of $P$, denoted $\CStr(P)$,  is built as follows. 
  The root is labeled $\epsilon$, and has $n$ children: 
   the subtrees recursively built from processes $P'_{1}, \dots, P'_{n}$ 
   with roots labeled $a_1, \dots, a_n$ (instead of $\epsilon$), respectively.
When $n=1$, we say that the containment structure
$\CStr(P)$ is \emph{single-child}.
\end{defi}
 \begin{exa}\label{ex:cstr}
 Consider the processes $P$, $Q$, and $R$ defined as 
 $$
 P = P_{1} \parallel \component{b}{P_{2}} \parallel \component{a}{P_{3} \parallel \component{c}{S}}
 \qquad
 Q =  a.P_{2} \parallel \component{b}{P_{3}} \parallel \component{a}{\component{c}{S}}
 \qquad R = P_{1} \parallel \component{d}{S}
 $$ where 
 $P_{1}, P_{2}, P_{3}$ do not contain adaptable processes.
 Then, $P$ and $Q$ have the same 
 containment structure denotation; it is depicted in 
 Figure \ref{f:csd} (left). 
 As for $R$, the
 containment structure denotation $\CStr(R)$, that is single-child,
is depicted in Figure \ref{f:csd} (right).
 \begin{figure}[t]
 \begin{center}
\rowheight=1.0pc
\xytree{
       & \xynode[-1,1]{$\epsilon$} \\
       \xyterminal{$b$} & & \xynode[0]{$a$} \\
       & & \xytrinode{$c$}\\
       & & \xyterminal{$\CStr(S)$} 
}
\qquad \qquad
\xytree{
        \xynode[0]{$\epsilon$} \\
        \xytrinode{$d$}\\
        \xyterminal{$\CStr(S)$} 
}
\end{center}
\caption{Containment structure denotation for $P =  P_{1} \parallel \component{b}{P_{2}} \parallel \component{a}{P_{3} \parallel \component{c}{S}}$ and $R = P_{1} \parallel \component{d}{S}$, as in Example \ref{ex:cstr}.} \label{f:csd}
 \end{figure}
  \end{exa}
 
 Given an update pattern $U$, 
the following two definitions on 
\evols{} processes
indicate the number of holes 
and adaptable processes 
which syntactically occur in $U$, 
respectively.
In both cases, we do not consider occurrences inside nested update prefixes.


 \begin{defi}\label{d:numap}
Let $U$ denote an \evols{}  update pattern or an \evols{}  process.
The number of adaptable processes which occur in $U$, denoted
\numap{U}, is inductively defined as follows:
\begin{align*}
\numap{\bullet} & = 0 & \numap{U_{1} \parallel U_{2}} & =  \numap{U_{1}} + \numap{U_{2}} &  \numap{! \pi.U} &= 0 \\
\numap{\component{a}{U}} &= 1 + \numap{U} & \numap{\sum_{i \in I}\pi_{i}.U_{i}} &= 0\, 
\end{align*}

\end{defi}

\noindent Notice that in the above definition, as we are considering \evols{}  processes, the number of adaptable processes after a prefix is necessarily 0.

\begin{defi}
Let $U$ be an \evols{}  update pattern.
The number of holes which occur in $U$, denoted
\numholes{U}, is inductively defined as follows:
\begin{align*}
\numholes{\bullet} & = 1 & \numholes{U_{1} \parallel U_{2}} & =  \numholes{U_{1}} + \numholes{U_{2}} &  \numholes{! \pi.U} &= \numholes{U}\\
\numholes{\component{a}{U}} &= \numholes{U} &
\numholes{\sum_{i \in I}\pi_{i}.U_{i}} &= \sum_{i \in I}\numholes{U_{i}\,}  & &
 \end{align*}
\end{defi}
\noindent The following auxiliary notation will be useful to formalize the properties of  $\evols{}$ processes along reductions.\nobreak%
\begin{defi}
Let $U$ be an \evols{} update pattern. The number of prefixed holes occurring in $U$, denoted
\numph{U}, is inductively defined as follows:
\begin{align*}
\numph{\bullet} &= 0 & \numph{\component{a}{U}} &= \numph{U} & \numph{U_{1} \parallel U_{2}}  &=  \numph{U_{1}} + \numph{U_{2}} \\
 \numph{\sum_{i \in I}\pi_{i}.U_{i}} & =   \sum_{i \in I}\numholes{U_{i}} & \numph{! \pi.U} & =\numholes{U}
\end{align*}
\end{defi}

\begin{exa}
Let $P$ and $U$ be an \evols{} process and an \evols{} update pattern  defined as 
$$
P = 
\componentbbig{a}{\component{b}{Q_{1}} \parallel P_{1}} \parallel
\updatebig{a}{U}.P_{2} 
\qquad 
U = \outC{b}.\update{d}{\bullet \parallel U_{1}}.\nil \parallel \component{b}{a.\bullet \parallel \bullet}
$$
Then we have: 
\begin{enumerate}[$\bullet$]
\item $\numap{P} = 2 + \numap{Q_{1}} + \numap{P_{1}}  + 0$ and $\numap{U} = 1$
\item $\numholes{U} = 2$ and  $\numph{U} = 1$.
\end{enumerate}
\end{exa}


We are now ready to define an LTS semantics for \evols{} and another one for \evold{}.
Both LTSs are generated by the set of rules in Figure \ref{fig:ltswithalpha}; they 
only differ on a condition associated to update actions.
This is the content of the following definition.

\begin{defi}[LTS for $\evold{}$ and $\evols{}$]\label{d:lts}
Given transition labels 
\[
\alpha    ::=  ~~ a \sepr \outC{a} \sepr \component{a}{P} \sepr \update{a}{U} \sepr \tau
\]
the LTS for \evold{}, denoted $\arro{~\alpha~}_{d}$,  
is defined by the rules in Figure \ref{fig:ltswithalpha} in which, in rules \rulename{Tau3} and \rulename{Tau4},
we decree $\mathsf{cond}(U,Q){=}\mathtt{true}$. 

Similarly, 
the LTS for \evols{}, denoted $\arro{~\alpha~}_{s}$,  
is defined by the rules in Figure \ref{fig:ltswithalpha} in which,
 in rules \rulename{Tau3} and \rulename{Tau4},
  we decree that $\mathsf{cond}(U,Q)$ holds if 
we have:
  \begin{enumerate}[(1)]
  \item $\CStr(\component{a}{Q}) = \CStr(\component{a}{\fillcon{U'}{Q}} \parallel A)$  where $U=\component{a}{U'}\parallel A$, for some $U', A$,  and 
  \item $\numph{U} > 0 \Rightarrow \numap{Q} = 0$.
  \end{enumerate}
\end{defi}

\begin{rem}\label{r:fbranch}
The LTS for \evols{} and \evold{} are finitely branching. 
The proof proceeds by induction on the syntactic structure of terms; the base cases are $\sum_{i\in I}\pi_i.U_i$ and $!\pi.U$.
\end{rem}

\begin{figure}[t]
$$
\inferrule[\rulename{Comp}]{}{\component{a}{P} \arro{~\component{a}{P}~}  \star}
\qquad 
\inferrule[\rulename{Sum}]{}{\sum_{i\in I} \pi_i.P_i \arro{~\pi_j~}  P_j  ~~(j \in I)}  
\qquad
\inferrule[\textsc{(Repl)}]{}{!\pi.P \arro{~\pi~}  P \parallel !\pi.P }
$$
$$
\inferrule[\rulename{Loc}]{P \arro{~\alpha~} P'}{\component{a}{P} \arro{~\alpha~}  \component{a}{P'}}
\quad 
\inferrule[\rulename{Act1}]{P_1 \arro{~\alpha~} P_1'}{P_1 \parallel P_2 \arro{~\alpha~} P'_1 \parallel P_2}			
\quad
\inferrule[\rulename{Tau1}]{P_1 \arro{~a~} P_1' \andalso P_2 \arro{~\outC{a}~} P'_2}{P_1 \parallel P_2 \arro{~\tau~}  P'_1 \parallel P'_2}
$$
$$
\inferrule[\rulename{Tau3}]{P_1 \arro{~\component{a}{Q}~} P_1'\andalso P_2 \arro{~\update{a}{U}~} P_2'  \andalso \mathsf{cond}(U,Q)}{P_1 \parallel P_2 \arro{~\tau~} P_1'\sub{ \fillcon{U}{Q}  }{\star} \parallel P_2'}
$$

\caption{LTS for \evols{} and \evold{}.
Rules \rulename{Act2}, \rulename{Tau2}, and \rulename{Tau4}---the symmetric counterparts of 
\rulename{Act1}, \rulename{Tau1}, and \rulename{Tau3}---have been omitted.} \label{fig:ltswithalpha}
\end{figure}

We give intuitions on  both LTSs.
In addition to the standard CCS actions (input, output, $\tau$), we consider two
complementary actions for process update: 
$\update{a}{U}$ and $\component{a}{P}$.
The former represents the availability of an update pattern $U$ for the adaptable process at $a$;
the latter expresses the fact that the adaptable process at $a$, with current state $P$, is ready to update. 
We often write $\arro{~\alpha~}$ instead of $\arro{~\alpha~}_{d}$ and $\arro{~\alpha~}_{s}$; 
the actual LTS used in each case will be clear from the context.
Similarly, we define $\arro{~~~}$ as $\arro{~\tau~}$.

In Figure \ref{fig:ltswithalpha}, 
rule \rulename{Comp} represents the contribution of a process at $a$ in an update operation; 
we use  $\star$ to denote a unique placeholder. 
Rule \rulename{Loc}
formalizes
transparency of localities.
Rules \rulename{Sum}, \rulename{Repl}, 
\rulename{Act1},
and \rulename{Tau1} 
 are standard.
 Rule \rulename{Tau3} formalizes process evolvability.
To realize the 
evolution of an adaptable process at $a$, it requires: 
(i)  a process $Q$---which represents its current state; 
(ii) an update action offering an update pattern  $U$ for updating the process at $a$---which is represented in $P'_{1}$ by $\star$ (cf. rule \rulename{Comp});
(iii) that $\mathsf{cond}(U,Q)$ holds (cf. Definition \ref{d:lts}).
As a result, 
$\star$ in $P'_{1}$ is replaced with process $\fillcon{U}{Q}$ (cf. Definition \ref{d:fillit}).
 
It is useful to elaborate on the definition of $\mathsf{cond}(U,Q)$---the only point in which the LTS of \evold{} and that of \evols{} differ.
While 
$\mathsf{cond}(U,Q)$  does not have influence on the update actions of 
\evold{} processes, it does ensure that the syntactic restrictions associated to \evols{} processes are preserved along transitions. 
As specified in Definition \ref{d:lts}, 
the condition 
for \evols{} processes 
is given in two parts.
The first part ensures that the current structure of nested adaptable processes---the containment structure denotation from $\component{a}{Q}$---is preserved once $Q$ is substituted into $U$ as a result of the transition.
The second part of the condition ensures that no new adaptable processes will appear behind prefixes as a result of 
the update operation. 
Recall that by the syntactic restrictions enforced by 
Definition \ref{d:eccsstatic}, adaptable processes cannot occur behind prefixes.
In fact, and  using the terminology introduced in that definition, 
the syntax of \evols{}  decrees that only processes in process category $A$ (which do not contain adaptable processes)
can occur behind prefixes.
The second part of the condition ensures precisely this.  
As a simple example, 
this  part of the condition rules out the synchronization of 
adaptable process $\component{b}{\component{a}{\nil}}$
with  update prefix
$\update{b}{\component{b}{\component{a}{b.\bullet}}}.Q$,
as 
it would lead to the non static process 
$\component{b}{\component{a}{b.\component{a}{\nil}}} \parallel Q$.

By considering the syntactic restrictions associated to \evols{} processes,
the following lemma characterizes  the conditions under which $\mathsf{cond}(U_{0},Q)$ holds for them.



\begin{lem}
\label{lem:statvsdyn}
Let $Q$ and $U_{0} = \component{a}{U} \parallel A$   be an $\evols{}$ process and an $\evols{}$ update pattern, respectively. Also, let $A$ be as in Definition \ref{d:eccsstatic}. We have
\begin{equation}
\CStr(\component{a}{Q}) = \CStr(\component{a}{\fillcon{U}{Q}} \parallel A) \land  (\numph{U} > 0 \Rightarrow \numap{Q} = 0)\label{eq:st}
\end{equation}
if and only if one of the following holds:
\begin{enumerate}[\em(0)]
\item[\em(0)] $\numholes{U}=0 \wedge \CStr(Q) = \CStr(U)$.
\item[\em(1)] $\numholes{U}=1  \wedge \numap{U} = 0 \wedge (\numph{U} > 0 \Rightarrow \numap{Q} = 0)$ 
\item[\em(2)] $\numholes{U}>1 \wedge \numap{U} = 0\wedge \numap{Q} = 0$.
\end{enumerate}
\end{lem}
\proof
The ``if'' direction is straightforward by observing that
by definition $\numap{A}=0$.
Therefore, 
(\ref{eq:st}) reduces to 
$$\CStr(Q) = \CStr(\fillcon{U}{Q}) \land  (\numph{U} > 0 \Rightarrow \numap{Q} = 0)$$
and the analysis focuses on the structure of $U$. Hence if $\numholes{U}=0$ then immediately from~(\ref{eq:st}) we have $\CStr(Q) = \CStr(U)$.
If $\numholes{U}=1$ then as  $\CStr(Q) = \CStr(\fillcon{U}{Q})$ we have that $\numap{U} = 0$ and from the second part of (\ref{eq:st}) we conclude $(\numph{U} > 0 \Rightarrow \numap{Q} = 0)$.
Finally, if $\numholes{U}>1$ following from   $\CStr(Q) = \CStr(\fillcon{U}{Q})$ we conclude $\numap{U} = 0$ and $\numap{Q} = 0$.

As for the "only if" direction, we consider each item separately:
\begin{enumerate}[$\bullet$]
\item Item (0):  
Then $Q$ occurs exactly once only at the left-hand side of the desired equality.
Using the first part of the item (i.e., $\numholes{U}=0$) we infer that  $\fillcon{U}{Q} = U$.
Since by definition $\numap{A}=0$, we have that 
the second part of the item (i.e., $\CStr(Q) = \CStr(U)$) is enough to obtain
$\CStr(\component{a}{Q}) = \CStr(\component{a}{U} \parallel A)$, as wanted.

\item Item (1): Then $Q$ occurs exactly once in both sides of the desired equality. 
The second part of the item (i.e., $\numap{U} = 0 $) guarantees that 
$\fillcon{U}{Q}$ does not involve any adaptable processes different from those in $Q$.
The third condition ensures that no adaptable processes occur behind prefixes: 
if $Q$ has adaptable processes then it should necessarily occur at the top level in $\fillcon{U}{Q}$.
Hence, the thesis follows.

\item Item (2): Then $Q$ occurs exactly once in the right-hand side of the equality, and arbitrarily many times in the left-hand side.
The second part of the condition, on the number of adaptable processes in $U$, follows the same motivations as in the previous case.
Given the possibility of arbitrarily many occurrences of $Q$ in the left-hand side, 
the only option 
to ensure identical containment structure denotations in both sides
is to forbid adaptable processes inside $Q$, hence the third part of the condition. \qed
\end{enumerate}


\noindent The following lemma is standard:
\begin{lem}
Let $P$ be an \evol{} process.
Structural congruence is preserved by reduction: 
if $P \equiv Q$ and $P \pired P'$, then also $Q \pired Q'$ for some $P' \equiv Q'$.\qed
\end{lem}

The following lemma states that \evols{} processes are closed under reduction.
Hence, the operational semantics of \evols{} preserves the syntactic conditions of Definition \ref{d:eccsstatic}.

 \begin{lem}[Static topologies are preserved by reduction]\label{l:esred}
 Let $P$ be an \evols{} process.
 If $P \pired P'$ then
 also $P'$ is an  \evols{} process. Moreover, $\CStr(P)=\CStr(P')$.
 \end{lem}
 
\proof
 By induction on the derivation of $P \arro{~\tau~} P'$.
 See Appendix \ref{ap:esred}, Page \pageref{ap:esred}.\qed

\newcommand{\proj}[2]{\ensuremath{#1\downarrow#2}}

\subsection{From Static to Dynamic Topologies}\label{ss:stdyn}
%
%
We have already remarked that from a syntactic
point of view every \evols{} process is also an \evold{} process.
As far as the operational semantics is concerned, 
an \evols{} process could have less possible computations
due to the additional constraint $\mathsf{cond}(U,Q)$ 
of the rules \rulename{Tau3} and \rulename{Tau4}. Nevertheless, 
in this section we show that it is always possible to translate 
a process with static topology into a 
process with dynamic topology which has the same 
semantics (the two LTSs are isomorphic). More precisely, we will
define an encoding 
$\dyn{\cdot}: \evols{} \to \evold{} $
such that the following holds: 
$$P \pired_s P'\text{ if and only if }\dyn{P} \pired_d \dyns{P'}{S}$$


We start by presenting some auxiliary definitions.
\begin{defi}\label{d:cstrs}
Let $P$ be an \evols{} process.
We define the set
$$\CStrs(P) = \{ \CStr(a[P']) \mid a[P'] \text{ is a subterm of } P \}$$\smallskip 
\end{defi}

\noindent Hence, $\CStrs(P)$ is a set of trees: 
it contains the 
containment structure denotations of  the adaptable processes occurring in  $P$. 
Notice that by construction $\CStrs(P)$ is a set of single-child containment structure denotations.

\begin{exa}
Let $P$ be as  in Example \ref{ex:cstr}. 
Then, we have $$\CStrs(P) = \{\CStr(\component{a}{P_{3} \parallel \component{c}{S}}), \CStr(\component{b}{P_{2}}), \CStr(\component{c}{S})\} \cup \CStrs(S)$$
\end{exa}




\begin{conv}\label{not:enc} Below $P$ and $a$ stand for an \evols{} process and a name, respectively.
\begin{enumerate}[$\bullet$]
\item Let $S$ be   
a set of 
containment structure denotations. We write $\proj{S}{a}$  
 to represent the 
 subset of single-child containment structure denotations of $S$ in which the label of the only child of the root is $a$. 

\item We assume an injective function $\varphi$ 
 that associates
containment structure denotations
to 
names in $\mathcal{N}$.
We use $\kappa,\kappa',\ldots$ to range over the codomain of  $\varphi$.
Moreover, 
for every $P$ such that $\numap{P}= 0$,  we fix $\ecs{\CStr(\component{a}{P})} = \kappa_{a}$.
The definition of $\varphi$ extends to sets of containment structure denotations as expected; 
 this way, e.g., $\ecs{\proj{S}{a}}$ stands for the set of names associated to those single-child
 containment structure denotations in $S$ with label $a$. 
  With a slight abuse of notation, we sometimes write $\ecs{\component{a}{P}}$ instead
 of $\ecs{\CStr(\component{a}{P})}$.\smallskip
 \end{enumerate}
\end{conv}
 
\noindent  We are now ready to present the definition of $\dyn{\cdot}$.
 
\begin{defi}
\label{def:din}
Adopting the notations in Convention \ref{not:enc}, 
let $P $ and $U$ be an \evols{} process and an \evols{} update pattern, respectively.
Also, let $S$ be a set of containment structure denotations such that $\CStrs(P) \subseteq S$. 
Moreover, assume $\mathrm{err} \notin \ecs{S}$. 
The encoding of $P$ into an \evold{} process over $\mathcal{N}$, denoted $\dyn{P}$, 
is inductively defined in Figure \ref{f:encsd}, where
\begin{enumerate}[$\bullet$]
\item $C1$ stands for $\numholes{U}=0$; 
\item $C2$ stands for $(\numholes{U}=  1\wedge \numph{U} > 0 \land \numap{U} = 0) \lor (\numholes{U}> 1 \land \numap{U} = 0)$; 
\item $C3$ stands for $\numholes{U}= 1 \wedge \numph{U} = 0 \land \numap{U} = 0$; 
\item $C4$ stands for $\numholes{U} \neq 0 \wedge \numap{U} \neq 0$.
\end{enumerate}
\end{defi}

We now comment on the definition in Figure \ref{f:encsd}.
Unsurprisingly, the encoding only concerns adaptable processes and update prefixes; input and output prefixes are not modified (cf. line $(6)$), 
 and guarded sum, parallel composition, and holes are treated homomorphically (cf. lines $(7)$, $(8)$, and $(9)$, respectively).
 Intuitively, the encoding 
 captures correct update actions by 
 renaming every adaptable process and update prefix according to their containment denotation structure.
 This way, an adaptable process  $\component{a}{P}$ 
 is translated into an adaptable process on name $\kappa$, 
 which depends on its containment structure denotation (cf. line $(1)$).
The intention of this renaming is to allow synchronization only with update prefixes on name $\kappa$, 
that is, with update prefixes having the same containment structure denotation; 
this way, condition $\CStr(U) = \CStr(Q)$ in the LTS of \evols{} is enforced via name equality.
As for the encoding of a process $P$ at $a$, it is important to observe that 
 the definition of \evols{} ensures that holes syntactically occurring in $P$ do not occur at top level---they can only appear inside an update prefix.
As such, they are handled by lines $(2)$--$(7)$ in recursive applications of the encoding.

Clearly, update prefixes must be modified accordingly; 
there are four different possibilities,  represented by conditions $C1$--$C4$ of Definition~\ref{def:din}:
\begin{enumerate}[$\bullet$]
\item $C1$ captures the cases in which the update pattern $U$  does not contain holes, i.e., $U$ is a process. 
Update prefixes with update patterns of this kind are encoded homomorphically, renaming the prefix accordingly (cf. line $(2)$).
Together with the above explained renaming of adaptable processes with respect to  the structure of their contents, 
this condition corresponds to Lemma \ref{lem:statvsdyn}(0).

\item $C2$ captures the cases in which the update prefix is only meant to interact with adaptable processes whose content have no adaptable processes. 
As explained before, 
and by the definition of $\varphi$,
these are adaptable processes of the form $\component{\kappa_{a}}{P}$ (with $\numap{P}=0$); this explains the encoding described in line $(3)$.
According to Lemma~\ref{lem:statvsdyn}, this is the case
when (i) 
the update pattern $U$ of the update prefix has exactly one hole that occurs behind a prefix (cf. Lemma \ref{lem:statvsdyn}(1) when $\numph{U} > 0$)
or (ii) $U$ has more than one hole (cf. Lemma \ref{lem:statvsdyn}(2)).

\item $C3$ captures the cases in which the update pattern $U$  has exactly one hole which does not occur behind a prefix.
These are update prefixes that may synchronize with any adaptable process at name $a$. 
In order to account for all the possibilities, 
each non replicated update prefix at $a$ is encoded as a sum of prefixed processes,
each summand corresponding to an update prefix on a name $\kappa_{i} \in \ecs{\proj{S}{a}}$. 
The only difference between the summands 
is the name of the update prefix; the update pattern within the update prefix and  its continuation is the same for all of them  (cf. line $(3a)$).
When the update prefix is replicated, rather than the sum  of all possible adaptable processes, 
we consider their product (cf. line $(3b)$).
This condition corresponds to Lemma \ref{lem:statvsdyn}(1) when $\numph{U} = 0$.

\item $C4$ captures those update patterns that do not adhere to any of the conditions of Lemma \ref{lem:statvsdyn}.
Hence, interaction with these prefixes may lead to ill-formed \evols{} processes. 
To prevent such undesirable interactions, these update prefixes are renamed into $\mathrm{err}$---a distinguished name signaling error  (cf. line $(5)$).\smallskip

\end{enumerate}

\begin{figure}
\begin{align*}
(1)~& \dyn{\component{a}{P}}&=&~\componentbbig{\kappa}{\dyn{P}}  \quad\text{with $\kappa = \ecs{\component{a}{P}}$} & \\
(2)~& \dyn{\xi \, \update{a}{\component{a}{U} \parallel A}.U_{1}}&=&~ \xi \, \updatebig{\kappa}{\componentbbig{\kappa}{\dyn{U}} \parallel \dyn{A}}. \dyn{U_{1}}\\
&&&\text{with $\kappa = \ecs{\component{a}{U}}$}& \text{if $C1$}  \\
(3)~& \dyn{\xi \, \update{a}{\component{a}{U} \parallel A}.U_{1}}&=&~ \xi \, \updatebig{\kappa_{a}}{\componentbbig{\kappa_{a}}{\dyn{U}} \parallel \dyn{A}}. \dyn{U_{1}}&  \text{if $C2$} \\
(3a)~& \dyn{\update{a}{\component{a}{U} \parallel A}.U_{1}} & = &~ \sum_{\kappa_{i} \in \ecs{\proj{S}{a}} } 
\updatebig{\kappa_{i}}{\componentbbig{\kappa_{i}}{\dyn{U}} \parallel \dyn{A}}.\dyn{U_{1}} & \text{if $C3$}\\
(3b)~& \dyn{! \, \update{a}{\component{a}{U} \parallel A}.U_{1}} & = &~ \prod_{\kappa_{i} \in \ecs{\proj{S}{a}}}
! \, \updatebig{\kappa_{i}}{\componentbbig{\kappa_{i}}{\dyn{U}} \parallel \dyn{A}}.\dyn{U_{1}} &  \text{if $C3$} \\
(5)~& \dyn{\xi \, \update{a}{\component{a}{U} \parallel A}.U_{1}}&=&~ \xi \, \update{\mathrm{err}}{\nil}. \dyn{U_{1}} & \text{if $C4$} \\
(6)~& \dyn{\xi \, \pi . U} & = &~ \xi \, \pi.\dyn{U}\quad \text{if $\pi = a$ or $\pi = \outC{a}$} & \\
(7)~& \dyn{\sum_{i \in I} \pi_i.U_{i}} & = & ~\sum_{i \in I} \dyn{\pi_i. U_{i}} & \\
(8)~& \dyn{U_{1} \parallel U_{2}} &= &~\dyn{U_{1}} \parallel \dyn{U_{2}} & \\
(9)~& \dyn{\bullet} &= &~\bullet &
\end{align*}
\small{Above, $\xi \pi.P$ denotes a possibly replicated prefixed process:
$\xi \pi.P$ is either $!\pi.P$ or $\pi.P$, with $\xi$ being the same on both sides of the definition.}
\caption{The encoding $\dyn{\cdot}: \evols{}~\to~\evold{}$ given in Definition \ref{def:din}. 
}\label{f:encsd}
\end{figure}


\noindent Before stating the correctness of the encoding, 
we illustrate it further  through a series of examples.

\begin{exa} Below, notice that by virtue of Definition \ref{d:eccsstatic},  $\numap{A_{i}} = 0$ for every $A_{i}$.
\begin{enumerate}[(1)]
\item Given the \evols{} process
$$P_{1} = \componentbbig{b}{\component{c}{A_{1} \parallel A_{2}}} \parallel \componentbbig{b}{\component{d}{e.A_{3}}} \parallel  \updatebig{b}{\component{c}{A_{4}}}.Q_{2}$$
we have the \evold{} process 
$$\dyn{P_{1}} = \componentbbig{\kappa_{1}}{\dyn{\component{c}{A_{1} \parallel A_{2}}}} \parallel \componentbbig{\kappa_{2}}{\dyn{\component{d}{e.A_{3}}}} \parallel  \updatebig{\kappa_{1}}{\dyn{\component{c}{A_{4}}}}.\dyn{Q_{2}}$$
with $\kappa_{1} = \ecs{\component{b}{\component{c}{A_{1} \parallel A_{2}}} }$, $\kappa_{2} = \ecs{\component{b}{\component{d}{e.A_{3}}}}$.
 Notice how the renaming to $\kappa_{2}$
rules out the possibility of an update action for the second adaptable processes on $b$.

\item Given the \evols{} process
\begin{align*}
P_{2} = &\component{c}{A_{1}} \parallel \component{c}{A_{2}} \parallel \component{d}{A_{3}} \parallel \componentbbig{d}{\component{e}{A_{4}}} \parallel \\
& \update{c}{\component{c}{\bullet \parallel \bullet} \parallel A_{5}}.Q_{1} \parallel \update{d}{\component{d}{A_{6} \parallel a.\bullet}}.Q_{2}
\end{align*}
we have the \evold{} process 
\begin{align*}
\dyn{P_{2}} = &
\componentbbig{\kappa_c}{\dyn{A_{1}}} \parallel \componentbbig{\kappa_c}{\dyn{A_{2}}} \parallel \componentbbig{\kappa_d}{\dyn{A_{3}}} \parallel \componentbbig{\kappa_{1}}{\dyn{\component{e}{A_{4}}}} \parallel \\
& \updatebig{\kappa_c}{\component{\kappa_c}{\bullet \parallel \bullet} \parallel \dyn{A_{5}}}.\dyn{Q_{1}} \parallel \updatebig{\kappa_d}{\component{\kappa_d}{\dyn{A_{6}} \parallel a.\bullet}}.\dyn{Q_{2}}
\end{align*}
with $\kappa_{1} = \ecs{\component{d}{\component{e}{A_{4}}}}$. Notice how the renaming to $\kappa_{1}$
rules out the possibility of an update action for the second adaptable processes on $d$.

\item 
Given the \evols{} process $P_3$ defined as:
$$
\componentbbig{e}{\component{f}{A_{1}}} \parallel \componentbbig{e}{\component{g}{\component{h}{A_{2}} \parallel A_{3}}} \parallel (\updatebig{e}{\component{e}{\bullet} \parallel A_{4}}.Q_{1} \, + \, \updatebig{e}{\component{e}{\component{f}{\bullet \parallel \bullet}} \parallel A_{5}}.Q_{2})$$
we have (assuming $S$ to be minimal) the $\evold{}$ process
\begin{align*}
\dyn{P_{3}}  = & ~\componentbbig{\kappa_{1}}{\dyn{\component{f}{A_{1}}}} \parallel \componentbbig{\kappa_{2}}{\dyn{\component{g}{\component{h}{A_{2}} \parallel A_{3}}}} \parallel \\
& (\updatebig{\kappa_{1}}{\component{\kappa_{1}}{\bullet} \parallel \dyn{A_{4}}}.\dyn{Q_{1}} + 
 \updatebig{\kappa_{2}}{\component{\kappa_{2}}{\bullet} \parallel \dyn{A_{4}}}.\dyn{Q_{1}} + 
\update{\mathrm{err}}{\nil}.\dyn{Q_{2}})
\end{align*}
with $\kappa_{1}  = \ecs{\component{e}{\component{f}{A_{1}}}}$ and 
$\kappa_{2} = \ecs{\component{e}{\component{g}{\component{h}{A_{2}} \parallel A_{3}}} }$.\\
Observe how the first summand in $P_{3}$ has been duplicated in $\dyn{P_{3}}$, so as to account for the two possible update actions on $e$.
\end{enumerate}
\end{exa}

We are in place to state the promised correspondence between \evols{} and \evold{} processes:

\begin{thm}\label{stdynequiv}
Let $P$  be an \evols{} process.
Also, let $S$ be 
a set of containment structure denotations, such that  $\CStrs(P) \subseteq S$.
 Then we have:
$$P \pired_s P'\text{ if and only if }\dyn{P} \pired_d \dyns{P'}{S}$$
\end{thm}
\proof[Proof (Sketch)]
The proof is in two parts, one for the ``if'' direction and another other for the ``only if'' direction.
In both cases, we proceed 
by induction on the height of the derivation tree for $P \pired_{s} P'$ (resp. $\dyn{P} \pired_{d} \dyn{P'}$), with a case
analysis on the last applied rule.
For the former, we rely on the characterization of reduction for \evols{} processes given by Lemma~\ref{lem:statvsdyn}
so as to show that a reduction in the static side is preserved in the dynamic side.
As for the latter, the proof is similar, and exploits the fact that the encoding transforms update prefixes that may lead to incorrect update actions
into ``error'' update prefixes which are unable to participate in reductions. This ensures that for every dynamic reduction there is also a static reduction.
See Appendix \ref{ap:stdynequiv} in Page \pageref{ap:stdynequiv} for details.\qed


 \section{Correctness Properties: Bounded and Eventual Adaptation}\label{s:prop}



Here we define the correctness problems that we consider throughout the paper.
We would like adaptation properties defined in the most general way possible; 
this would allows us to analyze models of evolvable systems in different settings. 
For this purpose, 
our correctness properties are stated in terms of observability predicates, or  \emph{barbs}.
The definition of barbs is parameterized on the number of repetitions of a given signal.
We thus obtain a uniform definition for \emph{bounded} 
 and \emph{repeated} weak  barbs.

\begin{defi}[Barbs]\label{d:barb}
Let $P$ be an $\mathcal{E}$ process, and let $\alpha$ be an action in $\{a, \outC{a} \mid a \in \mathcal{N}  \}$.
We write $P \downarrow_{\alpha}$ if there exists a $P'$ such that $P \arro{~\alpha~} P'$. Moreover:
\begin{enumerate}[$\bullet$]
\item Given  $k > 0$, we write $P\barb{\alpha}^{k}$ 
iff there exist $Q_{1},\ldots, Q_{k}$ such that
$P \pired^{*} Q_1 \pired \ldots \pired Q_k$
with $Q_i \downarrow_\alpha$, for every $i \in \{1,\ldots, k\}$.
\item We write $P\barb{\alpha}^{\omega}$ iff
there exists an infinite computation
$P \pired^* Q_1 \pired Q_2 \pired  \ldots$
with $Q_i \downarrow_\alpha$ for every $i \in \mathbb{N}^+$.
\end{enumerate}
Furthermore, we  use $\negbarbk{\alpha}$ and $\negbarbw{\alpha}$ to denote the negation of $\barbk{\alpha}$ and $\barbw{\alpha}$.
\end{defi}


We shall consider two instances of the problem of reaching an error configuration in an aggregation of terms, or \emph{cluster}.
A \emph{cluster} is a process obtained as the parallel composition of an initial process $P$
with an arbitrary set of processes $M$ representing its 
possible subsequent
modifications.
That is, 
processes in $M$
may contain update actions on the names of the adaptable processes  in $P$, 
and therefore may potentially lead to its modification (evolution).

\begin{defi}[Cluster] \label{d:cluster}
Let $P, P_1,\ldots,P_n$ 
be \evol{} processes and $M=\{P_1,\ldots,P_n\}$. 
The set of clusters $\BC_P^M$ is defined as:
$$\BC_P^M = \big\{P \parallel \prod^{m_1} P_1 \parallel \dots \parallel \prod^{m_n} P_n \mid m_1, \dots, m_n \in \mathbb{N}\cup \{ 0 \} \big\}$$
\end{defi}

The  \emph{adaptation} problems 
below
formalize correctness of clusters
with respect to their ability for recovering from errors by means of update actions.
More precisely, 
given a set of clusters $\BC_P^M$
and a 
barb $e$ (signaling an error), 
we would like to know if all computations of processes in $\BC_P^M$
\begin{enumerate}[(1)]
\item have \underline{\emph{at most} $k$} consecutive states exhibiting $e$, or
\item have a \underline{\emph{finite}} number of consecutive states exhibiting $e$.
\end{enumerate}
We thus have the following definition:

\begin{defi}[Adaptation Problems]\label{def:adaptprob}
Suppose an initial process $P$,
a set of processes $M$, and a barb $e$.

\begin{enumerate}[$\bullet$]
\item  Given $k>0$, the \emph{bounded adaptation} problem (\OG) consists in checking  
whether for all  processes 
$R \in \BC_P^M$, $R \, \negbarbk{e}$ holds.

\item Similarly, the \emph{eventual adaptation} problem (\LG) consists in checking  
whether   
for all processes 
$R \in \BC_P^M$,  $R \, \negbarbw{e}$ holds.\smallskip
\end{enumerate}
\end{defi}

%

\noindent Similarly as processes, static clusters can be encoded into equivalent dynamic ones.

\begin{defi}
Let $P, P_{1},\ldots, P_{n}$ 
be \evols{} processes 
such that $M = \{P_1,\ldots,P_n\}$.
The static cluster set $\BC_P^M$ 
is transformed 
into a dynamic cluster set $\dyn{\BC_{P}^{M}}=\BC_{P'}^{M'}$ by taking $P' = \dyn{P}$ and $M' = \{\dyn{P_1},\ldots,\dyn{P_n}\}$, where $S = \CStrs(P) \cup \bigcup_{1 \leq i \leq n} \CStrs (P_i)$.
\end{defi}

\begin{thm}\label{th:clusterstat}
Let $P, P_1,\ldots,P_n$ 
be \evols{} processes such that $M=\{P_1,\ldots,P_n\}$. 
Then 
we have
$\dyn{\BC_{P}^{M}} = \{ \dyn{C} \mid C \in \BC_{P}^{M} \}$, where $S = \CStrs(P) \cup \bigcup_{1 \leq i \leq n} \CStrs (P_i)$.
\end{thm}
\proof
Immediate by observing that 
by Definition \ref{def:din},  
$\dyn{P}$ is an 
homomorphism with respect to parallel composition, i.e., $\dyn{P\parallel Q}= \dyn{P}\parallel \dyn{Q}$. \qed

Notice that, for every cluster $C$ in $\dyn{\BC_{P}^{M}}$ 
by construction 
we have $\CStrs(C) \subseteq S$.
Hence, the operational correspondence given by  Theorem~\ref{stdynequiv} is individually applicable to each cluster.


\section{Adaptable Processes, By Examples}\label{s:examp}

Next we present some concrete scenarios of adaptable processes 
and discuss their representation as \evol{} processes.
We also comment on how the adaptation properties proposed in the paper (and their associated decidability results) 
relate to such scenarios.


\subsection{Mode Transfer Operators}
In \cite{Baeten00}, 
dynamic behavior at the process level is defined by means of two so-called \emph{mode transfer} operators.
Given processes $P$ and $Q$, the \emph{disrupt} operator 
starts executing $P$ but at any moment it may abandon $P$ and execute $Q$ instead.
The \emph{interrupt} operator is similar,  but 
it returns to execute what is left of $P$ once $Q$ emits a termination signal.
We can represent similar mechanisms in \evol{} as follows:
$$
\mathsf{disrupt}_{a}(P,Q)  \midef \component{a}{P} \parallel \update{a}{Q} \qquad
\mathsf{interrupt}_{a}(P,Q)  \midef  \component{a}{P} \parallel \update{a}{Q \parallel t_{Q}.\bullet} 
$$
Assuming that $P$ can evolve on its own to $P'$, 
the semantics of \evol{} 
decrees that 
$\mathsf{disrupt}_{a}(P,Q)$ 
may evolve either to $\component{a}{P'} \parallel \update{a}{Q}$ 
(as locality $a$ is transparent)
or to $Q$ (which represents disruption at $a$).
Similarly, 
by assuming that $P$ was able to evolve into $P''$ just before being interrupted, process $\mathsf{interrupt}_{a}(P,Q)$ evolves
to $Q \parallel t_{Q}.P''$. 
Above, we assume that $a$ is  not used in $P$ and $Q$, and that 
termination of $Q$ is signaled at the designated name $t_{Q}$. 

These simple definitions show how
defining $P$ as an adaptable process at $a$ is enough to formalize its potential 
disruption/interruption.
It is worth observing that the encoding of 
$\mathsf{interrupt}_{a}(P,Q)$ can only be an \evold{1} process:
in the update action at $a$, 
there is a hole occurring behind a prefix (hence, it is not a \evol{2} process) and the 
topology of adaptable process is dynamic (since $a$ does not occur in $Q$, the adaptable process cannot be rebuilt after interruption).
In contrast, the encoding of 
$\mathsf{disrupt}_{a}(P,Q)$ is both an $\evold{1}$ and an $\evold{2}$ process, 
as in the update pattern there are no holes in the scope of prefixes (in fact, the update pattern does not have any holes).

\subsection{Dynamic Update in Workflow Applications}
Designing business/enterprise applications in terms of \emph{workflows} is a common practice nowadays.
A workflow is a conceptual unit that describes how  a 
number of \emph{activities} coordinate to achieve a particular task.
A workflow can be seen as a container of activities; such 
activities 
are usually defined in terms of simpler ones, and 
may be software-based (such as, e.g., ``retrieve credit information from the database'') or may depend on human intervention  
(such as, e.g., ``obtain the signed authorization from the credit supervisor'').  
As such, workflows  are typically long-running and have a transactional character. 
A workflow-based application usually consists of a \emph{workflow runtime engine} 
that contains a number of workflows running concurrently on top of it; a \emph{workflow base library} on which activities may rely on; and of a number of
 \emph{runtime services}, which are
 application dependent and implement things such as transaction handling and communication with other applications.
 A simple abstraction of a workflow application is the following \evol{} process:
 $$
  App \midef \componentbig{\nm{wfa}}{\, \componentbbig{\nm{we}}{\nmu{WE} \parallel \nm{W}_{1} \parallel \cdots \parallel \nm{W}_{k} \parallel \component{\nm{\nm{wbl}}}{\nmu{BL}}} \parallel \nmu{S}_{1} \parallel \cdots \parallel \nmu{S}_{j}\,}
 $$
 where the application is modeled as an adaptable process \nm{wfa} 
 which contains a workflow engine \nm{we} and a number of 
 runtime services $\nmu{S}_{1}, \ldots, \nmu{S}_{j}$.
In turn,  the workflow engine contains 
a number of workflows $\nm{W}_{1}, \ldots, \nm{W}_{k}$, a
process \nmu{WE} (which represents the engine's behavior and is left unspecified), 
and an adaptable process \nm{wbl} representing the base library (also left unspecified).
As described before, each workflow is composed of a number of activities.
We model each $\nm{W}_{i}$ as an adaptable process $\nm{w}_{i}$
containing 
a process $\nmu{WL}_{i}$
---representing the workflow's logic---, 
and $n$ activities. Each of them is formalized as an adaptable process $\nm{a}_{j}$ 
and an \emph{execution environment} $\nm{env}_{j}$:
$$
\nm{W}_{i} = \componentbig{\nm{w}_{i}}{\,  \nmu{WL}_{i} \parallel \prod_{j=1}^n \big(\component{\nm{env}_j}{\nmu{P}_{j}} \parallel \componentbbig{\nm{a}_{j}}{ !u_j. \update{\nm{env}_j}{\component{\nm{env}_j}{\bullet \parallel \nmu{A}_{j}}}}\big) \,}
$$
The current state of the activity $j$ is 
represented by 
process $\nmu{P}_{j}$ running in $\nm{env}_{j}$.
Locality $\nm{a}_{j}$ contains an update action for $\nm{env}_{j}$, which is guarded by $u_{j}$ and always available. 
As defined above, such an update action allows to add process $\nmu{A}_{j}$ to the current state of the execution environment of $j$.
%
It can also be seen as a procedure that is yet not active, and that becomes active only upon reception
of an output at  $u_{j}$ from, e.g., $\nmu{WL}_{i}$.
Notice that by defining  
update actions on $\nm{a}_{j}$ 
(inside $\nmu{WL}_{i}$, for instance)
we can describe the evolution of the execution environment.
An example of this added flexibility is the process
$$
U_{1} = !\,\nm{replace}_{j}.\updatebig{\nm{a}_{j}}{\componentbbig{\nm{a}_{j}}{!u_{j}.\updatebig{\nm{env}_j}{\component{\nm{env}_j}{\bullet \parallel \nmu{A}_{j}^{2}}}}}
$$
Hence,  
given an output  at $\nm{replace}_{j}$, process 
$ \component{\nm{a}_{j}}{!u_{j}.\update{\nm{env}_j}{\component{\nm{env}_j}{\bullet \parallel \nmu{A}_{j}}}} \parallel U_{1}$
evolves to 
$\component{\nm{a}_{j}}{!u_{j}.\update{\nm{env}_j}{\component{\nm{env}_j}{\bullet \parallel \nmu{A}_{j}^{2}}}}$
thus discarding $\nmu{A}_{j}$ in a \emph{future} evolution of $\nm{env}_{j}$.
This kind of \emph{dynamic update} is available in  commercial workflow engines, such as
the Windows Workflow Foundation (WWF) \cite{wwf}.
Above, 
for simplicity, 
we have abstracted from 
lock mechanisms that  
keep consistency between concurrent updates on $\nm{env}_{j}$ and $\nm{a}_{j}$.

In the above processes, it is worth observing that 
if processes $\nmu{A}_{j}$ and $\nmu{A}_{j}^{2}$ contain no adaptable processes, then $\nm{W}_{i}$ is an $\evols{3}$ process.
This is because the update action at $\nm{env}_j$ recreates the adaptable process, and preserves the previous state with a hole that is in parallel to $\nmu{A}_{j}$.
Otherwise,  $\nm{W}_{i}$ would be an $\evold{3}$ process, as the topology of adaptable processes would change as a result of an update action on  $\nm{env}_j$. 
For the sake of the example, suppose an emergency activity that executes inside the workflow:
process $\nmu{P}_{j}$ would emit a signal representing an urgent request, and 
an update action  at $\nm{env}_j$ would represent a response to the emergency,
implemented as process $\nmu{A}_{j}$.
The two adaptation problems are useful to represent the future state of the workflow in which the emergency has been controlled:
\LG refers to an \emph{undetermined} future state  in which the request signal disappears (meaning that the emergency will be eventually controlled); 
whereas \OG refers to a \emph{fixed} future state in which the request signal disappears (meaning that the emergency will be controlled within a certain  bound).
The 
topology of  $\nmu{A}_{j}$ is relevant in the light of our decidability results for these two properties:
if $\nm{W}_{i}$ is given as an $\evols{3}$ process, then both 
\LG and \OG
are decidable; otherwise, 
if $\nm{W}_{i}$ is given as an $\evold{3}$ process, then only 
\OG
would be decidable.

%

In the WWF, dynamic update can also take place at the
level of the workflow engine. 
This way, e.g., the engine may 
\emph{suspend} those workflows which have been inactive for a certain amount of time.
This optimizes resources at runtime, and favors active workflows. 
We can implement this policy as part of the process \nmu{WE} as follows:
$$
U_{2} = !\,\nm{suspend}_{i}.\updatebig{\nm{w}_{i}}{!\, \nm{resume}_{i}.\component{\nm{w}_{i}}{\bullet}}
$$
This way, given an output signal at $\nm{suspend}_{i}$, 
process 
$\component{\nm{w}_{i}}{\nmu{W}_{i}} \parallel U_{3}$
evolves to the persistent process $!\, \nm{resume}_{i}.\component{\nm{w}_{i}}{\nmu{W}_{i}}$
which can be reactivated at a later time.
Observe that, in case one considers policies such as $U_{2}$ then we would end up with an 
\evold{1} process, as the hole and an adaptable process occur guarded behind a prefix.


\subsection{Scaling in Cloud Computing Applications}

In the emerging cloud computing para\-digm, applications are deployed in 
the infrastructure offered by external providers.
Developers act as clients: they only pay for the resources they consume 
(usually measured as the processor time in remote \emph{instances})
and for associated services (e.g., performance metrics or automated load balancing).
Central to the paradigm is 
the goal of optimizing resources for both clients and provider.
An essential feature towards that goal is \emph{scaling}: 
the capability that cloud applications have for expanding themselves in times of high demand, 
and for reducing themselves when the demand is low. 
Scaling can be appreciated in, e.g., the number of running instances supporting the application, 
and may have important financial effects. 
Consequently, cloud providers such as Amazon's Elastic Cloud Computing (EC2) \cite{autoscaling} offer 
libraries and APIs and services for \emph{autoscaling}; 
also common are external tools which build on available APIs to implement sophisticated scaling policies.


Here we represent a cloud computing application as adaptable processes.
Our focus is in the formalization of scaling policies, 
drawing inspiration from 
the autoscaling library 
provided by 
EC2.
For scaling purposes, applications in EC2 are divided into \emph{groups}, each defining different scaling policies for 
different parts of the application. This way,  e.g., 
the part of the application deployed in Europe can have different scaling policies
from the part deployed in the US.
Each group is then composed of a number of identical instances implementing the web application, and of active processes implementing the scaling policies.
This scenario can be abstracted in \evol{} as the process
$App  \midef  G_{1} \parallel \cdots \parallel G_{n}$, with 
$$
G_{i}  =  \componentbbig{g_{i}}{\, I \parallel \cdots \parallel I \parallel S_{dw} \parallel S_{up} \parallel \nmu{CTRL}_{i} \,}
$$ 
where each group $G_{i}$ contains a fixed
number of running instances, each represented by 
$I = \component{\nmu{mid}}{\nm{A}}$, a process that abstracts an instance as an adaptable process 
with an identification $\nmu{mid}$ and state \nm{A}.
Also, $S_{dw}$ and $S_{up}$ stand for the processes implementing scaling down and scaling up policies, respectively.
Process $\nmu{CTRL}_{i}$ abstracts the part of the system which controls scaling policies
for group $i$.
In practice, this control relies on
external services (such as, e.g., services that monitor cloud usage and produce appropriate \emph{alerts}).
A simple way of abstracting scaling policies is the following:
$$
S_{dw}  =  \componentbbig{s_{d}}{\, !\,\nm{alert^{d}}.\prod^{j} \update{\nmu{mid}}{\nil}\, } \qquad
S_{up}  =  \componentbbig{s_{u}}{\, !\,\nm{alert^{u}}.\prod^{k} \updatebig{\nmu{mid}}{\component{\nmu{mid}}{\bullet} \parallel \component{\nmu{mid}}{\bullet}}\, }
%
$$
Given proper alerts  from $\nmu{CTRL}_{i}$, the above processes modify the number of running instances.
In fact, given an output at $\nm{alert^{d}}$ 
process $S_{dw}$ destroys $j$ instances.
This is achieved by leaving the inactive process as the new state of locality $\nmu{mid}$.
Similarly,  an output at $\nm{alert^{u}}$ 
process $S_{up}$ spawns $k$ update actions, each creating a new instance.

Observe that both $S_{dw}$ and $S_{up}$ are \evold{2} processes: since we represent instances as adaptable processes with state,
every modification enforced by the scaling policies will result in a different topology of adaptable processes. 
A correctness guarantee in this setting is that the cloud infrastructure satisfies the scaling requirements of client applications
within a fixed bound. 
More precisely, we would like to ensure that every scaling alert managed by $\nmu{CTRL}_{i}$ (requesting more instances, for instance) will disappear within a certain bound, meaning that the scaling request is promptly addressed by the cloud provider. This kind of reliability guarantees can be represented 
in terms of \OG, an adaptation problem which is decidable for \evold{2} processes.
Of course, the decidability of correctness guarantees depends much on their actual representations. Above, we have opted for simple, illustrative representations; clearly, different process abstractions may exploit other decidability results.


Autoscaling in  EC2 also allows to \emph{suspend} and \emph{resume}
the scaling policies themselves. 
To formalize this capability, we proceed similarly as  we did for process $U_{2}$ above.
This way, for the scale down policy,
one can assume that $\nmu{CTRL}_{i}$
includes a process 
$U_{dw} =  !\, \nm{susp_{down}}.\update{s_{d}}{! \, \nm{resume_{dw}}.\component{s_{d}}{\bullet}}$
which, provided an output signal on \nm{susp_{down}}, captures
the current 
policy, and evolves into a process that allows to 
resume it later on.
Using the same principle, other modifications to the policies are possible.
For instance, a natural request is to modify the scaling policies by 
changing the 
number of instances involved 
(i.e., $j$ in $S_{dw}$ and $k$ in $S_{up}$).
As before, if our specification includes the ability of suspending/resuming scaling policies as implemented by $U_{dw}$, then we would obtain an \evold{1} process.


\section{Preliminaries}\label{s:prelim}
We now introduce some background notions on Minsky machines, well-structured transition systems (WSTS), 
and Petri nets. 

\subsection{Minsky machines}
Our undecidability results will be obtained by encodings of \emph{Minsky machines} \cite{Minsky67}. 
A Minsky machine (\mm)  is a Turing complete 
model composed of a set of sequential, labeled   instructions, and two registers.   
Registers $r_j ~(j \in \{0,1\})$ can hold arbitrarily large natural numbers.   
Instructions $(1:I_1), \ldots, (n:I_n)$ can be of three kinds:  
$\mathtt{INC}(r_j)$ adds 1 to register $r_j$ and proceeds to the next instruction;  
$\mathtt{DECJ}(r_j,s)$ jumps to instruction $s$ if $r_j$ is zero, otherwise it decreases register $r_j$ by 1 and proceeds to the next instruction;
a $\mathtt{HALT}$ instruction stops the machine. 
A \mm includes a program counter $p$ indicating the label of the instruction  being executed.   

In its initial state, the machine has both registers set to $0$ and the program counter $p$ set to the first instruction.  
We assume that instructions are proper, in the sense that there is no program counter that refers to a non-existing instruction.
The \mm \emph{terminates} whenever the program counter is set to a $\mathtt{HALT}$ instruction.
%
A \emph{configuration} of a \mm is a tuple $(i,m_0,m_1)$; it consists of the current program counter and the values of the registers. Formally, the reduction relation  
over configurations of a \mm, denoted $\minskred$, is defined 
in Figure \ref{a:mm}.

Since \mmss are Turing complete, termination is undecidable.

\begin{thm}[Minsky
\cite{Minsky67}]
Minsky machines are Turing complete.
Hence, for a \mm it is undecidable whether it terminates.
\qed
\end{thm}

We shall exploit encodings into \mmss to prove undecidability of \LG and \OG. 
In our encodings, we sometimes make the unrestrictive assumption that 
at the beginning and at the end of the computation the
registers (must) contain the value zero.

 \begin{figure}[t]
  \begin{mathpar}  
 \inferrule[\textsc{(M-Inc)}]{i:\mathtt{INC}(r_j) \ \ m_j' = m_j + 1 \ \ m_{1-j}' = m_{1-j}}{(i,m_0,m_1)\minskred(i+1,m_0',m_1')}  
 \quad 
 \inferrule[\textsc{(M-Jmp)}]{i:\mathtt{DECJ}(r_j,s) \quad m_j = 0}{(i,m_0,m_1)\minskred(s,m_0,m_1)}  
 \and  
 \inferrule[\textsc{(M-Dec)}]{i:\mathtt{DECJ}(r_j,s) \quad m_j \neq 0 \quad  m_j' = m_j - 1 \quad m_{1-j}' = m_{1-j}}{(i,m_0,m_1)\minskred(i+1,m_0',m_1')}
 \end{mathpar}  
 \caption{Semantics of \mmss}\label{a:mm}
\end{figure}

\subsection{Well-Structured Transition Systems}\label{s:wsts}
The decidability of \OG for \evold{2} processes will be shown by appealing to the theory of
well-structured transition systems \cite{FinkelS01,AbdullaCJT00}.
The following results and definitions are from \cite{FinkelS01},  unless
differently specified. 


Recall that a \emph{quasi-order} (or, equivalently, preorder) is a reflexive
and transitive relation.

\begin{defi}[Well-quasi-order]
A \emph{well-quasi-order} (wqo) is a quasi-order $\leq$ over a
set $X$ such that, for any infinite sequence $x_0 , x_1 , x_2 \ldots \in X$, there exist indexes $i < j$
such that $x_i \leq x_j$.
\end{defi}

 Note that if $\leq$ is a wqo then any infinite sequence $x_0 , x_1 , x_2 , \ldots$ contains an infinite
increasing subsequence $x_{i_0} , x_{i_1} , x_{i_2} , \ldots$ (with $i_0 < i_1 < i_2 < \ldots$). 
Thus well-quasi-orders exclude the possibility of having infinite strictly decreasing sequences.

We also need a definition for (finitely branching) transition systems. Here and in the following 
$\rightarrow^*$ denotes the 
reflexive and transitive
closure of the relation $\rightarrow$.

\begin{defi}[Transition system]
\label{def:WSTS}
A \emph{transition system} is a structure $TS = (S, \rightarrow)$, where $S$ is a set of states
and $\rightarrow \subseteq  S \times S$ is a set of transitions. 
We define $Succ(s)$ as the set $\{s' \in S \mid s \rightarrow s' \}$
of immediate \emph{successors} of $s$. 
$TS$ is \emph{finitely branching} if, for each $s \in S$,
$Succ(s)$ is finite.
We also define $Pred(s)$ as the set $\{s' \in S \mid s' \rightarrow s\}$ of \emph{immediate predecessors} of $s$,
while $Pred^*(s)$ and $Pred^+(s)$ denote the sets $\{s \in S \mid s' \rightarrow^* s\}$ and  $\{s \in S \mid s' \rightarrow^+ s\}$, respectively, of \emph{predecessors} of $s$.
\end{defi}

\begin{conv}
In the rest of the paper, 
and with a slight abuse of notation,
we will  assume the expected point-wise extensions of definitions to sets. 
For instance, 
function $Succ$ just defined on states is extended to sets of states as:
$Succ(S) = \bigcup_{s \in S} Succ(s)$.  
\end{conv}

The key tool to 
the decidability of 
several properties of computations 
is the notion of \emph{well-structured transition system} \cite{FinkelS01,AbdullaCJT00}. 
This is a transition system equipped with a well-quasi-order on states
which is (upward) compatible with the transition relation. Here we will use a strong version 
of compatibility; hence the following definition.

\begin{defi}[Well-structured transition system] 
A \emph{well-struc\-tured transition system with strong compatibility} is a transition system 
$TS = (S, \rightarrow)$, equipped with a quasi-order $\leq$ on $S$, such that the two following conditions hold:
\begin{enumerate}[(1)]
\item 
 $\leq$ is a well-quasi-order;
\item 
 $\leq$ is strongly (upward) compatible with $\rightarrow$, that is, for all $s_1 \leq t_1$ and all transitions
   $s_1 \rightarrow s_2$ , there exists a state $t_2$ such that $t_1 \rightarrow t_2$ and $s_2 \leq t_2$ holds.\smallskip
\end{enumerate}
\end{defi}

%

\noindent Given a quasi-order $\leq$ over $X$, an {\em upward-closed set} is a subset $I\subseteq X$ such that
the following holds: $\forall x,y\in X: (x\in I\wedge x\leq y)\Rightarrow y\in I$. Given $x\in X$, we define its upward closure as $\uparrow x = \{y\in X\mid x\leq y\}$. This notion can be extended to sets as expected: given a set $Y\subseteq X$ we define its upward closure as $\uparrow Y = \bigcup_{y\in Y}\uparrow y$. 

\begin{defi}[Finite basis]\label{d:finbas}
A {\em finite basis} of an upward-closed set $I$ is a finite set $B$ such that $I = \bigcup_{x\in B}\uparrow x$.
\end{defi}

The notion of basis is particularly important when considering 
the basis of the predecessor of a state in a transition system. More precisely, we are interested in  \emph{effective} pred-basis as defined below.

\begin{defi}[Effective pred-basis]\label{d:efpb}
A well-structured transition system has {\em effective pred-basis} if there exists an algorithm such that, for any state $s\in S$, it returns the set $pb(s)$ which is a finite basis of $\uparrow Pred(\uparrow s)$.
\end{defi}

The following proposition is a special case of Proposition 3.5 
in \cite{FinkelS01}.
\begin{prop} \label{predcomp}
Let $TS = (S,\rightarrow,\leq)$ be a finitely branching,
well-structured transition system
with strong compatibility, decidable $\leq$ and effective pred-basis.
It is possible to compute a finite basis of $Pred^*(I)$ for any upward-closed set $I$
given via a finite basis.\qed
\end{prop}

%
%
%

Finally we will use the following proposition, whose proof is immediate.

\begin{prop}\label{prop:eqwqo}
Let $S$ be a finite set. Then the equality is a wqo over $S$.\qed
\end{prop}

We shall also appeal to the following result.
In \cite{kruskal60}, Kruskal proved that a wqo on a set $S$ can be extended to the set of finite trees whose nodes have labels ranging in $S$; we refer to this as the set of trees \emph{over} $S$.
We define how to extend a quasi order on a set $S$ to the trees over $S$.
If $t$ is a tree and $n$ a node in $t$, we denote with $label(n)$ the label of the node~$n$.

\begin{defi}\label{def:prectr}
  Let $S$ and $\leq$ be a set and a wqo over $S$, respectively. 
  The relation $\leq^{\mathsf{tr}}$ on the set of
trees over $S$ is defined as follows. Let $t, u$ be trees over $S$. We have that $t \leq^{\mathsf{tr}} u$ iff there
exists an injection $f$ from the nodes of $t$ to the ones of $u$ such that:
\begin{enumerate}[(1)]
 \item Let $m,n$ be nodes in $t$. If $m$ is an ancestor of $n$ then $f(m)$ is an ancestor of $f(n)$.
 \item Let $m,n,p$ be nodes in $t$. If $p$ is the minimal common ancestor of $m$ and $n$ then $f(p)$ is the minimal common ancestor of $f(m)$ and $f(n)$.
 \item Let $n$ be a node in $t$. Then $label(n) \leq label(f(n))$.
\end{enumerate}
\end{defi}

The relation $\leq^{tr}$ is  a quasi-order over the trees over $S$. 
It is also a wqo, since we have the following result.

\begin{thm}[Kruskal \cite{kruskal60}]\label{lem:kruskal}
   Let $S$ be a set and $\leq$ a wqo over $S$. Then, the relation $\leq^{tr}$ 
is a wqo on the set of trees over $S$.\qed
\end{thm}

%
%
%


\subsection{Petri Nets}\label{sec:petri}
We will use Petri nets to prove the decidability of \OG for \evols{3}. 
More precisely,
we will reduce \OG  for \evols{3} to a problem on Petri nets, that we call
\emph{infinite visit}, which can be easily reduced to place boundedness.

A \emph{Petri net}  
is a tuple $N = (S, T, m_0 )$, where $S$ and $T$
are finite sets of \emph{places} and \emph{transitions}, respectively.
A finite multiset over the set $S$ of places is called a \emph{marking}, and $m_0$ is the initial marking.
Given a marking $m$ and a place $p$, we say that the place $p$
contains $m(p)$ \emph{tokens} in the marking $m$ if there are $m(p)$ occurrences of $p$ in the multiset $m$.
A transition is a pair of markings
written in the form $m'\derriv{}m''$.
The marking $m$ of a Petri net can be modified
by means of transition firing: a transition
$m'\derriv{}m''$ can fire if
$m(p) \geq m'(p)$ for every place $p \in S$;
upon transition firing the
new marking of the net becomes $n=(m \setminus m') \uplus m''$
where $\setminus$ and $\uplus$ are the difference and
union operators for multisets, respectively.
This is written as $m \derivv n$.
We call \emph{computation} a sequence $m_{0}\derivv m_{1}\derivv \cdots \derivv m_{n}$.
A marking $m$ is \emph{reachable} if there exists 
a computation with final marking $m$.
A place $p \in S$ is \emph{bounded}
if there exists a natural number $k$ such that $m(p) \leq k$
for every reachable marking $m$.
The place boundedness problem is decidable for Petri nets~\cite{KarpM69}.

\begin{defi}[Infinite visit]\label{d:infvisit}
Given a Petri net $N = (S, T, m_0)$, a 
set of places to visit $V \subseteq S$, 
and a mandatory place $p \in S$,
we say that $N$
\emph{infinitely visits} $V$ with mandatory place $p$,
if there exists an infinite sequence 
$m_{0}\derivv m_{1}\derivv m_{2} \derivv \cdots$ and an index $i$ such that for every $j \geq i$ there exists a place $p_j \in V$
such that $m_{j}(p_j) \geq 1$, and moreover $m_{j}(p) \geq 1$.
\end{defi}

\begin{thm}\label{th:infiniteVisit}
Given a Petri net $N = (S, T, m_0)$, a 
set of places $V \subseteq S$, and a mandatory place $p \in S$,
it is decidable whether $N$
\emph{infinitely visits} $V$ with mandatory place $p$.
\end{thm}
\proof
By reduction to the place boundedness
problem. Given a Petri net $N = (S, T, m_0)$
and a set of places $V \subseteq S$,
we construct a Petri net $N' = (S\cup \{ph1,ph2,check\}, T', m_0\cup\{ph1\})$
such that $N$ infinitely visits $V$ with mandatory place $p$
if and only if $check$
is not bounded in $N'$.

The Petri net $N'$ reproduces the computations
in $N$ by (possibly) dividing them into two phases:
the first phase is witnessed by the presence of one
token in the additional place $ph1$, while
the second phase by one token in the additional
place $ph2$. During the second phase, a transition
can be mimicked only if there is at least one token 
in one of the places in $V$ and one token in the place
$p$. Moreover, during the second
phase, each transition puts one token in the additional place $check$.

Formally, we define the set $T'$ of the transitions of $N'$ as follows:
\begin{enumerate}[$\bullet$]
\item
for each transition $m' \derriv{} m''$ in $T$, $T'$ contains 
the transition $m'\uplus\{ph1\} \derriv{} m''\uplus\{ph1\}$;
\item
$T'$ contains the transition $\{ph1\} \derriv{} \{ph2\}$;
\item
for each transition $m' \derriv{} m''$ in $T$ and for each place 
$q \in V$, 
$T'$ contains the transition $m'\uplus\{p,q,ph2\} \derriv{} m''\uplus\{p,q,ph2,check\}$.
\end{enumerate}
The first group of transitions governs the first phase of the
simulation; the second transition implies the passage from the
first to the second phase; while the third group of transitions
is for the second phase.

First, assume that $N$ infinitely visits $V$ with mandatory place $p$. 
This means that in $N$
there
exists an infinite sequence 
$m_{0}\derivv m_{1}\derivv m_{2} \derivv \cdots$ and an index $i$ such that for every $j \geq i$ there exists a place $p_j \in V$
such that $m_{j}(p_j) \geq 1$ and $m_{j}(p) \geq 1$.
This implies that in $N'$ there is a corresponding 
computation that mimics the transition $m_{0}\derivv m_{1}\derivv m_{2} \derivv m_{i-1}$ during the first phase,
and the transitions $m_{i}\derivv m_{i+1}\derivv \cdots$ 
during the second one. The second phase is infinite, hence $check$
is not bounded because each transition in the second phase
puts one token in such a place.

Assume now that $check$ is unbounded in $N'$.
As tokens are introduced in $check$ only during the second
phase, this means that there exists no bound
to the length of the computations in $N'$ that
include the second phase. This implies the existence of at least one infinite
computation in $N'$ having both the first and the second phase.
Consider now the computation in $N$ composed of the
transitions simulated in such an infinite computation of $N'$.
This computation in $N$ has a 
suffix (the part corresponding to the second phase)
in which all the traversed markings have
at least one token in one of the places in $V$
as well as one token in $p$. \qed


 \section{Undecidability Results for \evol{1}}\label{s:ev1}

We prove that \OG and \LG are undecidable in both  \evold{1} and \evols{1}.
The result relies on an
encoding of \mmss into \evols{1} which
satisfies the following:
a \mm terminates if and only if its 
encoding into \evols{1}
evolves into a state that starts an infinite computation
that traverses states exhibiting a distinguished barb $e$.

The encoding, denoted $\encp{\cdot}{\mmn{1}}$, 
is given in Table \ref{t:encod-pas}. 
A register $j$ with value $m$ is represented by an adaptable process at $r_j$ that contains 
the encoding of number $m$, denoted $\encn{m}{j}$. 
In turn, $\encn{m}{j}$ consists of a sequence of $m$ 
output prefixes on name $u_j$, ending with
an output action on $z_j$, which represents zero.
Instructions are encoded as replicated processes guarded by $p_i$, which represents
the \mm when the program counter $p=i$.
Once $p_i$ is consumed, each instruction is ready to interact with the registers.
To encode the increment of register $r_j$, we 
enlarge the sequence of output prefixes it contains.
The adaptable process at  $r_j$ is updated with the encoding of the incremented value
(which results from putting the value of the register behind some prefixes) and then
the next instruction is invoked. 
The encoding of a decrement 
of register $j$
consists of an exclusive choice: the left side implements the decrement of the value of a register,
while the right one implements the jump to some given instruction.
This choice is indeed 
exclusive: 
the encoding of numbers as a chain of output prefixes ensures that 
both an
input prefix on $u_j$ and one on $z_j$ are never available at the same time. 
 When the \mm reaches the $\mathtt{HALT}$ instruction the encoding can either exhibit a barb on $e$,
or set the program counter again to the  $\mathtt{HALT}$ instruction so as to pass
through a state that exhibits $e$ at least $k >0$ times.
The encoding of a \mm into \evols{1} is defined as follows:

\begin{table}[t]

\begin{tabular}{l}   
$  
\mathrm{\textsc{Register}}~r_j \qquad
\encp{r_j = n}{\mmn{1}}   =  \component{r_j}{\encn{n}{j}} 
$  \\
\quad where 
\quad  $  
\encn{n}{j}=\left\{  
\begin{array}{ll}  
\overline{z_j}  & \textrm{if } n= 0 \\  
 \overline{u_j}.\encn{n-1}{j} & \textrm{if } n> 0 .  
\end{array}\right.  
$
\\ 
$
\begin{array}{ll}   
\multicolumn{2}{l}{\mathrm{\textsc{Instructions}}~(i:I_i)}\\  
\encp{(i: \mathtt{INC}(r_j))}{\mmn{1}} \!&= !p_i.\update{r_j}{\component{r_j}{\overline{u_j}.\bullet}}.\overline{p_{i+1}}\\
\encp{(i: \mathtt{DECJ}(r_j,s))}{\mmn{1}}\!&=  !p_i.(u_j.\overline{p_{i+1}} + z_j.\update{r_j}{\component{r_j}{\overline{z_j}}}.\overline{p_{s}})\\
\encp{(i: \mathtt{HALT})}{\mmn{1}}\!&=  !p_i.(e + 
\outC{p_i})
\end{array}   
$
\end{tabular}
\caption{Encoding of \mmss into \evols{1}.}  \label{t:encod-pas}  
\end{table}

\begin{defi}\label{d:mmev1}
Let $N$ be a \mm, with registers $r_0 = 0$, $r_1 = 0$ and instructions
$(1:I_1) \ldots (n:I_n)$. 
Given the encodings in Table \ref{t:encod-pas}, the encoding of $N$ 
in \evol{} (written $\encp{N}{\mmn{1}}$)
is defined as
$\encp{r_0 = 0}{\mmn{1}} \parallel \encp{r_1 = 0}{\mmn{1}} \parallel  \prod^{n}_{i=1} \encp{(i:I_i)}{\mmn{1}} \parallel \outC{p_1}~.
$
\end{defi}


Given this encoding, 
we have that a \mm $N$ terminates iff its encoding has at least $k$ consecutive barbs on the distinguished action $e$, for every $k \geq 1$.

\begin{lem}\label{th:corrE1}
Let $N$ be a \mm and $k \geq 1$. 
$N$ terminates iff $\encp{N}{1} \barbk{e}$.
\end{lem}
\proof
See Appendix \ref{app:e1}, Page \pageref{app:e1}.\qed

\begin{thm}\label{th.ev1}
\OG and \LG are undecidable in \evols{1}.
\end{thm}
\proof[Proof (Sketch)]
The proof 
proceeds by  considering 
a \mm  $N$ and  
its encoding $\encp{N}{\mmn{1}}$.
Taking  the cluster $\BC_{\encp{N}{\mmn{1}}}^{\emptyset}=\{\encp{N}{\mmn{1}}\}$, 
undecidability of \OG 
follows from undecidability of the termination problem in \mmss
and Lemma \ref{th:corrE1}.

Moreover,  the number of consecutive barbs on $e$ can be unbounded: once the machine reaches the $\mathtt{HALT}$ instruction 
then a barb $e$ will be continuously available by always choosing to synchronize on $\outC{p_i}$. Hence, there exists a computation where $\encp{N}{\mmn{1}} \barb{e}^{\omega}$ and  we can  conclude that \LG is undecidable.
\qed
Notice that  $\encp{N}{\mmn{1}}$ is an \evols{1} process without nested adaptable processes. Hence, even if we consider $\encp{N}{\mmn{1}}$ as an \evold{1} process, update prefixes cannot modify the topology of nested adaptable processes (that is, in the semantics of Figure \ref{fig:ltswithalpha} condition $\mathsf{cond}(U,Q)$ always holds true) and the generated transition system is the same.
Formally, this can be verified by using Lemma \ref{lem:statvsdyn}. 
As a consequence, the above undecidability result holds for \evold{1} processes as well: 

\begin{cor}
\OG and \LG are undecidable in \evold{1}.\qed
\end{cor}


\section{(Un)decidability Results for \evol{2}}\label{s:ev2}
\subsection{Decidability of Bounded Adaptation}\label{sec:decwsts}

Here we prove that despite the previous undecidability result, 
\OG is decidable for $\evold{2}$ processes. 
That is,  
given 
a process $P$, a set of processes $M$, and 
a barb $\alpha$, 
there exists an algorithm to determine whether there exists a process $R \in \BC_P^M$ such that $R \barb{\alpha}^{k}$ holds.
The proof  appeals to the theory of well-structured transition systems (see Section \ref{s:wsts}).
\new{The algorithm consists of five steps:}
\begin{enumerate}[(1)]
\item \label{uno} We restrict the set of terms under consideration  to those reachable by any $R \in \BC_P^M$.  
We characterize this set by 
\begin{enumerate}[(a)]
\item[(a)] considering the 
set of \emph{sequential subterms} in $\BC_P^M$, i.e., 
the subterms of $P$ and the processes in $M$ that do not have parallel composition or adaptable processes 
as their topmost operator---see Definition~\ref{d:seqst}---and 
\item[(b)] 
introducing the ordering $\procleq$ over 
a tree-like representation of the processes in such a set---see Definition~\ref{d:order}.
\end{enumerate}
\item \label{due}\new{Next, we prove that $\procleq$  is 
a well-quasi-ordering (cf. Theorem~\ref{th:wqoccs}) 
which is strongly compatible with respect to  $\pired$ (cf. Theorem~\ref{th:scccs}).
\item \label{tre}
These results enable us to compute a finite basis for the set of
processes exhibiting   $\alpha$; this set is \emph{upward-closed} with respect to  $\procleq$ (cf. Theorem~\ref{th:fb})}.

\item  \label{quattro} We then show that it is possible to compute the finite basis of the set of processes 
that expose $\alpha$ at least $k$ consecutive times (Lemma \ref{lem:decpred}).
\item \label{cinque} Finally, we show that it is possible to determine whether or not some process $R \in \BC_P^M$ is included in the set generated by the finite basis (Theorem \ref{th:badec}).
\end{enumerate}

\new{In what follows, we describe the definitions and results associated to these steps.
For the sake of clarity, each of these descriptions is presented separately, in Sections~\ref{ss:uno}---\ref{ss:cinco}.
}

Observe that the above strategy requires Kruskal's theorem (Theorem \ref{lem:kruskal}) on well-quasi-orde\-rings on trees.
Unlike similar previous results exploiting the theory of well-structured transition systems for obtaining decidability results (e.g., \cite{Busi}),
in the case of $\evold{2}$ it  is not possible to find a bound on the ``depth'' of processes.   
We illustrate this issue with a small example. Consider the process 
$R = \component{a}{P} \parallel ! \update{a}{\component{a}{\component{a}{\bullet}}}.\nil $.
One possible evolution of $R$ is the following:
$$R \pired \component{a}{\component{a}{P}} \parallel ! \update{a}{\component{a}{\component{a}{\bullet}}}.\nil \pired \component{a}{\component{a}{\component{a}{P}}} \parallel ! \update{a}{\component{a}{\component{a}{\bullet}}}.\nil  \pired \ldots$$ 
and thus one obtains a process with an unbounded number of nested adaptable processes. 
Nevertheless, not everything is lost and some regularity can be found also in our case. 
By mapping processes into particular forms of trees and then exploiting an ordering over those trees, 
it can be shown that this is indeed a well-quasi-ordering with strong compatibility, 
and that it has an effective pred-basis. This way, decidability of \OG  can be shown 
by following the five steps described above.

\subsubsection{Step (\ref{uno})}\label{ss:uno}
We start by introducing some 
auxiliary definitions.


\begin{defi}[Parallel Processes]\label{d:pps}
  Let $P = \prod_{i=1}^{m} P_i \parallel \prod_{j=1}^{n} \component{a_j}{P'_{j}}$ be an \evol{} process in normal form.
  The set of \emph{top-level, parallel processes} of $P$, is defined as 
$$\Par(P)= \{P_i \mid i \in [1..m] \} \cup \{ \component{a_j}{P'_{j}} \mid j\in [1..n]  \}$$ 
This definition extends to sets of processes in normal form  in the expected way.
\end{defi}

 \begin{defi}[Sequential Subprocesses]\label{d:seqst}
 Let $P$ be an \evold{2} process.
The set of  sequential subprocesses of $P$, denoted 
 $\subp(P)$, is defined inductively as follows:
$$
 \begin{array}{ll}
 \subp(\pi.P) & =  \{\pi.P\}\cup \subp(P) ~~\text{if $\pi = a$ or $\pi = \outC{a}$}\\
 \subp(\update{a}{U}.Q) & =  \{ \update{a}{U}.Q \} \cup \subp(U) \cup \subp(Q)\\  
 \subp(\sum_{i \in I} \pi_i.P_i) & =  \big\{  \sum_{i \in I} \pi_i.P_i \big\} \cup \bigcup_{i \in I} \subp(\pi_{i}.P_i)\\
 \subp(!\pi.P)  & =   \{!\pi.P\} \cup \subp(P) \\
\subp(P \parallel Q) &  =  \subp(P)\cup\subp(Q)  \\
 \subp(\component{a}{P}) & =   \subp(P) \\
 \subp(\bullet) & =  \emptyset 
 \end{array}
$$
Observe that $\subp(\nil) = \subp(\sum_{i \in \emptyset} \pi_i.P_i)~=~\{ \nil \}$.
The definition extends to sets of processes as expected.
 \end{defi}
 
Notice that, since we are considering processes $P \in \evold{2}$ (which make use of update patterns that cannot include $\bullet$ in the scope of prefixes), $\subp(P)$ is a set of processes that are not update patterns, that is they cannot have free occurrences of $\bullet$.

\begin{defi}
 Let $P$ be an \evold{2} process.
The set of  adaptable processes names occurring in $P$, denoted 
 $\cnames(P)$,  is inductively defined (by resorting, in general, to $\cnames(U)$ over \evold{2} update patterns $U$) as follows:
$$
 \begin{array}{ll}
\cnames(\component{a}{U}) &= \{a\}\cup \cnames(U) \\
 \cnames(\pi.P) & =  \cnames(P) ~~\text{if $\pi = a$ or $\pi = \outC{a}$}\\
 \cnames(\update{a}{U}.Q) & =  \cnames(U) \cup \cnames(Q)\\  
 \cnames(\sum_{i \in I}\pi_{i}.U_{i}) &= \bigcup_{i \in I}\cnames(\pi_{i}.U_{i}) \\  
  \cnames(! \pi.U) &= \cnames(\pi. U) \\
\cnames(U_{1} \parallel U_{2}) & =  \cnames(U_{1}) \cup \cnames(U_{2}) \\
\cnames(\bullet) & = \emptyset 
\end{array}
$$
 The definition extends to sets of processes as expected.
 \end{defi}


\begin{defi}
Given a set of \evold{2} processes $S$, 
we define:
$$\nset{S} = \subp(S) \cup   \{a[\ ] \mid  a \in \cnames(S) \}$$
\end{defi}

We are now ready to define the tree denotation of a process.



 \begin{defi}[Tree of a process]\label{def:tree}
  Let $P = \prod_{i=1}^{m} P_i \parallel \prod_{j=1}^{n} \component{a_j}{P'_{j}}$ be an \evold{2}  process in normal form.
The tree denotation of $P$, denoted $\Tree(P)$, is a tree over $\nset{\{P\}} \cup \{\varepsilon\}$ and it
  is built as follows.
  The root is labeled $\varepsilon$, and
  has $m+n$ children: the former $m$ are leaves labeled $P_1, \dots, P_m$, while the latter $n$ are subtrees recursively built from processes $P'_{1}, \dots, P'_{n}$, where the only difference is that their roots are labeled $a_1[\ ], \dots, a_n[\ ]$, respectively.

Given a set of \evold{2} processes $S$, ${\mathcal{T}}_{S}$ denotes the set of trees over $\nset{S} \cup \{\varepsilon\}$.

 \end{defi}

\begin{figure}[t]
\centering

\subfigure[A tree denotation]{\label{fig:uno}\includegraphics[width=0.33\textwidth]{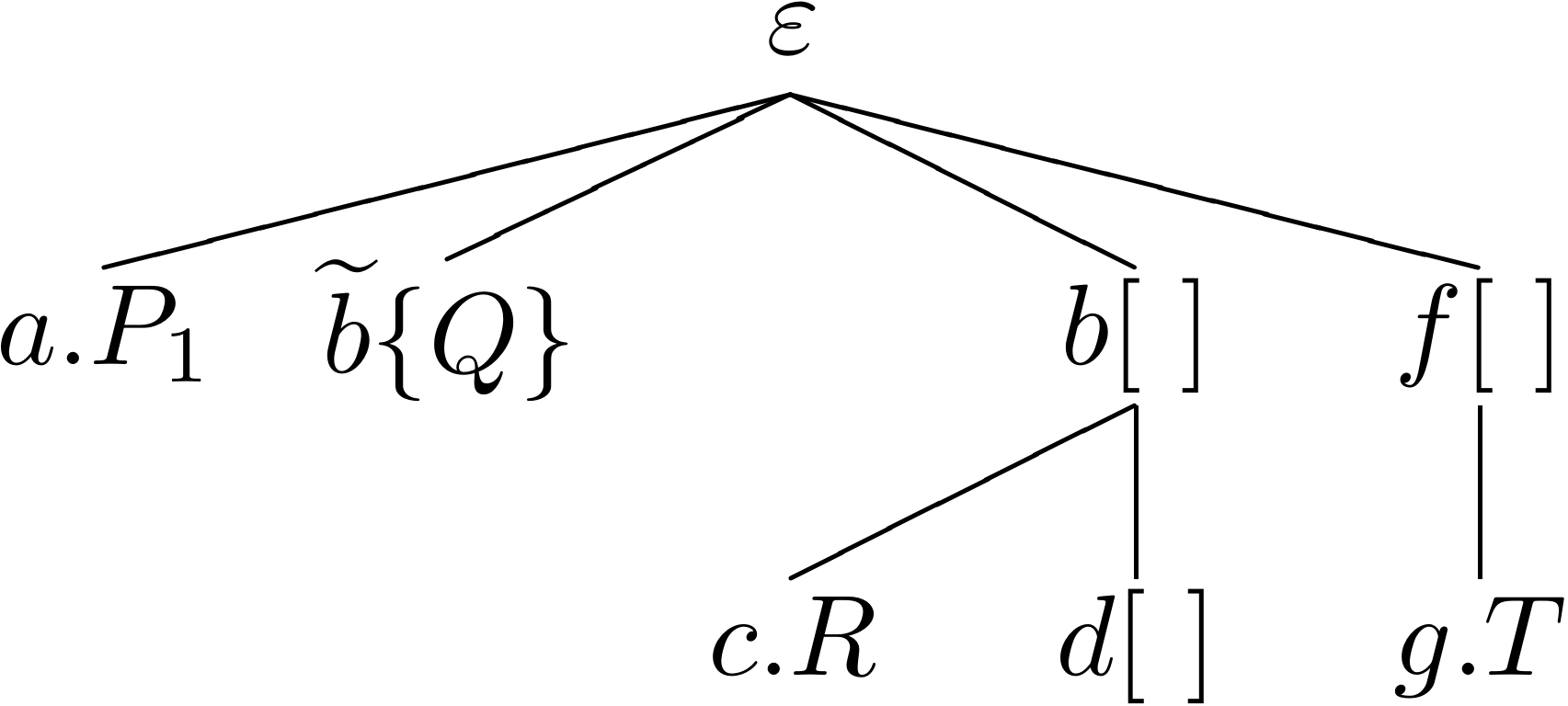}} 
\quad
\subfigure[Tree embedding]{\label{fig:due}\includegraphics[width=0.33\textwidth]{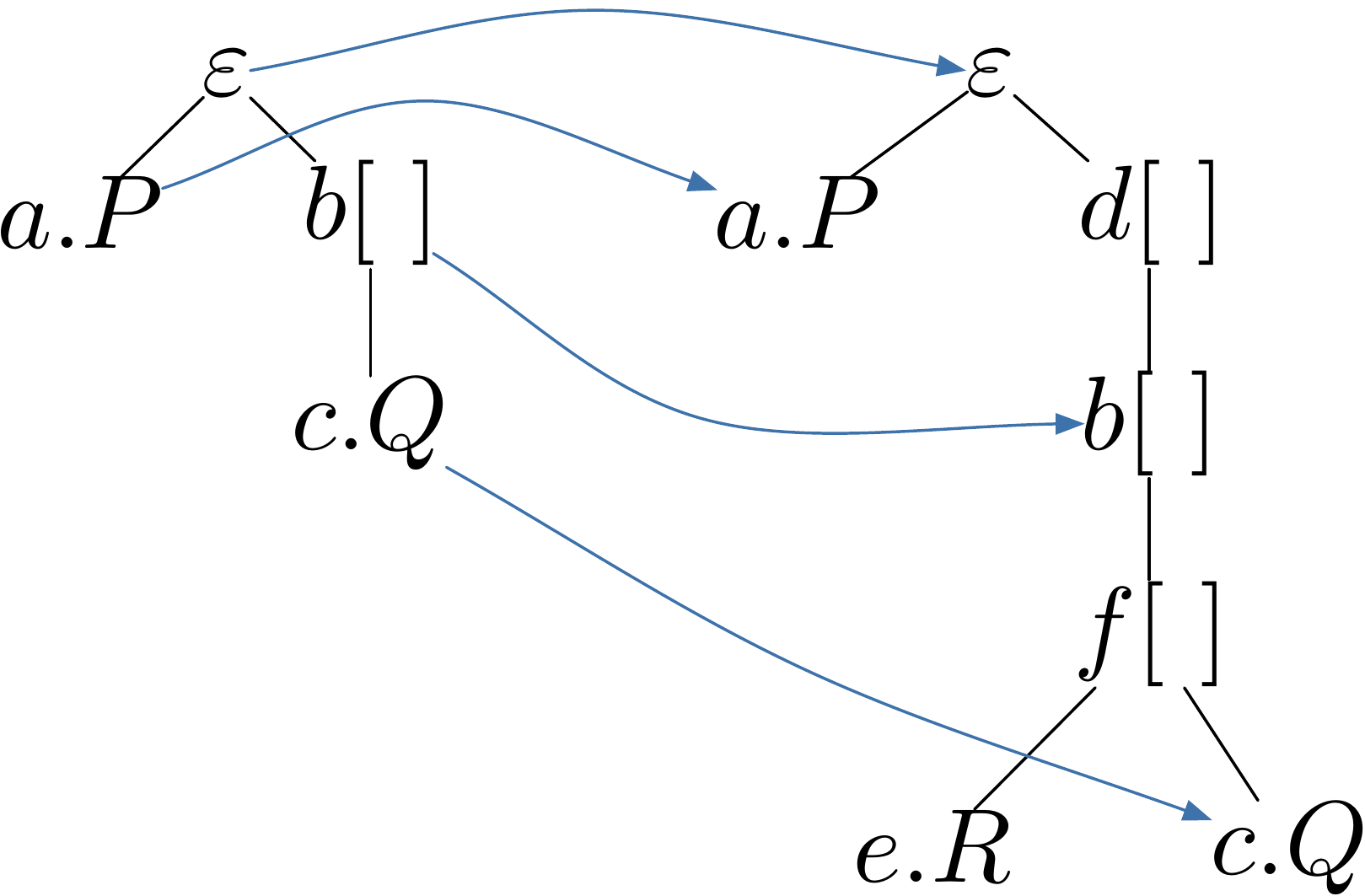}}

\caption{Tree denotations for \evold{2} processes.} 
\label{fig:proc2tree}
\end{figure}

%
\begin{exa}

Let $P$ be the process $a.P_{1} \parallel \update{b}{Q} \parallel \component{b}{c.R \parallel \component{d}{\ }}  \parallel \component{f}{g.T}$.
Given Definition~\ref{def:tree}, $\Tree(P)$ is depicted in Figure \ref{fig:uno}.
\end{exa}

We now define the ordering $\preceq$ on processes. 
It corresponds to the extension of $=$, 
 as described in Definition \ref{def:prectr}, to trees.
 Notice that when $=$ is extended to trees it is no longer a symmetric relation.
More precisely: 
 \begin{defi}[Ordering $\preceq$]\label{d:order}
 Let $P$ and  $Q$ be \evold{2} processes.
 Also, let $=^{\mathsf{tr}}$ stand for the extension of $=$ as in  Definition \ref{def:prectr}. 
 Then we decree:
$P \preceq Q$ iff 
$\Tree(P) =^{\mathsf{tr}} \Tree(Q)$.
 \end{defi}

In other words, given two processes $P$ and $Q$ such that $\Tree(P) =^{\mathsf{tr}} \Tree(Q)$, one simply checks if all the labels of $\Tree(P)$ occur in $\Tree(Q)$ and respect the ancestor relation. 

\begin{exa}
Let $S$ and $T$ be the processes defined as
\begin{align*}
 S & =  a.P \parallel \component{b}{c.Q}\\
 T & = a.P \parallel \component{d}{ \component{b}{\component{f}{e.R \parallel c.Q}}}
 \end{align*}
 Then we have $S \preceq T$; tree denotations for both processes (and the injection between them) are depicted in Figure \ref{fig:due}.

\end{exa}

We write $P \pired \succeq Q$ if there is some $P'$ such that $P \pired P'$ and $P' \succeq Q$.
We now define the set of all derivatives of a given \evold{2} process and show that $\preceq$ is a wqo over it.

\begin{defi}
Given an \evold{2} process $P$, we define $\deriv{P} =\{Q \mid P \pired ^* Q \}$.
This definition is extended to sets of processes in the expected way.
\end{defi}

\subsubsection{Step \ref{due}}

We start by showing that given a set of processes $S$, $=^{\mathsf{tr}}$ is  a wqo over ${\mathcal{T}}_{S}$.


 \begin{thm} \label{th:wqoccs}
  Let $S$ be a set of $\evold{2}$ processes. Then, relation $=^{\mathsf{tr}}$ is a wqo over ${\mathcal{ T}}_{S}$.
 \end{thm}
\proof
The set $\nset{S}$ is finite by construction. 
Hence, by Proposition \ref{prop:eqwqo}, equality is a wqo over $\nset{S} \cup \{\varepsilon\}$.
Finally, since $=$ is a wqo, 
using Kruskal's Theorem (Theorem \ref{lem:kruskal}) we infer that 
$=^{\mathsf{tr}}$ is a wqo over ${\mathcal{T}}_{S}$. 
\qed

We now prove that the trees constructed from processes contained in the set of all derivatives form a subset of ${\mathcal{ T}}_{S}$. 
The following notion of \emph{monadic and biadic contexts} will be useful in proofs.

\begin{defi}[Monadic and Biadic  Contexts]\label{d:mc}
A \emph{monadic  context} is a context with one hole (denoted ``$\cdot$'') and is defined according to the following grammar: 
$$
C[\cdot]::= [\cdot] \midd C[\cdot] \parallel P  \midd    \componentbbig{a}{C[\cdot]  } 
$$
where $P$ is  an \evol{} process. Similarly, a \emph{biadic context} is a context with two holes 
(denoted ``$\cdot_{1}$'' and ``$\cdot_{2}$'', respectively)
defined according to the following grammar: 
$$
D[\cdot_{1},\cdot_{2}]::= C[\cdot_{1}] \parallel C[\cdot_{2}]  \midd  \componentbbig{a}{D[\cdot_{1},\cdot_{2}] } \parallel P \midd \componentbbig{a}{D[\cdot_{1},\cdot_{2}] }
$$
where $P$ is an \evol{} process and $C$ is a monadic context.
As customary, $C[P]$ and $D[R,Q]$ represent the processes obtained by replacing the holes in contexts $C[\cdot]$ and $D[\cdot_{1},\cdot_{2}]$
with processes $P$ and $R,Q$, respectively.
\end{defi}

\begin{lem} \label{lem:deriv}
 Let $P$ be an \evold{2} process. If  $P \pired Q$ then $\Tree(Q) \in {\mathcal{ T}}_{\{P\}}$.
\end{lem}
\proof
By induction on 
the height of the derivation tree for $P \pired Q$, with a case analysis in the last rule used.
There are seven cases to check. 
We recall that $\Tree(Q) \in {\mathcal{ T}}_{\{P\}}$ iff $\Tree(Q)$ is over $\nset{P} \cup \{\varepsilon\}$.+
\begin{desCription}
\item\noindent{\hskip-12 pt\bf Case \rulename{Act1}:}\
Then $P = P_{1} \parallel P_{2}$ and $Q = P'_{1} \parallel P_{2}$, with $P_{1} \pired P'_{1}$.
By Definition \ref{def:tree} we have $\Tree(P)$ is over $\nset{P_1} \cup \nset{P_2} \cup \{\varepsilon\}$. By inductive hypothesis, we have that $\Tree(P'_{1})$ is over $\nset{P_1} \cup \{\varepsilon\}$. Hence we can conclude that $\Tree(Q)$ is over $\nset{P_1} \cup \nset{P_2} \cup \{\varepsilon\}$, thus $\Tree(Q) \in {\mathcal{T}}_{\{P\}}$.

\item\noindent{\hskip-12 pt\bf Case \rulename{Act2}:}\ Analogous to the case for \rulename{Act1} and omitted. 

\item\noindent{\hskip-12 pt\bf Case \rulename{Loc}:}\
hen $P = \component{a}{P_1}$ and $Q = \component{a}{P_1'}$, with $P_1 \pired P_1'$.
By Definition \ref{def:tree} we have $\Tree(P)$ is over $\nset{P_1} \cup \{a[\ ]\} \cup \{\varepsilon\}$. By inductive hypothesis, we have that $\Tree(P'_{1})$ is over $\nset{P_1} \cup \{\varepsilon\}$. Hence we can conclude that $\Tree(Q)$ is over $\nset{P_1} \cup \{a[\ ]\} \cup \{\varepsilon\}$, thus $\Tree(Q) \in {\mathcal{T}}_{\{P\}}$.

\item\noindent{\hskip-12 pt\bf Cases \rulename{Tau1}-\rulename{Tau2}:}\
Then $P \equiv  \fillcont{C_1}{A}\parallel \fillcont{C_2}{B}$, where $C_{1}$ and $C_{2}$ are monadic contexts as in Definition \ref{d:mc}.
Moreover, 
$A$ is either 
$!b.Q$ or 
$\sum_{i \in I} \pi_i.Q_i$ with $\pi_{l}=b$, for some $l\in I$, and 
$B$ is either 
$!\outC{b}.R$
or $\sum_{i \in I} \pi_i.R_i$ with $\pi_{l}=\outC{b}$, for some $l\in I$.

We consider only the case in which $A = \sum_{i \in I} \pi_i.Q_i$ with $\pi_{l}=b$ and $B = !\outC{b}.R$;  the other cases are similar. 
Then $Q \equiv \fillcont{C_1}{Q_l}\parallel \fillcont{C_2}{R \parallel!\outC{b}.R }$. We know that  $\Tree(P)$ is over $\nset{C_1} \cup \nset{A} \cup \nset{C_2} \cup \nset{B} \cup \{\varepsilon\}$ and by noticing that $\nset{Q_l} \subseteq \nset{A}$ and $\nset{R \parallel!\outC{b}.R} \subseteq \nset{B}$ we can conclude that $\Tree(Q) \in {\mathcal{T}}_{\{P\}}$.

\item\noindent{\hskip-12 pt\bf Cases \rulename{Tau3}-\rulename{Tau4}:}\
Then $P \equiv \fillcont{C_1}{A} \parallel \fillcont{C_2}{B}$ where:
\begin{enumerate}[$\bullet$]
\item $C_{1}$ and $C_{2}$ are monadic contexts, as in Definition \ref{d:mc}; 
\item $A = \component{b}{P_1}$, for some $P_{1}$;  
\item $B  =  \sum_{i \in I} \pi_i.R_i$ with $\pi_{l}=\update{b}{\component{b}{U} \parallel P_2}$ for $l\in I$, or $ B = !\update{b}{U}.R$, for some $R$.
\end{enumerate}

We consider the case in which $B = !\update{b}{\component{b}{U} \parallel P_2}.R$; the other case is similar. 
Then $Q \equiv \fillcont{C_1}{\fillcon{U}{P_1}}\parallel \fillcont{C_2}{R \parallel !\update{b}{U}.R }$.
 We know that  $\Tree(P)$ is over $\nset{C_1} \cup \nset{A} \cup \nset{C_2} \cup \nset{B} \cup \{\varepsilon\}$ and by noticing that $\nset{R \parallel !\update{b}{U}.R } \subseteq \nset{B}$ and that because of the restrictions on \evold{2} $P_3$ cannot occur behind a prefix, then  we can conclude that $\Tree(Q) \in {\mathcal{T}}_{\{P\}}$. \qed
\end{desCription}

\noindent Lemma \ref{lem:deriv} 
can be used to show that 
$\{\Tree(P) \mid P \in \deriv{S}\} \subseteq {\mathcal{T}}_{S}$, 
for some  set of processes $S$.
Then, using Theorem \ref{th:wqoccs} we can conclude that $\preceq$ is a wqo over it:

\begin{cor}
  Let $S$ be a set of $\evold{2}$ processes. Then, $\preceq$ is a wqo over $\deriv{S}$.\qed
\end{cor}

The next result states 
strong compatibility of $\preceq$
with respect to reductions of \evold{2}. 

\begin{thm}[Strong Compatibility]\label{th:scccs}
Let $P$ and $Q$ be \evold{2} processes such that $P \preceq Q$.
Then, $P \pired P'$ implies that there exists $Q'$ such that $Q \pired Q'$ and $P' \preceq Q'$. 
\end{thm}
\proof
By a case analysis on the reduction $P \pired P'$. 
It can be the result of either 
a \emph{input/output synchronization}---through rules \rulename{Tau1}/\rulename{Tau2}---
or 
an \emph{update synchronization}---through rules \rulename{Tau3}/\rulename{Tau4}).
In both cases, the reduction may be combined with uses of rule \rulename{Loc}, \rulename{Act1}, and \rulename{Act2}. 

We consider these two kinds of synchronizations  separately.
Let $n$ be a node with ancestor $m$, and let $\Tree(P)$ be a tree with root $\varepsilon$.
Below, when we say that $n$ \emph{is replaced by} $\Tree(P)$ we mean that: 
(i) $\varepsilon$ is merged with $m$; (ii) all children of $\varepsilon$ are added as siblings of $n$; and (iii) $n$ itself is removed.


\begin{desCription}
\item\noindent{\hskip-12 pt\bf Input/output synchronization:}\ Then we have 
$$P \equiv D[A, B]$$ 
where
$D$ is a biadic context as in Definition \ref{d:mc}, $A$ is either 
$!a.P_1$ or 
$\sum_{i \in I} \pi_i.Q_i$ with $\pi_{l}=a$ and $Q_l = P_1$ for some $l\in I$, and 
$B$ is either 
$!\outC{a}.P_2$
or $\sum_{i \in I} \pi_i.R_i$ with $\pi_{l}=\outC{a}$ and $R_l = P_2$ , for some $l\in I$.

Consider the tree
$\Tree(P)$, and let $m$ and $n$ be two of its nodes, 
labeled $A$ and $B$, respectively.

We first consider the modifications to $\Tree(P)$ when $P \pired P'$. The tree $\Tree(P')$ is obtained from $\Tree(P)$ in the following way:
\begin{enumerate}[(1)]
\item the node labeled $A$ is replaced with $\Tree(P_1)$; 
\item the node labeled $B$ is replaced with $\Tree(P_2)$.
\end{enumerate}

Since $P \preceq Q$, 
the definition of $\preceq$ 
ensures that there exists a mapping $f$ that associates nodes in $\Tree(P)$ to nodes in $\Tree(Q)$. 
In turn, this ensures the existence of a node $f(m)$ in $\Tree(Q)$ 
labeled $A$. It also ensures the existence of a node $f(n)$ labeled $B$ and 
which has a common ancestor with $f(m)$. 
Hence, the reduction can take place in $Q$ as well, and so $Q \pired Q'$.
Now, $\Tree(Q')$ is obtained from $\Tree(Q)$ by applying the same changes described above to the target nodes (of the input and the output) according to $f$.

The last thing to show is  $P' \preceq Q'$, which follows 
 by observing that the mapping between $\Tree(P')$ and $\Tree(Q')$ is necessarily the same mapping $f$  
 between $\Tree(P)$ and $\Tree(Q)$, for all the nodes that have not been modified by the reduction and 
 that there is a one-to-one correspondence for the other nodes, as the new trees $\Tree(P_1)$ and $\Tree(P_2)$ are added to both  $\Tree(P)$ and $\Tree(Q)$. 
Thus, $\Tree(P')=^{\mathsf{tr}} \Tree(Q')$. 

\item\noindent{\hskip-12 pt\bf Update synchronization:}\ Then we have 
$$P \equiv D[\component{a}{P_1}, A]$$ 
where $D$ is a biadic context as in Definition \ref{d:mc} and  $A$ is either 
$!\update{a}{P_2}.R$ or 
$\sum_{i \in I} \pi_i.Q_i$ with $\pi_{l}=\update{a}{P_2}$ and $Q_l = R$ for some $l\in I$ .   
Consider the tree
$\Tree(P)$, and let $m$ and $n$ be two of its nodes, 
labeled $A$ and $\component{a}{\,}$ (with subtree $\Tree(P_1)$), respectively.

We first consider the modifications to $\Tree(P)$ when $P \pired P'$. The tree $\Tree(P')$ is obtained from $\Tree(P)$ in the following way:
\begin{enumerate}[(1)]
\item the node labeled 
$A$ 
is replaced with $\Tree(R)$; 
%
\item as for the tree rooted in $a[\ ]$ 
($\Tree(P_1)$), it is replaced with 
$\Tree(\fillcon{P_2\,}{P_1})$. 
\end{enumerate}

Since $P \preceq Q$, 
the definition of $\preceq$ ensures that 
there exists a mapping $f$ that associates nodes in $\Tree(P)$ to nodes in $\Tree(Q)$. 
In turn, this ensures the existence of a node $f(m)$ in $\Tree(Q)$ 
labeled $A$. It also ensures the existence of a node $f(n)$ labeled $a[\ ]$ and 
which has a common ancestor with $f(m)$. 
Hence, the update synchronization above can take place in $Q$ as well, and so $Q \pired Q'$.
Now, $\Tree(Q')$ is obtained from $\Tree(Q)$ by applying the same changes described above to the target nodes (of the adaptable process $a$ and of the
update  in $A$) according to $f$.

The last thing to show is that $P' \preceq Q'$, which follows 
 by observing that the mapping between $\Tree(P')$ and $\Tree(Q')$ is the same mapping $f$  
 between $\Tree(P)$ and $\Tree(Q)$, for all the nodes that have not been modified by the reduction and 
 that there is a correspondence one to one for the other nodes. More precisely: 
\begin{enumerate}[(1)]
 \item Consider the label in node $f(m)$: all nodes removed in $\Tree(P')$ have been removed in $\Tree(Q')$, hence nodes $m$ and $f(m)$ are still in relation.
 \item Finally, we consider the two trees rooted in $n$ and $f(n)$, namely 
 $S= \Tree(\fillcon{P_2\,}{P_1})$ and 
 $T=\Tree(\fillcon{P_2\,}{Q_1})$, respectively. $S$ is the same subtree as $T$ apart from some subtrees of $P_2$ and $Q_2$ that can be put easily in relation 
as the subtrees $\Tree(P_1)$ and $\Tree(Q_1)$ are in relation with $f$.
 \end{enumerate}
Thus, $\Tree(P')=^{\mathsf{tr}} \Tree(Q')$. \qed
\end{desCription}

\subsubsection{Step (\ref{tre})}
We now move on to characterize the set of predecessors of a given process (cf. Definition \ref{def:WSTS}) by means of a finite basis (cf. Definition \ref{d:finbas}).
Given a set $S$ of processes, 
we are only interested in those predecessors whose tree is in $\bigset{S}$.
As it will be clear later on, $S$ is intended to represent all processes in a cluster (cf. Definition \ref{d:cluster}).

\begin{defi}
Let $P$ 
and $S$ be an \evold{2} process and a set of \evold{2} processes, respectively.
We define:
$$\Pred_{S}(P) = \{Q \mid Q \in Pred(P),~\Tree(Q) \in \bigset{S} \}.$$\smallskip
\end{defi}

\noindent As we have seen, reductions in \evol{} originate only from synchronizations between input and output prefixes or 
from synchronizations between an adaptable process  and a corresponding update prefix. 
Our characterization of $\Pred_{S}(P)$ as a finite basis relies, intuitively, 
 on the formalization of the 
``parts'' of $P$ that might have been  involved in a reduction leading to $P$. 
We introduce the notion of \emph{syntactic context}: it allows us to reason about the \emph{decompositions} of $P$, which are useful to 
describe the subprocesses that have been involved in the reduction to $P$; 
such subprocesses may be contained in $P$ or they can be found in $S$. 
In the latter case, we must appeal to \emph{parallel extensions} of the syntactic context defining the given decomposition, as we give next:

\begin{defi}[Syntactic Contexts, Decompositions, Extensions]\label{def:SDE}
\emph{Syntactic contexts}, ranged over $K, K', \ldots$, are defined by the following syntax:
\begin{align*}
K ::= [\cdot] ~\mid ~\component{a}{K} ~\mid~ K \parallel K ~\mid~ P
\end{align*}
where $P$ is as in Definition \ref{d:finiteccs} using contexts as in Definition \ref{def:varianti} (2).

Given a process $P$, a syntactic context $K$, and processes $\til R$, we say 
that $K[\tilde{R}]$ is a \emph{decomposition} of $P$ if $P = K[\tilde{R}]$.
We assume processes $\tilde{R}$ fill the holes in $K$ preserving the order in which they appear.

A \emph{parallel extension} of $K$ is a syntactic context with exactly two holes obtained in the following way:
$$\mathsf{Ext}(K) = \{K, \, K \parallel  [\cdot], \, K \parallel  [\cdot] \parallel  [\cdot] \} \cap SC_2$$
where $SC_2$ is the set of all syntactic contexts with exactly two holes.
\end{defi}

We move on to define the pred-basis function 
for processes; it is defined with respect to a set of processes $S$ and 
noted $pb_{S}(\cdot)$. First, we present some intuitions and auxiliary definitions.
Given a process $P$, 
the set $pb_{S}(P)$ represents the basis for the set $\uparrow \Pred_{S}(\uparrow P)$; 
in other words, it is a finite representation of those processes that reduce  to $P$,
up to $\preceq$, i.e.,  a basis for all those $Q$ such that $Q \pired \succeq P$. 
To this aim,  we consider all the decompositions of 
$P$ as  $K[\til R]$, for some syntactic context $K$ and processes $\til R$, with ${|}\til R{|} \leq 2$.
There are finitely many such decompositions. 
The idea is to characterize a predecessor $Q$ of $P$ by 
suitably 
filling in the holes in (possibly an extension of) $K$ so that the  $Q$ is such that $\Tree(Q) \in {\mathcal{T}}_S$. 
Now, each $K$ can have two, one, or even zero holes (as a process can be a decomposition of itself).
In case ${|}\til R{|} < 2$, the syntactic context must be extended so as to contain exactly two holes; this is defined by $\mathsf{Ext}(K)$ above. 

Let us analyze the possibilities for such an extended context. 
As we have seen, reductions in \evold{2} arise from the synchronization of two complementary prefixes occurring (i) inside two sums, or (ii) one  inside a sum and the other  in a replicated process;  or (iii) both prefixes  in  two replicated processes. For the sake of readability, and with a little abuse of notation, in the explanation below we use biadic contexts filled in with the interacting prefixes, rather than with the processes in which such prefixes occur. That is, we write $D[\alpha.P, \beta.Q]$ rather than, e.g., $D[\alpha.P + M, !\beta.Q]$.
There are six cases. 
If $K$ has exactly two holes then 
it means that the reduction is ``internal'' to process $P$. 
That is, the reduction can be traced back by looking at  subprocesses of $P$.
Then $P = K[P_1, P_2]$ and 
no parallel extension is needed. There are two possible cases:
\begin{enumerate}[(1)]
 \item $P$ is the result of an input/output synchronization and so its predecessors are of the form 
 $Q = K[a.P_1, \outC{a}.P_2]$, for some $a \in  \cnames(S)$ and where $a.P_1$ and $\outC{a}.P_2$ are processes in $\subp(S)$.
 
 \item $P$ is the result of a synchronization between an update prefix and some corresponding adaptable process, and so its predecessors are of the form $Q = K[\update{a}{Q'}.P_1, \component{a}{Q''}]$, where $P_2 = \fillcon{Q'}{Q''}$ and $a \in  \cnames(S)$.  
Also, process $\update{a}{Q'}.P_1$ should belong to $\subp(S)$. Moreover, depending on the number of holes in $Q'$ there are two possible situations: (1) if $\numholes{Q'} = 0$ then $P_2 = Q'$ and $Q''$ can be any process in $\subp(S)$; (2) if  $\numholes{Q'} > 0$ then $Q''$ is taken in such a way that $P_2 = \fillcon{Q'}{Q''}$.\smallskip
\end{enumerate}

\noindent In case $K$ has one hole only then we extend the context with a hole so as to accommodate  some process not originally present in $P$. 
That is,  $P = K[P_1]$ and the reduction to $P$ is characterized by  the interaction between 
a prefix guarding subprocess $P_{1}$ 
and some other subprocess external to $P$ (cases (3) and (4) below).
It can also be the case that the reduction is an update synchronization leading to $P_{1}$ (case (5)).
We thus consider the extended context  $D[\cdot,\cdot] \equiv K[\cdot] \parallel [\cdot]$. There are three possible cases:


\begin{enumerate}[(3)]
 \item[(3)] $P$ is the result of an input/output synchronization, and so its predecessors are either of the form $Q \equiv D[a.P_1 , \outC{a}.Q_2$] 
 or $Q \equiv D[\outC{a}.P_1, a.Q_2]$, for some $a \in  \cnames(S)$ and processes $a.P_1$ and $\outC{a}.Q_2$ ($\outC{a}.P_1$ and $a.Q_2$, respectively) belong to $\subp(S)$. 
 \item[(4)] $P$ is the result of a synchronization between an update prefix guarding $P_{1}$ and some corresponding adaptable process.
 Hence, for some $a \in  \cnames(S)$, its predecessors are of the form $Q \equiv D[\update{a}{Q'}.P_1, \component{a}{Q''}]$,  with processes $\update{a}{Q'}.P_1$ and $Q''$ in $\subp(S)$.
 \item[(5)] $P$ is the result of a synchronization between an update prefix and some corresponding adaptable process, 
 in such a way that their synchronization leads to $P_{1}$. This way, 
the predecessors of $P$ are of the form $Q \equiv D[\update{a}{Q'}.Q_2, \component{a}{Q''}]$ or $Q \equiv D[\component{a}{Q''}, \update{a}{Q'}.Q_2]$ where $P_1 = \fillcon{Q'}{Q''}$, for some $a \in  \cnames(S)$. Similarly as in case (2) above, process $\update{a}{Q'}.Q_2$ should belong to $\subp(S)$. Moreover, depending on the number of holes in $Q'$ there are two possible situations: (1) if $\numholes{Q'} = 0$ then $P_1 = Q'$ and $Q''$ can be any process in $\subp(S)$; (2) if  $\numholes{Q'} > 0$ then $Q''$ is taken in such a way that $P_1 = \fillcon{Q'}{Q''}$.\smallskip
\end{enumerate}

\noindent The last case to consider is when $K$ has no holes, i.e., the trivial decomposition of $P$ as itself.
Then $D[\cdot,\cdot] \equiv P \parallel [\cdot] \parallel [\cdot] $ and we have:

\begin{enumerate}[(6)]
\item $P$ is the result of a synchronization between the subprocesses in the two added holes.
That is, its predecessors are of one of the following:  
(1) $Q \equiv P \parallel a.R_1 \parallel \outC{a}.R_2 $ and 
(2) $Q \equiv P \parallel \update{a}{Q'}.R_1 \parallel \component{a}{R_{2}} $.
In both cases,  $a \in  \cnames(S)$ and the holes are filled in with processes in $\subp(S)$.\smallskip
\end{enumerate}

\noindent Before giving the definition of $pb_{S}(Q)$, we introduce an auxiliary notion.

\begin{defi}
 Let $P$ be an \evold{2} process.
The set of  update patterns occurring in $P$, denoted 
 $\upd{P}$,  is inductively defined as follows:
$$
\begin{array}{ll}
 \upd{\update{a}{U}.Q} & =  \{U\} \cup \upd{U} \cup \upd{Q}\\  
\upd{\component{a}{P}} &= \upd{P} \\
 \upd{\pi.P} & =  \upd{P} ~~\text{if $\pi = a$ or $\pi = \outC{a}$}\\
 \upd{\sum_{i \in I}\pi_{i}.U_{i}} &= \bigcup_{i \in I}\upd{\pi_{i}.U_{i}} \\  
  \upd{! \pi.U} &= \upd{\pi. U} \\
\upd{U_{1} \parallel U_{2}} & =  \upd{U_{1}} \cup \upd{U_{2}} \\
\upd{\bullet} & = \emptyset 
\end{array}
$$
 This definition extends to sets of processes as expected.
 \end{defi}




\begin{defi}[Pred-basis]\label{d:predbasis}
 Let $S$ be a set of  \evold{2} processes and $P$ be an \evold{2} process such that $\Tree(P)\in \bigset{S}$. 
Given the set 
\begin{align*}
\mathcal{G}_{S, \widetilde{R}}  = & ~~\subp(S) \,  \cup \{  \component{a}{H} \mid a\in \cnames(S), ~ H \in \subp(S)\} \, \cup \\
& \qquad ~~~~~\qquad \qquad \{\component{a}{H} \mid R=\fillcon{U}{H}, ~R \in \widetilde{R}, ~U \in \mathsf{Upd}(S) , ~\numholes{U} \geq 1 \}
\end{align*}
the \emph{pred-basis of $P$ with respect to $S$}, denoted 
$pb_{S}(P)$, is defined as the set: 
\begin{align*}
pb_{S}(P) & =  
            \bigcup_{P  = K[\til R]} \big\{  Q \mid   Q \pired \succeq P, ~ Q= D[\til G], ~ D \in \mathsf{Ext}(K),  ~\til G \subseteq   \mathcal{G}_{S, \til{R}} \big\}
\end{align*}\smallskip
\end{defi}

\noindent We show that the well structured transition system given above has an effective pred-basis (cf. Definition \ref{d:efpb}).

\begin{thm}\label{th:pbs}
 Let $P$ and $S$ be a \evold{2} process   and a set of  \evold{2} processes, respectively.
 We then have that $\uparrow pb_{S}(P) = \uparrow \Pred_{S}(\uparrow P)$.
Moreover, $pb_{S}(\cdot)$ is effective.
\end{thm}
\proof
The inclusion $\uparrow pb_{S}(P) \subseteq \uparrow \Pred_{S}(\uparrow P)$
follows by construction. We consider the other inclusion, i.e.,
$\uparrow \Pred_{S}(\uparrow P) \subseteq \uparrow pb_{S}(P)$.
Given some $R \in \uparrow \Pred_{S}(\uparrow P)$, then 
we show that there is
a $Q \in pb_{S}(P)$ such that $Q \preceq  R$. 
 As hinted at above, depending on the kind of reduction that can occur to reach process $P$ we should consider six cases. 
Below,  $K, K_1$ and $K_2$ are syntactic contexts as in Definition \ref{def:SDE}:

\paragraph{\bf Reduction is ``internal'' to $P$} Then we have one of the following cases:
\begin{enumerate}[(1)]
 \item $P$ is obtained as an input/output synchronization. Then, $R = K_1[A, B]$ (or $R = K_1[B,A]$) where $A$ is either 
$!a.Q_1$ or 
$\sum_{i \in I} \pi_i.P_i$ with $\pi_{l}=a$ and $P_l = Q_1$ for some $l\in I$, and 
$B$ is either 
$!\outC{a}.Q_2$
or $\sum_{i \in I} \pi_i.R_i$ with $\pi_{l}=\outC{a}$ and $R_l = Q_2$ , for some $l\in I$.
   There exists $K_2$ such that  $P = K_2[Q_1, Q_2]$ and $R \pired \succeq P$. Since $R \in \uparrow \Pred_{S}(\uparrow P)$ then $A,B \in \subp(S)$  
and we can conclude $R \succeq Q = K_2[A,B] \in pb_S(P)$.
 \item if $P$ is the result of an update of an adaptable process then $$R = K_1[A, \component{a}{Q''}]$$  where $A$ is either 
$!\update{a}{Q'}.Q_1$ or 
$\sum_{i \in I} \pi_i.R_i$ with $\pi_{l}=\update{a}{Q'}$ and $R_l = Q_1$ for some $l\in I$, and there exists $K_2$ such that $P = K_2[Q_1, Q_2]$,  $R \pired \succeq P$ where we have that $Q_2 = \fillcon{Q'}{Q''}$. If $\numholes{Q'} =0$ then $Q_2 = Q'$ and as $R \in \uparrow \Pred_{S}(\uparrow P)$ we have 
$A, Q'' \in \subp(S)$ and  therefore $R \succeq Q =  K_2[A, \component{a}{\nil}] \in  pb_S(P)$. Otherwise if $\numholes{Q'} >0$ then $A \in \subp(S)$ and we can immediately conclude $R \succeq Q =  K_2[A, \component{a}{Q''} \in  pb_S(P)$.
\end{enumerate}

\paragraph{\bf Reduction partially present in $P$} Then we have one of the following cases:
\begin{enumerate}[(3)]
\setcounter{enumi}{2} 

\item[(3)] if $R = K_1[A, B]$ (or $R = K_1[B,A]$) where $A$ is either 
$!a.Q_1$ or 
$\sum_{i \in I} \pi_i.P_i$ with $\pi_{l}=a$ and $P_l = Q_1$ for some $l\in I$, and 
$B$ is either 
$!\outC{a}.Q_2$
or $\sum_{i \in I} \pi_i.R_i$ with $\pi_{l}=\outC{a}$ and $R_l = Q_2$ , for some $l\in I$.
Then there exists $K_2$ such that  $P = K_2[Q_1]$ and $R \pired \succeq P$. As $A, B \in \subp(S)$ (respectively $\outC{a}.Q_1, a.Q_2 \in \subp(S)$) we can conclude $R \succeq Q =  K_2[A] \parallel B \in  pb_S(P)$ (respectively $Q = K_2[B] \parallel A$). 

\item[(4)] if $R = K_1[A, \component{a}{Q_2}]$ where $A$ is either 
$!\update{a}{Q'}.Q_1$ or 
$\sum_{i \in I} \pi_i.R_i$ with $\pi_{l}=\update{a}{Q'}$ and $R_l = Q_1$ for some $l\in I$. Then there exists $K_2$ such that  $P = K_2[Q_1]$ and $R \pired \succeq P$. As  $A \in \subp(S)$ then  $R \succeq Q =  K_2[A] \parallel \component{a}{\nil} \in  pb_S(P)$.

\item[(5)] if $R = K_1[\component{a}{Q''},A$  where $A$ is either 
$!\update{a}{Q'}.Q_2$ or 
$\sum_{i \in I} \pi_i.R_i$ with $\pi_{l}=\update{a}{Q'}$ and $R_l = Q_2$ for some $l\in I$. 
Then there exists $K_2$ such that  $P = K_2[Q_1]$, $R \pired \succeq P$. If $\numholes{Q'} =0$ then $Q_1 = Q'$, $A \in \subp(S)$ and  we can conclude $R \succeq Q =  K_2[\component{a}{\nil}] \parallel A  \in  pb_S(P)$. Otherwise if $\numholes{Q'} >0$ then $Q_1 = \fillcon{Q'}{Q''}$ and we can conclude $R \succeq Q =  K_2[\component{a}{Q''}] \parallel A  \in  pb_S(P)$.
\end{enumerate}

\paragraph{\bf Reduction external to $P$}  Then we have:
\begin{enumerate}[(6)]
\setcounter{enumi}{5}
 
 \item  $R = K_1[P,  A, B]$ or $R = K_1[P, C, \component{a}{Q_3}]$ where $A$ is either 
$!a.Q_1$ or 
$\sum_{i \in I} \pi_i.P_i$ with $\pi_{l}=a$ and $P_l = Q_1$ for some $l\in I$,  
$B$ is either 
$!\outC{a}.Q_2$
or $\sum_{i \in I} \pi_i.R_i$ with $\pi_{l}=\outC{a}$ and $R_l = Q_2$ , for some $l\in I$ and $C$ is either 
$!\update{a}{Q_1}.Q_2$ or 
$\sum_{i \in I} \pi_i.P_i$ with $\pi_{l}=\update{a}{Q_1}$ and $P_l = Q_2$ for some $l\in I$.
 As all processes $A, B, C, Q_3$ are taken from $\subp(S)$ we can conclude $R \succeq Q =  P \parallel A \parallel B \in  pb_S(P)$ (respectively $Q =  P \parallel C \parallel \component{a}{Q_3} )$.\smallskip

\end{enumerate}

\noindent Moreover, the construction of $pb_{S}(Q)$ is effective. In particular, given a syntactic context $K$, there are finitely many ways of extending it with one or two holes so as to obtain a parallel extension  $D \in \mathsf{Ext}(K)$.
In Definition \ref{d:predbasis},
notice that when filling in the contexts with terms in $\til G$, both the set of sequential subprocesses and the ways of constructing an update pattern $U$
are finite.
This concludes the proof.\qed

\begin{thm}\label{th:fb}
Let $S$ be a set of \evold{2} processes.
$ (\deriv{S}, \pired, \preceq)$  is a finitely branching,
well-structured transition system
with strong compatibility, decidable $\preceq$,  and effective pred-basis $pb_S$.
Hence, it is possible to compute a finite basis \new{$\pbstar$} of $\Pred_{S}^*(I)$ (and $\Pred_{S}^+(I)$) for any upward-closed set $I$
which is given via a finite basis.
\end{thm}

\proof
Follows from 
Proposition \ref{predcomp}, using 
Remark \ref{r:fbranch}, and
Theorems \ref{th:scccs} and \ref{th:pbs}.\qed

\subsubsection{Step (\ref{quattro})}

Next, we define the basis of the set of processes that immediately 
exhibit a barb $\alpha$.

\begin{defi}\label{d:fb1}
 Let $S$ 
 and $\alpha$ 
 be a set of  \evold{2} processes  and a name $\alpha \in \{a, \outC{a} \mid a \in \mathcal{N}\}$, respectively.
Then,  we define:
$$\mathsf{fb}_{\alpha}(S)= \{R \in \subp(S) \mid R \downarrow_{\alpha} \}$$
\end{defi}\smallskip

\noindent Given an initial process $P$, a set of processes $M$,  and a barb $\alpha$, 
to determine whether \OG is decidable, we check if  there exists a process $R \in \BC_P^M$ such that  $R \barb{\alpha}^{k}$. 
It is sufficient to check if $R$ appears in the set of the predecessors of the processes that can exhibit $\alpha$ at least $k$ consecutive times. Since $\preceq$ imposes a well-quasi order on $\evold{2}$ processes, it is enough to characterize the set of predecessors by means of its finite basis, as shown by Theorem~\ref{th:fb}.
More precisely, if $k=1$ then it is  sufficient to check if $R$ is in the
set of predecessors of the processes 
in $\mathsf{fb}_{\alpha}(S)$, where $S = M \cup \{P\}$. 
Otherwise, if $k>1$ then we need 
we need to check for the existence of 
processes $R_{1}, \ldots, R_{k}$ such that 
$R \pired^{*} R_1 \pired \dots \pired R_{k}$, with $R_i \xrightarrow{\alpha}$ for $i \in [1..k]$. 
To do this, we proceed backwards. 
We begin by computing the finite basis 
 $\mathsf{fb}_{\alpha}(S)$; process 
$R_{k}$ should be in its upward closure. 
Then, 
we compute 
a finite basis for the set of processes in 
$\Pred_{S}(\mathsf{fb}_{\alpha}(S))$ 
which exhibit $\alpha$ immediately; $R_{k-1}$ should be in the upward closure of this finite basis, which 
is constructed as follows.
Notice by virtue of Theorem \ref{th:pbs},
 we can rely on the pred-basis given by Definition \ref{d:predbasis}, i.e., $pb_S\big(\mathsf{fb}_{\alpha}(S)\big)$, in this case. 
We consider two classes of elements of $pb_S\big(\mathsf{fb}_{\alpha}(S)\big)$: 
the first one is composed of those processes that can immediately 
perform $\alpha$, while the second contains the rest.
The desired finite basis is obtained by taking 
the set of processes containing 
(i) every process in the first class and 
(ii) every $Q$ in the second class (but with a minimal modification,
with respect to the ordering $\preceq$, 
in such a way it can exhibit $\alpha$ immediately).
The latter is achieved by 
function $\mathsf{Add}_B(Q)$ (cf. Definition \ref{def:fbk})
which ``plugs'' 
into every $Q$ 
a process in $\mathsf{fb}_{\alpha}(S)$ either in parallel at the top level or inside an adaptable process.
This procedure iterates 
as expected; 
each iteration considers the predecessors of the elements of the finite basis obtained in the previous one.
In the last step, in order to calculate all the predecessors of process $R_1$ we apply Theorem \ref{th:fb}, thus obtaining a finite basis \new{$\pbstar$} where  
it is
sufficient to check whether $R$ belongs to 
its upward closure. 
More formally:

\begin{defi}\label{def:fbk}
 Let $S$ be a  set of  \evold{2} processes. 
Given the following set definitions (with $C$ being a monadic context as in Definition \ref{d:mc})
\begin{align*}
\mathsf{Add}_B(Q) & =  \{Q \parallel R \mid R \in B \} \cup \{C\big[\component{a}{R \parallel Q_{i}}\big] \mid Q = C\big[\component{a}{Q_{i}}\big], ~R \in B\}\\
\mathsf{Ib}_{\alpha}(A, B) & =  \{Q \in A \mid Q \xrightarrow{\alpha} \} \cup \{ \mathsf{Add}_B(Q) \mid Q\in A \text{ and } Q \not \xrightarrow{\alpha}\} 
\end{align*}
we define the finite basis $\mathsf{FB}_{\alpha,k}(S)= \pbstar(\mathsf{B}_{\alpha,k}(S))$ where $k \geq 1$ and
$$
\mathsf{B}_{\alpha,k}(S) =
\begin{cases}
\mathsf{fb}_{\alpha}(S) & \text{if } k=1\\
\mathsf{Ib}_{\alpha}\Big(pb_S\big(\mathsf{B}_{\alpha, k-1}(S)\big), \mathsf{fb}_{\alpha}\big(S\big)\Big)
& \text{otherwise}
\end{cases}
$$\smallskip
\end{defi}

%

\noindent The effectiveness of $\mathsf{FB}_{\alpha,k}$ will allow us to prove the decidability of \OG. 

\begin{lem}\label{lem:decpred}
 Let $S$ be a set of  \evold{2} processes, 
and let $\alpha \in \{a, \outC{a} \mid a \in \mathcal{N}\}$. 
Then, $\mathsf{FB}_{\alpha, k}(S)$ is effective.
\end{lem}
\proof
The effectiveness of the calculation of the finite basis of $\Pred_S^*(\cdot)$ follows from Theorem~\ref{th:fb}. 
The set $\mathsf{Ib}_{\alpha}(\cdot,\cdot )$ is finite and hence can be computed as defined above.
Moreover, it is easy to see that it is a finite basis representing all the predecessors of $\mathsf{fb}_{\alpha}(S)$, which in turn can immediately exhibit $\alpha$. \qed 

\subsubsection{Step (\ref{cinque})}\label{ss:cinco}

\new{We conclude by showing how to determine whether there exists a process $R$ in $\BC_P^M$ that exhibits $\alpha$.}
Recall that $\Par(P)$ 
is the set  of all 
processes 
 and all adaptable processes in $P$ which are 
in parallel at top level (see Definition \ref{d:pps}).
We can finally state:

\begin{thm}\label{th:badec}
\OG  is decidable for \evold{2}.
\end{thm}
\proof
Let $P$ and $M=\{T_1, \dots, T_n\}$ be an initial process and a set of \evold{2} processes, respectively.
In order to show that \OG is decidable, it suffices to check 
that, given some $\alpha$ and $k \geq 1$, 
there exists a process $R \in \BC_P^M$ such that $R \barb{\alpha}^{k}$. 
More precisely, letting $S = \{P\} \cup M$, we have to check if there exists a process $Q \in \mathsf{FB}_{\alpha,k}(S)$ such that $Q \preceq R$.
From Lemma \ref{lem:decpred}, we know that it is possible to compute the set $\mathsf{FB}_{\alpha,k}(S)$. 
%
Then, for each $Q_{i}\in \mathsf{FB}_{\alpha,k}(S)$ we analyze the processes in $\mathsf{Par}(Q_{i})$ (cf. Definition \ref{d:pps}). 
Let $V$ be the set of the processes $Q'_{j}$ in $\mathsf{Par}(Q_{i})$
such that $Q_{j}' \preceq T$, for some $T \in M$.
We now consider $Q_{i}^{*}$, the process obtained
by $Q_{i}$ by removing all the occurrences of the parallel
processes in $V$. 
At this point, it is enough to check whether $Q_{i}^{*}\preceq P$. 
If this is the case, for at least one $Q_{i} \in \mathsf{FB}_{\alpha,k}(S)$,
then we can conclude that there exists
$R \in \BC_P^M$ such that $R \barb{\alpha}^{k}$;
otherwise there exists no $R \in \BC_P^M$ such that $R \barb{\alpha}^{k}$.
\qed

\noindent Note that 
the decidability result extends to \evold{3}, as it is a subcalculus of  \evold{2}.
Moreover, 
by virtue of Theorems \ref{stdynequiv} and  \ref{th:clusterstat}, 
decidability of \OG extends also to 
\evols{2} and \evols{3}. We have:

\begin{cor}
\OG is decidable for \evold{3}, \evols{2}, and \evols{3}.\qed
\end{cor}

\subsection{Undecidability of Eventual Adaptation} \label{sec:unde2d}
Here we show that  \LG  is undecidable in \evols{2} by relating it to termination  in \mmss;
this result carries over to \evols{1}, \evold{1},  and \evold{2}---see Corollary \ref{cor:ba-undec}.
This relationship is obtained by defining an encoding tailored to the features of the property.
In contrast to the encoding given in Section \ref{s:ev1}, 
the encoding presented here is \emph{non faithful} as it may 
perform erroneous tests for zero on the registers (i.e.
in the simulation of the \mm a register
is assumed to contain the value zero even if this is not the case). Nevertheless, we are able to define encodings that repeatedly simulate finite computations of the \mm, and if the number of repeated simulations is infinite, then we have the guarantee that the number of erroneous steps is finite. Thus infinitely
many of the performed simulations are correct.
This way, the \mm terminates iff its encoding has a non terminating computation. As during its execution the encoding continuously exhibits a barb on $e$, 
it 
then follows that \LG is undecidable for \evols{2} processes.

The encoding relies on finitely many output prefixes acting as \emph{resources} on which instructions of the \mm depend in order to be executed.
To repeatedly simulate finite runs of the \mm,
at the beginning of the simulation
the encoding produces finitely many instances of these resources. 
When $\mathtt{HALT}$ is reached, 
the registers are reset, some of the consumed resources 
are restored, and a new simulation is restarted from the  first instruction.
In order to guarantee that an infinite computation
of the encoding 
contains only finitely many
erroneous jumps, 
finitely many instances of 
a second kind of
resource (different from that required to execute instructions) 
are produced. 
Such a resource is consumed 
by increment instructions and restored by 
decrement instructions.
When the simulation performs a jump, the tested register is reset: 
if it was not empty (i.e., an erroneous test) then some resources are permanently lost. 
When 
the encoding runs out of 
resources, the simulation will
eventually block as increment instructions can no longer be simulated.
We make two non restrictive assumptions. First, we assume that a
\mm computation contains at least one increment instruction.
Second, in order to avoid resource loss at the end
of a correct simulation run, we
 assume that \mm computations terminate
with both the registers empty.

\begin{table}[t]
%
 \begin{tabular}{l} 
$\controll ~= ~!a.(\outC{f} \parallel \outC{b} \parallel  \outC{a}) \parallel \outC{a}.a.(\outC{p_1} \parallel e) \parallel \
!h.(g.\outC{f} \parallel \outC{h})$
  \\
$\mathrm{\textsc{Register}}~r_j $ \\
\ \  $  
\encp{r_j = m}{\mmn{2}}=\left\{  
\begin{array}{ll}  
 \component{r_j}{!inc_j.\outC{u_j} \parallel \outC{z_j}}  & \textrm{if } m= 0 \\  
 \component{r_j}{!inc_j.\outC{u_j} \parallel \prod^{m}\overline{u_j} \parallel \outC{z_j} }  
~~& \textrm{if } m> 0 .  
\end{array}\right.  
$ \\ 
$   
\begin{array}{ll}   
\multicolumn{2}{l}{\mathrm{\textsc{\!\!Instructions}}~(i:I_i)}\\  
\encp{(i: \mathtt{INC}(r_j))}{\mmn{2}}&  =   !p_i.f.(\outC{g} \parallel b.\outC{inc_j}.
\outC{p_{i+1}})\\  
\encp{(i: \mathtt{DECJ}(r_j,s))}{\mmn{2}} \! \! \! \! \!&  =  !p_i.f.\big(\outC{g} \parallel (u_j.(\outC{b} \parallel
           \outC{p_{i+1}}) \\
& \qquad +   z_j.\update{r_j}{\component{r_j}{!inc_j.\outC{u_j} \parallel \outC{z_j}}}. \outC{p_s})\big) \\
\encp{(i: \mathtt{HALT})}{\mmn{2}}&  =  !p_i.
                  \outC{h}.h.\update{r_0}{\component{r_0}{!inc_0.\outC{u_0} \parallel \outC{z_0}}}.\update{r_1}{\component{r_1}{!inc_1.\outC{u_1} \parallel \outC{z_1}}}.\outC{p_1}
\end{array}   
$ 
\end{tabular}
\caption{Encoding of \mmss into $\evols{2}$} \label{tab:minskyccsbs}
\end{table}

We now discuss the encoding defined in 
Table~\ref{tab:minskyccsbs}. 
We first comment on \controll, 
the process that manages  the resources.
It is composed of three processes in parallel.
The first replicated process is able  
to produce an unbounded amount of processes $\outC{f}$ and 
$\outC{b}$, which represent the two kinds of resources
described above. The second process starts and stops
a resource production phase by performing $\outC{a}$
and $a$, respectively. Then, it starts the \mm simulation
by emitting the program counter $\outC{p_{1}}$.
The third process is used at the end of the simulation
to restore some of the consumed resources $\outC{f}$
(that are transformed in $\outC{g}$, see below).

A register $r_j$ that stores number $m$ is encoded as 
an adaptable process at $r_{j}$ containing 
$m$ copies of the unit process $\outC{u_j}$. 
It also contains process $!inc_j.\outC{u_j}$ which allows to create further copies of $\outC{u_j}$
when an increment instruction is executed. 
Instructions are encoded as replicated processes guarded by $p_i$.
Once $p_i$ is consumed, 
increment and decrement instructions
consume one of the resources 
$\outC{f}$. If such a resource is available then it is renamed as
$\outC{g}$, otherwise the simulation blocks.
The simulation of an increment instruction also consumes an instance of resource $\outC{b}$.

The encoding of a decrement-and-jump instruction is 
slightly more involved.
It is implemented as a choice:  
the process can either perform a decrement and proceed 
with the next instruction, or to jump.
In case the decrement can be executed (the input $u_{j}$
is performed) then a resource $\outC{b}$ is restored.
The jump branch can be taken even if the register is not
empty. In this case, the register is reset via an
update that restores the initial state of the adaptable process at $r_{j}$.
Note that if the register was not empty, then some processes $\outC{u_{j}}$ are lost.
Crucially, this causes a permanent loss of a corresponding amount of resources $\outC{b}$, 
as these are only restored when process $\outC{u_{j}}$  
are consumed during the simulation of a decrement.

The simulation of the $\mathtt{HALT}$ instruction
performs two tasks before restarting 
the execution of the encoding  by reproducing the program counter $p_{1}$.
The first one is to restore some of the consumed
resources $\outC{f}$: this is achieved by  the third process of 
\controll, 
which 
repeatedly consumes one instance of $\outC{g}$
and produces one instance of $\outC{f}$. 
This process is started/stopped by executing the two prefixes $\outC{h}.h$.
The second task is to reset the registers by  
updating the adaptable processes at $r_{j}$ with their initial state.

The full definition of the encoding is as follows.

\begin{defi}\label{def:minskyconfs}
 Let $N$ be a \mm, with registers $r_0$, $r_1 $ and instructions
$(1:I_1) \ldots (n:I_n)$. 
Given the \controll process and the encodings in 
Table~\ref{tab:minskyccsbs}, the encoding of $N$ in \evols{2} (written $\encp{N}{\mmn{2}}$)
is defined as
$
\encp{r_0 = 0}{\mmn{2}} \parallel \encp{r_1 = 0}{\mmn{2}} \parallel \prod^{n}_{i=1} \encp{(i:I_i)}{\mmn{2}}  
 \parallel \controll$.
\end{defi}

As discussed above, the encoding has 
an infinite sequence of simulation runs if and only 
if the corresponding \mm terminates. As the barb $e$ is continuously
exposed during the computation (the process $e$ is spawn 
with the initial program counter and is never consumed), we can conclude
that a \mm terminates if and only if its encoding
does not eventually terminate the simulation runs. As
during the simulation runs the barb $e$ is always exhibited,
this coincides with checking whether the encoding
does not eventually adapt.

\begin{lem} \label{th:corrE2}
 Let $N$ be a \mm. $N$ terminates iff $\encp{N}{\mmn{2}} \barbw{e}$.
\end{lem}

\proof
See Appendix \ref{app:e2}, Page \pageref{app:e2}.\qed

Exploiting  Lemma \ref{th:corrE2}
and proceeding exactly as the proof of
Theorem \ref{th.ev1} for \evols{1}, 
we can state the following.

\begin{thm}\label{th:ev2}
\LG is undecidable in \evols{2}.\qed
\end{thm}

Similarly as in that case, 
undecidability  extends also to \evold{2}, \evols{1} and \evold{1}. This easily follows from the fact that \evols{2} is a subcalculus of \evols{1} and from Lemma \ref{lem:statvsdyn}, since $\encp{N}{\mms}$ is a process in \evols{2} that does not contain any nested adaptable processes.

\begin{cor}\label{cor:ba-undec}
\LG is undecidable in \evols{1}, \evold{1},  and \evold{2}.\qed
\end{cor}

Note that
the encoding $\encp{\cdot}{\mmn{2}}$ 
uses processes
that do not modify the topology 
of nested adaptable processes;  
update prefixes 
do not remove nor create adaptable processes:
they simply remove the processes currently in the updated 
locations and replace them with the predefined initial content.
One may wonder whether the ability to remove processes
is necessary for the undecidability result: 
next 
we show that this is not the case.


\section{(Un)decidability Results for \evol{3}}\label{s:ev3}

\subsection{Undecidability of Eventual Adaptation in \evold{3}}
Here we prove that \LG 
is undecidable for \evold{3} processes.
We obtain this result by means of a non-faithful encoding of \mmss similar to the one presented before.

In that encoding, Definition \ref{def:minskyconfs}, process deletion was used
to restore the initial state inside the adaptable processes representing
the registers. In the absence of process deletion, 
we use a more involved technique based on the
possibility of moving processes to a different context:
processes to be removed are guarded by an update
prefix $\update{c_j}{\component{c_j}{\bullet}}$
that simply tests for the presence of a parallel adaptable process at $c_{j}$;
when a process must be deleted, it is 
``collected'' inside $c_{j}$, thus
disallowing the possibility to execute such an update prefix.

\begin{table}[t] 
%
\centering  
{ 
\begin{tabular}{l}   
$\controll ~= ~!a.(\outC{f} \parallel \outC{b} \parallel  \outC{a}) \parallel \outC{a}.a.(\outC{p_1} \parallel e) \parallel \
!h.(g.\outC{f} \parallel \outC{h})$
  \\
$\mathrm{\textsc{Register}}~r_j $ \\ 
\ \ $  
\encp{r_j = 0 }{\mmn{3}}= \component{r_j}{Reg_j \parallel \component{c_j}{\nil}} 
$ \\
\ \ $\mbox{with~} Reg_j = !inc_j.\update{c_j}{\component{c_j}{\bullet}}.\outC{ack}.u_j.\update{c_j}{\component{c_j}{\bullet}}.\outC{ack}$
\\
$
\begin{array}{lll}
\multicolumn{3}{l}{\! \! \mathrm{\textsc{Instructions}}~(i:I_i)}\\  
\encp{(i: \mathtt{INC}(r_j))}{\mmn{3}}&  = &  !p_i.f.(\outC{g} \parallel
b.\outC{inc_j}.
ack.\outC{p_{i+1}}) \\
\encp{(i: \mathtt{DECJ}(r_j,s))}{\mmn{3}}&  = & !p_i.f.\big(\outC{g} \parallel(\outC{u_j}.
ack.(\outC{b} \parallel 
\outC{p_{i+1}})  + \\
&& \qquad \qquad \qquad \update{c_j}{\bullet}.\update{r_j}{\component{r_j}{Reg_j \parallel
\component{c_j}{\bullet}}}.\outC{p_s})\big) \\
\encp{(i: \mathtt{HALT})}{\mmn{3}}&  = & !p_i.
\outC{h}.h.\update{c_0}{\bullet}.\update{r_0}{\component{r_0}{Reg_0 \parallel \component{c_0}{\bullet}}}. \\
&& \qquad \qquad \qquad \update{c_1}{\bullet}.\update{r_1}{\component{r_1}{Reg_1 \parallel
\component{c_1}{\bullet}}}.\outC{p_1}
\end{array}   
$
\end{tabular}
}  
\caption{Encoding of \mmss into \evold{3}.}  
\label{t:encod-evold3d}  
\end{table}

The encoding is as in 
Definition~\ref{def:minskyconfs}, with registers
and instructions  as  in Table~\ref{t:encod-evold3d}:

\begin{defi}\label{def:minskyconfsevol3d}
 Let $N$ be a \mm, with registers $r_0$, $r_1 $ and instructions
$(1:I_1) \ldots (n:I_n)$. 
Given the \controll process and the encodings in 
Table~\ref{t:encod-evold3d}, the encoding of $N$ in \evold{3} (written $\encp{N}{\mmn{3}}$)
is defined as
$
\encp{r_0 = 0}{\mmn{3}} \parallel \encp{r_1 = 0}{\mmn{3}} \parallel \prod^{n}_{i=1} \encp{(i:I_i)}{\mmn{3}}  
 \parallel \controll$.
\end{defi}

A register $r_j$ that stores number $m$ is encoded as an adaptable process at $r_j$ that
contains $m$ copies of the unit process $u_j.\update{c_j}{\component{c_j}{\bullet}}.\outC{ack}$. 
It also contains
process $Reg_j$, which creates further copies of the unit
process when an increment instruction is invoked, as well as 
the collector $c_j$, which  is used to store the
processes to be removed.

An increment instruction adds an occurrence of $u_j.\update{c_j}{\component{c_j}{\bullet}}.\outC{ack}$. Note that 
an output 
$\outC{inc}$ could synchronize with the corresponding input inside a collected process. This immediately leads to deadlock as the containment induced by $c_j$ prevents further interactions.
The encoding of a decrement-and-jump instruction is implemented as a
choice, following the idea discussed for the static case.
If the process guesses that the register is zero then, before jumping to the given instruction, it proceeds at disabling its current content: 
this is done by 
(i) removing the boundary of the collector $c_{j}$ 
leaving its content at the top-level, 
and (ii) updating the register placing its previous state in the collector.
A decrement simply consumes one occurrence of $u_j.\update{c_j}{\component{c_j}{\bullet}}.\outC{ack}$. 
Note that as before 
the output 
$\outC{u_j}$ could synchronize with the corresponding input inside a collected process. 
Again, this immediately leads to deadlock.
The encoding of $\mathtt{HALT}$ exploits the same mechanism of
collecting processes to simulate the reset of the registers.

This encoding has the same properties of the one discussed
for the static case. In fact, in an infinite simulation the collected
processes are never involved, otherwise the computation would block.

\begin{lem}\label{th:corrE3}
 Let $N$ be a \mm. $N$ terminates iff $\encp{N}{\mms} \barbw{e}$.
\end{lem}
\proof
See Appendix \ref{app:e31}, Page~\pageref{app:e31}.\qed

Lemma \ref{th:corrE3} allows to conclude 
that $\LG$ is undecidable for processes in \evold{3}. The proof of the following theorem proceeds  as the proofs of Theorems \ref{th.ev1} and \ref {th:ev2}.

\begin{thm}
\LG is undecidable in \evold{3}.\qed
\end{thm}

We can conclude that process deletion is not necessary for proving 
the undecidability of $\LG$ in \evold{3}. Nevertheless, in the encoding
in Table~\ref{t:encod-evold3d} we need to use the possibility to 
remove and create adaptable processes (namely, the collectors $c_{j}$
are removed and then reproduced when the registers must be reset).
One could therefore wonder whether $\LG$ is still undecidable if we 
remove from \evold{3}  the possibility to remove 
processes. Next we  show that this is not the case.




\subsection{Decidability of Eventual Adaptation in \evols{3}}\label{sec:evol3}

We prove the decidability of \LG in \evols{3} 
by resorting to Petri nets. Namely, we reduce the 
eventual adaptation problem for \evols{3}
to the 
infinite visit problem (cf. Definition \ref{d:infvisit}).
%
%


Before formally defining the encoding of \evols{3} processes into Petri nets, we give some intuitions.
The idea is to use the markings of the Petri net
to represent the active sequential subprocesses and the available
adaptable processes. Transitions are used to model the execution of
actions. More precisely, each active sequential subprocess is represented
by one token. Two tokens corresponding to two sequential subprocesses
able to execute complementary actions can fire a transition,
whose effect is to produce tokens representing the two continuations. 
As for update actions,
they are represented by transitions that consume (at least) two 
tokens: one token corresponding to the process executing the update and another token
representing the adaptable process target of the update operations. 
In order to ensure that update actions take place between processes which are in parallel, 
we keep track of the adaptable processes in which
a process is included: we do so by decorating its place with a list of outer adaptable processes.
Intuitively, this list represents the ``address'' of a single adaptable process within the nested structure 
of adaptable processes.

We now present some auxiliary notations required by the definition.
Let $P$ be a process of $\evols{3}$ and 
$M=\{P_{1},\ldots,P_{n}\}$ be a set of processes of $\evols{3}$.
It is not restrictive to assume that all the update actions on a given adaptable process  can be executed:
even if the static semantics decrees that update actions should satisfy conditions on the nesting structure of adaptable processes,
Theorem~\ref{stdynequiv} ensures the existence of an \evold{} process with the same behavior for which such conditions are always true.
Let $\pseq(P,M)$ be the set of sequential subprocesses in 
$P,P_{1},\ldots,P_{n}$ and let $\ambpaths(P,M)$ be the 
set of location names nestings, i.e. strings composed
of names of nested locations, starting from the outermost
adaptable process, occurring in one of the processes $P,P_{1},\ldots,P_{n}$.
We use $\sigma, \theta$ to range over strings in $\ambpaths(P,M)$, 
and write $\sigma a$ for the string obtained from concatenating $\sigma$ and $a$.

\begin{defi}\label{d:pn}
Let $P$ and $M=\{P_{1},\ldots,P_{n}\}$ be \evols{3} processes. Its
associated Petri net is defined as the triple
$$\pnr{P}{M}=(\places{P,M},\transit{P,M},\initMark{P})$$
where
\begin{enumerate}[$\bullet$]
\item $\places{P,M}=  \{\coppia{P}{\sigma}\ |\  P \in 
\pseq(P,M), \, \sigma \in \ambpaths(P,M)\} \cup \ambpaths(P,M) \cup
\{start, go\}$,  with $start$ and $go$ being two distinguished auxiliary places.


\item $\transit{P,M}$ contains all the instances of the transition schemata reported in Table~\ref{tab:pn} over the set of places $\places{P,M}$.


\item $\initMark{P} = \decc{\varepsilon}{P} \uplus \{start\}$,
with $\decc{\sigma}{P}$
defined inductively as follows: 
\begin{align*}
\decc{\sigma}{\component{a}{P}} & =  \decc{\sigma a}{P} \uplus \{\sigma a\} \\
\decc{\sigma}{P \parallel P'} & =  \decc{\sigma}{P} \uplus \decc{\sigma}{P'}\\ 
\decc{\sigma}{P} & =  \{\coppia{P}{\sigma}\} \quad \mbox{otherwise} 
\end{align*}
where 
$\varepsilon$
corresponds to the empty string and
$\uplus$ denotes multiset union.\smallskip
\end{enumerate}

\end{defi}

\noindent We now describe the Petri net computation by giving intuitions on the transitions presented in Table~\ref{tab:pn}.
The initial marking includes one token in the place $start$
plus the tokens corresponding to the active 
sequential subprocesses of $P$. 
%
The token in $start$ allows to generate an arbitrary amount 
of copies of the processes $P_{1},\ldots,P_{n} \in M$ (Transition (1)). This is simply achieved
by considering $n$ transitions, such that the $i$-th transition
tests for the presence of the token in $start$ and then produces the 
sequential subprocesses of $P_{i}$.
Nondeterministically, the token is moved from $start$ to $go$ (Transition (2)).
At this point, the simulation of the evolution of the generated
configuration is started. As described above, synchronizations 
between complementary actions are modeled by transitions that consume
the tokens corresponding to the two synchronizing processes
and then produce the sequential subprocesses in the continuations.
Transitions (3)--(5) cover the different cases in which an input/output synchronization
can arise (namely, interaction between two guarded processes, between a replicated processes and a guarded process, and
between two replicated processes), while
Transitions (6)--(9) cover the cases 
in which a synchronization corresponds to an update action.
In the latter kind of transitions, we need to check the availability
of a target adaptable process, but this adaptable process should not enclose
the updating process (as in, e.g., $\component{a}{\update{a}{U} \parallel P}$).
More precisely, 
suppose there is a process $Q$ executing an update action on name $a$, and let $\sigma$ 
be the string of the names of the adaptable processes enclosing $Q$.
The availability of a target adaptable process can be checked by verifying 
the presence of  a token in a place $\theta a$ which is not a prefix of $\sigma$
(see Transitions (6) and (8)).
If $\theta a$ is a prefix of $\sigma$, then the adaptable process at $\theta a$ could enclose $Q$.
In such a  case, it is sufficient to check that the place $\theta a$ contains at least two tokens, 
thus indicating the
existence of a different adaptable process with the same path but that does not enclose $Q$
(see Transitions (7) and (9)).


\begin{table}[t]
\[
\begin{array}{ll}
(1) & \{start\} \derriv \{start\} \uplus \decc{\varepsilon}{P_{i}}\ \ \ \ \ \text{with}\ P_{i} \in M 
\\
\\
(2)& \{start\} \derriv \{go\}
\\
\\
(3)&\{go,  \coppia{\sum_{i \in I} \pi_i.A_i}{\sigma},
           \coppia{\sum_{j \in J} \rho_j.B_j}{\theta}\}
          \derriv
          \{go\}\uplus \decc{\sigma}{A_{l}}
\uplus \decc{\theta}{B_{m}}
          \\
          &
          \mbox{if $\pi_{l}=a$ and $\rho_{m}=\overline{a}$
          (for $l \in I$, $m \in J$)}
          \\
          \\
          
(4)&\{go, \coppia{!\pi.A}{\sigma},
           \coppia{ \sum_{j \in J} \rho_j.B_j}{\theta}\}
          \derriv \\
& \qquad \qquad \quad
          \{go, \coppia{!\pi.A}{\sigma}\} \uplus
\decc{\sigma}{A} \uplus \decc{\theta}{B_{m}}
          \\
          &
          \mbox{if $\pi=a$ (resp. $\overline{a}$) and $\rho_{m}=\overline{a}$ (resp. $a$)
          (for $m \in J$)}
          \\
          \\
          
(5)& \{go, \coppia{!\pi.A}{\sigma},
          \coppia{!\rho.B}{\theta}\}
          \derriv\\
& \qquad \qquad \quad
          \{go, \coppia{!\pi.A}{\sigma}, 
\coppia{!\rho.B}{\theta} \} \uplus
\decc{\sigma}{A}  \uplus \decc{\theta}{B_{m}}
          \\
          &
          \mbox{if $\pi=a$ (resp. $\overline{a}$) and $\rho=\overline{a}$ (resp. $a$)
          }
          \\
          \\

(6)& \{go, \coppia{\sum_{i \in I} \pi_i.A_i}{\sigma},\theta a\}
 \derriv\\
& \qquad \qquad \quad
 \{go\} \uplus \decc{\sigma}{A_{l}} \uplus 
 \decc{\theta}{A} 
 \uplus \decc{\theta a}{U}
 \uplus \{\theta a\}
           \\
           &
           \mbox{if $\theta a$ is not a prefix of $\sigma$,
           $\pi_{l}=
           \update{a}{\component{a}{U} \parallel A}$
           (for $l \in I$)}    
\\
\\
(7)& \{go, \coppia{\sum_{i \in I} \pi_i.A_i}{\sigma},\theta a,\theta a\}
 \derriv \\
& \qquad \qquad \quad
 \{go\} \uplus \decc{\sigma}{A_{l}} \uplus 
 \decc{\theta}{A}
 \uplus \decc{\theta a}{U}
 \uplus \{\theta a,\theta a\}
           \\
           &
           \mbox{if $\theta a$ is a prefix of $\sigma$,
           $\pi_{l}=
           \update{a}{\component{a}{U}\parallel A}$
           (for $l \in I$)}   

\\
\\
(8)& \{go, \coppia{!\pi.A'}{\sigma},\theta a\}
 \derriv \\
& \qquad \qquad \quad
 \{go, \coppia{!\pi.A'}{\sigma}\} \uplus
 \decc{\sigma}{A'} \uplus 
 \decc{\theta}{A} 
 \uplus \decc{\theta a}{U}
 \uplus \{\theta a\}
           \\
           &
           \mbox{if $\theta a$ is not a prefix of $\sigma$,
           $\pi=
           \update{a}{\component{a}{U} \parallel A}$
           }    
\\
\\
(9)& \{go, \coppia{!\pi.A'}{\sigma},\theta a,\theta a\}
 \derriv \\
 & \qquad \qquad  \quad \{go,  \coppia{!\pi.A'}{\sigma}\} \uplus \decc{\sigma}{A'} \uplus 
 \decc{\theta}{A} 
 \uplus \decc{\theta a}{U}
 \uplus \{\theta a,\theta a\}
           \\
           &
           \mbox{if $\theta a$ is a prefix of $\sigma$,
           $\pi=
           \update{a}{\component{a}{U} \parallel A}$
           }                          
\end{array}
\]
\caption{Transition schemata for the Petri net representation of $\evols{3}$ processes in Definition~\ref{d:pn}.}\label{tab:pn}
\end{table}

We now state the correspondence between 
processes
and their associated Petri net.

\begin{lem}\label{lem:initpetri}
 Let $P$ be a process of $\evols{3}$,
and $M$ be the set $\{P_{1},\cdots,P_{n}\}$ and $(\places{P,$ $M}, \transit{P,M}, \initMark{P})$
 be their associated Petri net,  as in Definition \ref{d:pn}. Then, given a marking $m$, we have
$\initMark{P} \rightarrow^* \{start\} \uplus m \rightarrow   \{go\} \uplus m$ iff  $m = \decc{\varepsilon}{R}$, for some $R \in \BC_P^M$.
\end{lem}
\proof
 Follows by construction of the Petri net.\qed

\begin{lem}\label{l:pn}
 Let $P$ and $(\places{P,\emptyset}, \transit{P,\emptyset}, \initMark{P})$
 be an \evols{3} process and its associated Petri net,  as in Definition \ref{d:pn}.
 Then we have:
$$P \pired P' \textrm{ iff } \decc{\varepsilon}{P} \uplus \{go\} \rightarrow  \decc{\varepsilon}{P'} \uplus \{go\}.$$
\end{lem}
\proof
See  Appendix \ref{app:e32}, Page~\pageref{app:e32}.\qed

The decidability of \LG for $\evols{3}$ follows
from the decidability of the existence of a suffix
of an infinite computation composed of markings with at least
one token in some given places. 

\begin{thm}\label{th:pnev3}
Let $P$ be a process of $\evols{3}$,
and let $M$ be the set $\{P_{1},\cdots,P_{n}\}$ of 
processes of $\evols{3}$.
Consider $S = \CStrs(P) \cup \CStrs(P_{1}) \cup \cdots \cup \CStrs(P_{n})$,
and let $P' = \dyn{P}$ and $M' = \{\dyn{P_{1}},\cdots,\dyn{P_{n}}\}$.
Let $\alpha$ be a barb.
We have that $P$ and $M$ satisfies \LG for the
barb $\alpha$ iff the 
Petri net $$(\places{P',M'}, \transit{P',M'}, \initMark{P'})$$
has an infinite computation with a suffix composed
of markings with one token in $go$ and 
with at least one token in one of the places 
$\coppia{\sum_{i \in I} \pi_i.A_i}{\theta}$, with 
$\pi_{l}=\alpha$ for some $l\in I$, or
$\coppia{!\alpha.A}{\theta}$. 
\end{thm}

\proof
Suppose that $P$ and $M$ satisfies \LG for the
barb $\alpha$ then there exists a process $R \in \BC_P^M$ such that $R \barb{\alpha}^{\omega}$.
Following from Lemma \ref{lem:initpetri} there exists an initial computation of the Petri net that reaches the marking $\decc{\varepsilon}{R} \uplus \{ go \}$. Then following from Lemma \ref{l:pn} there exists an  infinite computation with a suffix composed
of markings with at least one token in one of the places 
$\coppia{\sum_{i \in I} \pi_i.A_i}{\theta}$, with 
$\pi_{l}=\alpha$ for some $l\in I$, or
$\coppia{!\alpha.A}{\theta}$. Notice that in all of these
markings, the place $go$ contains one token.

Similarly if there exists   an  infinite computation with a suffix composed
of markings with one token in $go$ and at least one token in one of the places 
$\coppia{\sum_{i \in I} \pi_i.A_i}{\theta}$, with 
$\pi_{l}=\alpha$ for some $l\in I$, or
$\coppia{!\alpha.A}{\theta}$ then for Lemma \ref{lem:initpetri} and Lemma \ref{l:pn}  we know that there exists a process $R \in \BC_P^M$ such that $R \barb{\alpha}^{\omega}$.\qed

The check 
of the existence of an infinite computation
with a suffix composed of markings with one token in $go$ and
with at least one
token in some given places 
corresponds to the infinite visit problem (Definition~\ref{d:infvisit}).
Thus since this problem is decidable (Theorem \ref{th:infiniteVisit}) it follows that \LG is decidable in \evols{3}.


\section{Related Work and Discussion}\label{s:rw}

We now comment on the 
origin and 
motivations for the constructs of \evol{}, review some related works, 
describe a modeling technique derived from \OG and \LG, 
and discuss variants of the adaptation problems considered here.

\subsection{On the Constructs for Evolvability}\label{ss:facs}


The origins of  the \evol{} calculus can be traced back to our own previous work on expressiveness and decidability results for core 
 \emph{higher-order process calculi} (see, e.g., \cite{LaneseIC10,GiustoPZ09,Perez10}). 
Below, we overview these previous works, and discuss the motivations 
that led us from higher-order communication to adaptable processes.
 
Higher-order (or \emph{process-passing}) concurrency is often presented as an alternative paradigm to the first-order (or \emph{name-passing})
concurrency of the $\pi$-calculus for the description of mobile systems. 
As in the 
$\lambda$-calculus, 
higher-order process calculi 
 involve \emph{term instantiation}: 
a computational step 
results in the instantiation of a variable with a term, which is copied as many times as there are occurrences of the variable.
The basic  operators of these calculi are usually those of
CCS: parallel composition, input and output prefix, and restriction.
Replication and recursion 
can be encoded.
Proposals of higher-order process calculi include
the higher-order $\pi$-calculus~\cite{San923}, 
Homer~\cite{HilBun04}, and Kell~\cite{SchmittS04}.

With the purpose of investigating expressiveness and decidability issues in 
the hi\-gher-order paradigm, 
a \emph{core} higher-order process calculus, called \hocore, was introduced \cite{LaneseIC10}. 
\hocore  is \emph{minimal}, in that only the operators strictly necessary to obtain
 higher-order communications are retained. 
Most notably, \hocore has no restriction
   operator.
   Thus all names are global, and
    dynamic creation of new names is impossible.
The grammar of \hocore processes is:
\[P :: = \inp a x. P \midd \Ho{a}{P} \midd P \parallel P \midd x \midd \nil \]  
An input process 
$\inp a x . P$ can receive  on
name $a$ a process to be substituted in the place of $x$ in the body $P$;
an output message  $\Ho{a}{P}$ sends the output object $P$ on $a$;  parallel composition 
allows processes to interact. 
As in CCS, in \hocore processes evolve from the interaction of 
complementary actions; this way, e.g., 
$$\Ho{a}{P} \parallel a(x).Q \arro{~~~} Q \sub P x$$ 
is a sample reduction.
(See \cite{LaneseIC10,Perez10} for complete accounts on the  theory of \hocore.)

While considerably expressive, \hocore is far from a specification language for settings involving (forms of) higher-order communication.
For instance, it lacks primitives for describing the \emph{localities} into which distributed systems are typically abstracted. 
Similarly, \hocore also lacks constructs for 
expressing
forms of evolvability and/or dynamic reconfiguration.
In order to deal with these aspects, higher-order process calculi such as Homer and Kell provide 
mechanisms that allow to \emph{suspend} running processes. 
Such mechanisms rely on a form of named localities for processes, so called \emph{suspension (or passivation) units}.
Inside a suspension unit, a process may execute and freely interact with their environment, but it may also be stopped at any time.
More precisely, let us consider the extension of \hocore with process suspension. 
Let $a[P]$ denote the process $P$ inside the suspension unit $a$.
Assuming  an LTS with actions of the form $P \arro{~\alpha~} P'$, 
the semantics of suspension is formalized 
by the following two rules:
\[
\inferrule[\rulename{Trans}]{P \arro{~\alpha~} P' }{ a[P] \arro{~\alpha~} a[P']}  \qquad \quad \inferrule[\rulename{Susp}]{}{a[P] \arro{~a\langle P \rangle~} \nil }
\]
where $a\langle P \rangle$ corresponds to the output action in the LTS of \hocore (see \cite{LaneseIC10}).
While rule \rulename{Trans} defines the transparency of suspension units, rule 
\rulename{Susp} implements suspension: the current state of a located process is ``frozen'' as an output action, in which it can no longer evolve.
Hence, in this semantics
input prefixes may interact not only with output actions but also with suspension units; in fact, 
suspension of a running process is assimilated to regular process communication.
\new{As a simple example, consider the following process~$S$: 
$$
S \triangleq a[P] \parallel a\langle Q \rangle \parallel a(x).R
$$
It is easy to see that two possible evolutions for $S$ 
are $S' \triangleq a[P] \parallel R\sub{Q}{x}$  and $S'' \triangleq a\langle Q \rangle  \parallel R\sub{P}{x}$. 
Other evolutions, related to  the behavior of $P$,  are also possible.}
While the semantics for suspension just described allows for a straightforward definition, 
\new{we observe two potential drawbacks:}
First, the \emph{dual r\^{o}le} of input prefixes 
induces a form of non determinism that one may regard as unnatural.
\new{Consider $a(x).R$ in $S$ above: in the first evolution, it acts as a communication endpoint, 
whereas in the second it acts as a suspension realizer.}
Second, such a semantics is only possible for calculi which already feature process passing in communications.
\new{That is, the possibility of suspending/reconfiguring processes at runtime is somehow tied to the calculus being higher-order.}

\new{With these drawbacks in mind,}
in the definition of $\evol{}$ we have opted for a different approach: 
we do not assume higher-order communication, and rely instead on a restricted form of term instantiation for defining 
update actions. 
\new{That is, we exploit a very particular form of higher-order interaction 
to define process suspension for calculi which may well be first-order.
Here, in order to focus on the novel features of adaptable processes, we have considered a variant of CCS.
Moreover, as we elaborate below, update in $\evol{}$ can be seen as \emph{objective} rather than as \emph{subjective}: 
an adaptable process may evolve independently until it is updated by a prefix in its surrounding context.}
\new{Furthermore, by featuring update prefixes $\update{a}{U}$---a dedicated construct for representing the runtime
reconfiguration of located processes---}$\evol{}$ enforces a separation of concerns, which allows to distinguish
interaction/communication from actions of dynamic reconfiguration.
\new{We believe these are all reasonable design choices, which allow}
us to focus on the fundamental aspects of evolvability for concurrent processes.
\new{In fact, they could provide a basis for 
developing new formalisms with adaptation concerns, such as, e.g., 
an adaptable extension of the $\pi$-calculus 
or a variant of $\evol{}$ with the nested locations of Homer.}

\subsection{Related Work}
We have already discussed 
related works from the point of view of proof techniques in the Introduction.
Below, we comment on some languages/formalisms  related to~\evol{}.

Loosely related to \evol{} are process calculi for fault tolerance  (see, e.g., \cite{BergerH00,NestmannFM03,RielyH01,FrancalanzaH07}).
These are variants of the $\pi$-calculus 
tailored for describing algorithms on distributed systems; hence, they include 
explicit notions of sites/locations, network, and failures.
A series of extensions to the asynchronous $\pi$-calculus 
so as to model distributed algorithms is proposed in \cite{BergerH00}.
One such extensions, aimed at representing process failure,
is a higher-order operation that defines \emph{savepoints}:
process $\mathsf{save}\langle P \rangle.Q$ defines the savepoint $P$ for the current location; 
if such a location crashes, then it will be restarted with state $P$. 
A value-passing calculus to represent and formalize algorithms of distributed consensus is introduced in \cite{NestmannFM03}; it 
includes a \emph{failure detector} construct $\mathcal{S}(k).P$ which 
executes $P$ if locality $k$ is \emph{suspected} to have failed.
The \emph{partial failure} languages of  \cite{RielyH01,FrancalanzaH07} feature similar constructs; 
such works aim at  developing bisimulation-based proof techniques for distributed algorithms.
Crucially, in the constructs for failure 
proposed in the above works (savepoints, failure detectors), 
the \emph{post-failure} behavior is defined statically, and does not depend on 
some runtime behavior. 
Hence, as discussed in Section \ref{s:examp}, these constructs are easily representable in \evol{}.
None of the above works addresses
adaptation properties related to failures nor studies 
decidability/expressiveness issues 
for the languages they work on.

\evol{} relies on transparent localities as a way of structuring communicating processes for update purposes.
The hierarchies induced by transparent localities are rather weak; this is in contrast to 
process hierarchies  in calculi such as Ambients~\cite{CardelliG00} or Seal~\cite{CastagnaVN05}.
The ambients in the Ambient calculus represent \emph{administrative domains} and 
act as containers of concurrent processes.
Ambients may be dissolved using the \textbf{open} primitive; 
transparent localities can only be eliminated in \evold{} by an explicit synchronization with 
a suitable update prefix.
Movement across the ambient hierarchy is achieved via the  \textbf{in}/\textbf{out} primitives; 
it is said to be  \emph{subjective} 
rather than \emph{objective}, as ambients move themselves and are not  moved by their context.
Adapting this distinction to our setting, it is fair to say that \evol{} features a form of \emph{objective update}, 
as an adaptable process
does not contain information on its future update actions: 
it evolves autonomously until it is updated by a suitable update prefix in its context.
A fundamental difference of Ambients 
with respect to higher-order process calculi 
is that movement is \emph{linear}: it is not possible to duplicate an ambient through its movement.
This aspect is one of the main differences between Ambients and Seal, in which process duplication is possible.
A main design guideline in Seal is security; in fact, it is intended as a calculus of sealed objects.
Within the hierarchy of seals, only parent/child communication is allowed, thus establishing a noticeable difference
with respect to the hierarchies of transparent localities in \evol{}.

A suspension-like construct is at the heart of  
MECo \cite{MontesiS10},  
a model for evolvable components.
It is
defined as a process calculus in which 
components 
feature a hierarchical structure, rich input/output interfaces, as well as channel communication.
Evolvability in MECo is enforced by a 
suspension-like 
construct that 
stops a component and \emph{extracts} its ``skeleton''.
Because of its focus on components, adaptation in MECo 
is mostly concerned about consistent changes in input/output interfaces; 
in our case, adaptation is defined in terms of some distinguished observables of the system, thus constituting a rather general way of characterizing correctness.
\textsc{Comp} \cite{LienhardtFMCO10} is another process calculus for component models. It 
is intended to be the component model for the ABS modeling language; 
as such, it aims at providing a unified definition of evolvability for objects, components, and runtime modifications of programs. 
In \textsc{Comp}, 
 constructs for evolvability
 are based on the movement  primitives of the Ambient calculus 
 rather than on suspension-based constructs, 
 as in \evol{} and MECo.
 Hence, the semantics of reconfiguration in \textsc{Comp} is quite different from that in \evol{}, which prevents more detailed comparisons.
 
In a broader setting, related to \evol{}
are formalisms for the specification of (dynamic) software architectures.
While some of them are based on process calculi, none of them relies on suspension-like constructs to 
formalize evolu\-tion/adap\-ta\-tion.
Below we review some of them; we refer the reader to \cite{BradburyCDW04,Bradbury04,Kell:jucs,CuestaFBG05} for more extensive reviews.

One of the earliest proposals for formal grounds to dynamic architectures is \cite{AllenG97}, where a formal system for architectural components which relies on (a fragment of) Hoare's CSP is introduced. The approach in \cite{AllenG97}, however, does not consider dynamic architectures. 
Darwin \cite{Magee96} is an Architecture Description Language (ADL)  for distributed systems; it aims at describing 
the \emph{structure} of static and dynamic component
architectures which may evolve at runtime.
The focus is then on the bindings of interacting components; 
the operational semantics of Darwin relies on a $\pi$-calculus model for handling such bindings.
Darwin features a mechanism of dynamic instantiation which allows 
arbitrary changes in the system architecture. 
Associated techniques for analyzing dynamic change in Darwin have been proposed in \cite{KramerM98,Kramer90}.
In comparison to \evol{}, the kind of changes possible in Darwin 
concern the system topology rather than the ``state'' of the 
interconnected entities, as in our case.
$\pi$-ADL \cite{Oquendo2004} is an ADL for dynamic and mobile architectures.
Formally defined as a typed variant of the higher-order $\pi$-calculus, $\pi$-ADL focuses on  a combination of 
 structural and behavioral perspectives: while the former describes the architecture in terms of components, connectors, and their configurations, 
the latter describes it in terms of actions and behaviors. $\pi$-ADL is at the heart of ArchWare-ADL \cite{MorrisonBOWG07,archware}, a layered ADL for active architectures.
ArchWare-ADL complements $\pi$-ADL with a style layer that allows the specification of components and connectors, and with an analysis layer which enables the specification of constraints on the styles. 
In contrast to \evol{}, $\pi$-ADL does not offer any construct for supporting system evolvability. 
In fact, 
while ArchWare-ADL 
supports forms of evolution (via 
mechanisms for stopping running programs and decomposing them into its main constituents)
these are not provided by the formal framework of $\pi$-ADL but by technologies on top of it \cite{MorrisonKBMOCWSG04}.
Pilar~\cite{CuestaFBG05,CuestaFBB02,CuestaFBB01}
is an algebraic, reflective ADL.
Reflection in Pilar (defined as the capability of a system to reason and act upon itself) relies on 
the notion of reification which, roughly speaking, relates between entities in different levels of a specification for defining introspection capabilities.
The semantic foundation of Pilar is a first-order, polymorphic typed variant of the $\pi$-calculus; no constructs for dynamic update such as those in \evol{} are included in Pilar.

We conclude this review  by mentioning other works on formal approaches to dynamic 
update~\cite{Gupta96,Biyani2008,WermelingerF02,StoyleHBSN07,CansadoCSC10}.
They all rely on different approaches  from ours.

In~\cite{Gupta96}, an investigation on  \emph{on-line software version change} is presented.
There, an on-line change is said to be \emph{valid} 
if the updated program eventually exhibits behavior of the new version.
The problem of determining validity of an on-line change is shown to be 
undecidable by relating it to the halting problem. 
The study in \cite{Gupta96}, however, 
limits to restricted instances of imperative languages. 
Moreover, the notion of validity says very little about correctness and adaptation.
A formal model for adaptation in asynchronous programs in distributed systems is introduced in~\cite{Biyani2008}.
Programs are expressed as guarded commands, and represented 
as automata; adaptation can be then described as transforming one automaton to another automaton.
The focus of \cite{Biyani2008} is the verification of the behavior of system during adaptation, 
considering the interaction between the new program and the old one. 
The use of graph rewriting/category theory to formalize software architecture reconfiguration
has been studied in~\cite{WermelingerF02}.
In \cite{Bierman}, the \emph{update calculus}, a typed $\lambda$-calculus with a primitive operation for updating modules, is proposed.
A development of this idea was carried out in \cite{StoyleHBSN07}, where  
a calculus for \emph{dynamic update} in typed, imperative  languages is proposed.
There, the focus is on \emph{type-safe} updates---intuitively, the consistent update of type $\tau$ with some new type $\tau'$.
There is no knowledge about future software updates;
type coercions mechanisms
are then used to recast new (in principle, unknown) types to old types.
In contrast, in our case ``update code'' is defined in advance. 
In fact, this is a conceptual difference between 
\emph{update} (as in works such as \cite{StoyleHBSN07})
and \emph{dynamic adaptation}, as we have considered it here.
A framework for structural component reconfiguration with
behavioral adaptation considerations is introduced in~\cite{CansadoCSC10}, 
where component architectures are given by \emph{nets}
of interacting components represented by 
LTSs.
Notice that the concept of 
``behavioral adaptation''  in \cite{CansadoCSC10} is different from our notion of adaptation.
The former refers to the changes required in component interfaces so as to achieve effective compositions.
Instead, our notion of adaptation concerns a higher abstraction level, 
as we address the evolution of running processes through built-it adaptation mechanisms.

\subsection{Applying the Verification Problems}
\new{
In the examples given in Section~\ref{s:examp}, both \OG and \LG  were used  to check whether a process can reach a state without errors. 
In general, however, one may be interested in both solving errors and preserving the correct behavior of the system. 
In particular, one could be interested in checking whether certain states of the systems are still reachable after correcting an error. 
We now discuss a 
modeling technique which allows us to 
express such a property as an \emph{instance} of the \OG and \LG problems.  
The key idea is to extend the given system with parallel behaviors, 
defined in accordance with the observable events 
associated to errors and adaptation in the system. 
}

\new{We illustrate the technique by considering the particular case of the \LG problem;
the use of the \OG problem is analogous.
We consider a system abstracted as a process $P$, with  
the following observable actions: 
\begin{enumerate}[(i)]
\item $\outC{a}$ -- which signals that the system has reached the state we are interested in; 
\item $\outC{e_s}$ -- which is emitted as soon as the system enters in an error phase; 
\item  $\outC{e_f}$ -- which signals that the error has been corrected.
\end{enumerate}
We define a process $P^*$ as an extension of $P$ with parallel behaviors which, roughly speaking, ``complement'' the above actions.
Intuitively, by checking whether  for such a $P^*$ and a barb $e$ property \LG is satisfied, then we will be able to guarantee that after having corrected an error in $P$ the distinguished state signaled by $a$ is still reachable.
Process $P^*$ is defined as follows:
$$P^* \triangleq P \parallel  \outC{c} \parallel !c.a.\outC{c} \parallel  e_s.( e + c.(e+ e_f.(e + a.\outC{c})))$$ 
Above, we assume that $c$ and $e$ do not occur in $P$. 
In $P^*$,  we can identify four parts: the process $P$ which is kept unchanged; 
process $C \triangleq !c.a.\outC{c}$, which is used to check that the state signaled by $a$ has been reached; 
process $\outC{c}$, which is used to spawn the first copy of $a$;
finally, we have process $R \triangleq e_s.( e + c.(e+ e_f.(e + a.\outC{c})))$. 
}

\new{We explain the behavior of $P^*$. When $P$ enters in an error phase (as signaled by $\outC{e_s}$), a synchronization takes place and $R$ reaches the process $R_1 \triangleq e + c.(e+ e_f.(e + a.\outC{c}))$. This is the first point in which barb $e$ becomes available; the only way to satisfy the \LG property is to make $e$ disappear. Then, process $R_1$  synchronizes with $\outC{c}$, as this is the only possible evolution, thus obtaining  $R_2 \triangleq e+ e_f.(e + a.\outC{c})$. Notice that at this point the process $P$ cannot evolve by consuming $\outC{a}$, as the occurrence of $a$ in the process $C$ is guarded by a prefix $c$, and no copy of $\outC{c}$ is available.  In $R_2$, barb $e$ is available again and the process can evolve only when  $P$ corrects the error (i.e., when an action $\outC{e_f}$ is observed).  As soon as the error phase is completed, $P$ can  synchronize on $e_f$, thus reaching the process  $R_3 \triangleq e + a.\outC{c}$. In $R_3$, barb $e$ will finally disappear as soon as the system  $P$ performs again action $\outC{a}$.}

\new{Clearly, the specific definition of $P^*$ will depend on the features of the given $P$.
Still, the above example is already useful to illustrate how the two 
verification problems introduced in the paper can provide a suitable basis for 
reasoning about non-trivial properties of evolvable systems which may depend on the observables of the system under consideration.
}

\subsection{Variants of the Correctness Properties}

In this presentation, we have studied correctness of adaptable processes from a rather general perspective; 
in fact, the definition of \OG and \LG are based only on minimal observations on the behavior of the system.
This allows us to reason about the interplay between correctness and adaptation for diverse classes of concurrent systems.
More informative properties (relating correctness and the structure of the system, for instance)
can be devised according to the nature of some particular setting.

In this context, it is worth noticing that the technical machinery required for 
our (un)decidability results can be adapted to handle a slightly different definition of the adaptation problems stated
in Definition \ref{def:adaptprob}.
More precisely, 
such problems can be relaxed so as to consider \emph{non consecutive} error occurrences, rather than consecutive ones.
For this purpose, 
we modify the notion of barbs (cf. Definition \ref{d:barb}) 
by admitting an arbitrary number of reductions between the actual error barbs:
\begin{defi}[Barbs - Alternative Definition]\label{d:barb2}
Let $P$ be an $\mathcal{E}$ process, and let $\alpha$ be an action in $\{a, \outC{a} \mid a \in \mathcal{N}  \}$.
\begin{enumerate}[$\bullet$]
\item Given  $k > 0$, we write $P\barb{\alpha}^{k}$ 
iff there exist $Q_{1},\ldots, Q_{k}$ such that
$P \pired^{*} Q_1 \pired^{*} \ldots \pired^{*} Q_k$
with $Q_i \downarrow_\alpha$, for every $i \in \{1,\ldots, k\}$.
\item We write $P\barb{\alpha}^{\omega}$ iff
there exists an infinite computation
$P \pired^* Q_1 \pired^{*} Q_2 \pired^{*}  \ldots$
with $Q_i \downarrow_\alpha$ for every $i \in \mathbb{N}$.
\end{enumerate}
Furthermore, we  use $\negbarbk{\alpha}$ and $\negbarbw{\alpha}$ to denote the negation of $\barbk{\alpha}$ and $\barbw{\alpha}$, with the expected meaning.
\end{defi}

Variants of 
\LG and \OG can be then restated considering the new definition above. 
Thus, given a set of clusters $\BC_P^M$ and a
barb $e$ then   the \OG problems consists in checking whether all computations of processes in $\BC_P^M$ have at most $k$  states exhibiting $e$.
Similarly, \LG consists in checking whether there is
no computation in which $e$
is observable in infinitely many distinct states.
Given these alternative definitions of \LG and \OG, (un)decidability results can be easily derived from the ones presented here. 
In fact, Table \ref{t:results} remains unchanged under the alternative adaptation problems, and  
straightforwardly all undecidability results hold.
As for the decidability results, we should adapt the WSTS construction and the Petri net simulation.
In particular, to show decidability of the alternative definition of \OG  for \evold{2} processes, 
it is enough to slightly change the definition of $\mathsf{fb}_{\alpha}(S)$ (Definition \ref{def:fbk}) and substituting the occurrences of $pb_S$ with $\Pred_S^*$ whose effectiveness is guaranteed by Theorem~\ref{th:fb}.
Concerning the decidability of \LG for \evols{3}, the Petri net semantics presented in Section~\ref{sec:evol3}
reduces this alternative version of the property to the \emph{repeated coverability} problem. This problem
is known to be decidable for Petri nets, see e.g.~\cite{E03}.


\section{Concluding Remarks}\label{s:conc}

We have proposed the concept of \emph{adaptable process} 
as a way of describing complex evolvability patterns 
in models of concurrent systems.
We have introduced \evol{}, a process calculus of adaptable processes, in which located processes can be updated and relocated at runtime. 
\new{In our view, this ability improves the kind of reconfiguration that can be expressed in existing (higher-order) process calculi}. 
In the design of \evol{}, we aimed at isolating a small basis for representing 
reconfiguration of interacting processes:
we extended CCS without restriction and relabeling (a non Turing complete model), 
with transparent localities (arguably the simplest conceivable way of structuring processes into hierarchies) 
and with update prefixes. 
The interaction of adaptable processes with update prefixes 
constitutes a restricted form of higher-order communication that realizes process reconfiguration.

In order to formalize the correctness of evolvable processes, we  proposed 
the \emph{bounded} and \emph{eventual} adaptation problems.
We studied the (un)decidability  
of these problems in several variants of \evol{}, obtained by 
different evolvability patterns as well as static and dynamic topologies of adaptable processes.
Our results shed light on the expressive power of  \evol{} as well as on the 
nature of verification for 
concurrent processes that may evolve at runtime.

There are a number of \new{practical and technical} issues associated to adaptable processes that would be worth investigating  in future work.
\begin{enumerate}[$\bullet$]
\item \new{We would like to understand how to accommodate (a form of) restriction into $\evol{}$ while preserving our decidability results.
This is a delicate issue, as typically adding restriction causes decidability results to break (see, e.g,~\cite{Busi}).
Recently, higher-order calculi with \emph{name creation} (which replaces usual name restriction) have been put forward~\cite{DBLP:conf/lics/PierardS12}; a creationist treatment of names is claimed to be closer to distributed implementations and is 
shown to have benefits in the development of associated behavioral theories.
Exploring variants of $\evol{}$ with a name creation construct could be therefore insightful.}
\item
\new{In the definition of eventual adaptation we require the absence of computations with infinitely many successive error states. It would be interesting to investigate the impact of fairness on our results, in particular the decidability result for \evols{3}. In fact, the detected computation with infinitely many successive error state could be unfair, in the sense that a parallel process or thread able to solve the problem is available but never scheduled. 
In several concurrent models, properties like the existence of an infinite computation turn from decidable to undecidable when restricting to fair computations (see~\cite{Car87} for Petri nets or~\cite{ZCconcur08}
for CGF, a stochastic CCS-like process calculus for the modeling of chemical systems).
As a future work we intend to check whether a similar result applies also to our case.}
\item It would be interesting to study  the behavioral theory of \evol{} processes;
recent works on behavioral equivalences for higher-order process calculi with passivation (e.g.~\cite{LengletSS11,PierardS11,DBLP:conf/lics/PierardS12}) 
could provide a reasonable starting point.
\new{Also, it would be important to 
devise (logic-based) techniques for enhancing the verification of adaptable processes; 
in recent work~\cite{DBLP:conf/isola/BravettiGPZ12}, we have studied an alternative for tacking this challenging issue.}
\item From a practical standpoint, 
it would be interesting to 
develop extensions or variants of \evol{} tailored to concrete application settings,
to determine how 
the adaptation problems 
proposed here fit in such scenarios, and to study how to transfer our decidability results to such richer languages.
\new{For instance, it would be interesting to see how our adaptation problems fit in the context
of higher-order calculi such as Kell and Homer, which feature rich constructs  for structuring processes (kells in Kell, nested locations in Homer).}
\item Finally, it would be useful to address 
the complexity of \OG and \LG.
As far as \LG is concerned, we have presented its (polynomial) 
reduction to the Petri net place boundedness problem,
for which an EXPSPACE decision procedure exists \cite{Rackoff78}.
Concerning \OG, our proof of decidability does not give a precise indication
about the complexity, as only the termination of the procedure
is guaranteed by the well quasi-ordering we have defined.
We plan to investigate the complexity of the problem 
by comparing \OG to the coverability problem for reset Petri nets
which is known to be non primitive recursive (see, e.g.,~\cite{Schnoebelen10}). 
In fact, the possibility of atomically erasing the current contents of an adaptable
process is reminiscent of the ability that reset transitions have for removing
all the tokens in some given place.
Hence, a plausible direction of future work is to 
investigate suitable abstractions  that could help alleviating 
the state explosion problem.
\end{enumerate}

\section*{Acknowledgements}
We thank the anonymous reviewers for their useful remarks.
This work was partially supported by the French projects ANR-2010-SEGI-013 - AEOLUS, ANR-11-INSE-0007 - REVER, by the EU integrated project HATS, and by the Funda\c{c}\~{a}o para a Ci\^{e}ncia e a Tecnologia (Portuguese Foundation for Science and Technology) through the Carnegie Mellon Portugal Program, grant INTERFACES NGN-44 / 2009.


\bibliographystyle{abbrv}
 \bibliography{referen}
 

\appendix

\section{Proofs from Section \ref{s:calculi}}

\subsection{Proof of Lemma \ref{l:esred}}\label{ap:esred}

We need the following auxiliary definitions.

\begin{defi}
Given two \evol{} processes in normal form $P$ and $Q$, 
we define $\CStr(P \parallel Q)$ as follows.
The root is labeled $\epsilon$, and has $n+m$ children: 
the first $n$ sub-children correspond to the children of the root of $\CStr(P)$,
while the rest correspond to the $m$ children of the root of $\CStr(Q)$,
\end{defi}


\begin{prop}[Syntactic Closure for \evols{} processes]\label{prop:howstatic}
Let $P_{1}, P_{2}, \ldots$ be \evol{} processes.
\begin{enumerate}[\em(1)]
 \item $P_1, P_2 \in \evols{}$ iff $P_1 \parallel P_2 \in \evols{}$.
 \item $P \in \evols{}$ iff $\component{a}{P} \in \evols{}$.
 \item $P_i \in \evols{}$  and $\numap{P_i} =0$ for $i\in [1..n]$  iff $\sum_{i=1}^n \pi_i.P_i \in \evols{}.$
 \item $P \in \evols{}$  and $\numap{P} =0$  iff $! \pi.P \in \evols{}.$
\end{enumerate}
\end{prop}

\proof
Immediate from Definitions \ref{d:eccsstatic} and \ref{d:numap}.
In particular, items (3) and (4) follow  by observing that 
any process $P$ such that $\numap{P}=0$ belongs to the syntactic category $A$ in the grammar of \evols{} processes given in 
Definition \ref{d:eccsstatic}.\qed

We repeat the statement in Page \pageref{l:esred}:
\begin{lem}
Let $P$ be an \evols{} process.
If $P \pired P'$ then
also $P'$ is an  \evols{} process. Moreover, $\CStr(P)=\CStr(P')$.
\end{lem}
\proof
The proof proceeds by induction on the height of the derivation tree for $P \pired P'$, with a case analysis on the last applied rule. 
There are seven cases to check. 

\begin{desCription}
\item\noindent{\hskip-12 pt\bf Case \rulename{Act1}:}\
Then $P = P_{1} \parallel P_{2}$ and $P' = P'_{1} \parallel P_{2}$, with $P_{1} \pired P'_{1}$.
By inductive hypothesis, we have that $P'_{1}$ is an \evols{} process. By Proposition \ref{prop:howstatic} we have that $P_2 \in \evols{}$, and we can therefore conclude that $P' = P'_{1} \parallel P_{2}$ is an \evols{} process.

Moreover, by inductive hypothesis, we have that  $\CStr(P_{1}) = \CStr(P'_{1})$ and 
 by Definition~\ref{def:cstr} it is easy to see that  
 $\CStr(P_{1} \parallel P_{2}) = \CStr(P'_{1} \parallel P_{2})$ holds.

 \item\noindent{\hskip-12 pt\bf Case \rulename{Act2}:}\ Analogous to the case for \rulename{Act1} and omitted. 

\item\noindent{\hskip-12 pt\bf Case \rulename{Loc}:}\
Then $P = \component{a}{Q}$ and $P' = \component{a}{Q'}$, with $Q \pired Q'$.
By inductive hypothesis, we have that $Q'$ is an \evols{} process. For Proposition \ref{prop:howstatic} we have that $P' = \component{a}{Q'}$ is an \evols{} process. 

Moreover, by inductive hypothesis, we have that $\CStr(Q) = \CStr(Q')$.
Then, it is immediate to see that by Definition  \ref{def:cstr} 
$\CStr(\component{a}{Q}) = \CStr(\component{a}{Q'})$.

\item\noindent{\hskip-12 pt\bf Cases \rulename{Tau1}-\rulename{Tau2}:}\
Then $P \equiv  \fillcont{C_1}{A}\parallel \fillcont{C_2}{B}$, where $C_{1}, C_{2}$ are monadic contexts as in Definition \ref{d:mc}.
Moreover, 
$A$ is either 
$!b.Q$ or 
$\sum_{i \in I} \pi_i.Q_i$ with $\pi_{l}=b$, for some $l\in I$, and 
$B$ is either 
$!\outC{b}.R$
or $\sum_{i \in I} \pi_i.R_i$ with $\pi_{l}=\outC{b}$, for some $l\in I$.

We consider only the case in which $A = \sum_{i \in I} \pi_i.Q_i$ and $B = !\outC{b}.R$;  the other cases are similar. 
Then $P' \equiv \fillcont{C_1}{Q_l}\parallel \fillcont{C_2}{R \parallel!\outC{b}.R }$ and from Proposition \ref{prop:howstatic} we easily conclude that $P'$ is an \evols{} process.

By assumption and by Proposition \ref{prop:howstatic}  we have that $A$ and $B$ are \evols{} processes. In turn, this allows us to infer that 
 $\CStr(\sum_{i \in I} \pi_i.Q_i) = \CStr(Q'_{l})$
 and
  $\CStr( !\outC{b}.R) = \CStr(R)$, as well-formed \evols{} processes do not contain adaptable processes behind prefixes, 
  and therefore their component structure denotations are unaffected by input/output transitions.
  The thesis then follows by Definition \ref{def:cstr}: $\CStr(P) = \CStr(P')$.

\item\noindent{\hskip-12 pt\bf Cases \rulename{Tau3}-\rulename{Tau4}:}\
Then $P \equiv \fillcont{C_1}{A} \parallel \fillcont{C_2}{B}$ where:
\begin{enumerate}[$\bullet$]
\item $C_{1},C_{2}$ are monadic contexts, as in Definition \ref{d:mc}; 
\item $A = \component{b}{P_1}$, for some $P_{1}$;  
\item $B  =  \sum_{i \in I} \pi_i.R_i$ with $\pi_{l}=\update{b}{\component{b}{U} \parallel P_2}$ for $l\in I$, or $ B = !\update{b}{\component{b}{U} \parallel P_2}.R$, for some $P_{2}, R$.
\end{enumerate}

We consider the case in which $B = !\update{b}{\component{b}{U} \parallel P_2}.R$; the other case is similar. 
Then $P' \equiv \fillcont{C_1}{\component{a}{\fillcon{U}{P_1}} \parallel P_2}\parallel \fillcont{C_2}{R \parallel !\update{b}{\component{b}{U} \parallel P_2}.R }$.
For Proposition \ref{prop:howstatic} we have that $\fillcont{C_2}{R \parallel !\update{b}{\component{b}{U} \parallel P_2}.R }$  and $P_2$ are \evols{} processes. We now focus on process $\fillcon{U}{P_1}$, for Proposition \ref{prop:howstatic} we know that $P_1$ is an \evols{} process, if $\numph{U} = 0$ then it could not occur that an adaptable process in $P_1$ is prefixed.   Otherwise, if $\numph{U} > 0$ then the side condition (2) of rule \rulename{Tau3}(\rulename{Tau4}) ensures that $\numap{P_1} = 0$. As $U$ follows the syntax of \evols{} by means of Proposition \ref{prop:howstatic}
we can conclude that $\fillcon{U}{P_1} \in \evols{}$.

Moreover, the side condition (1) of rule \rulename{Tau3}(\rulename{Tau4}) implies that  $$\CStr( \component{b}{P_1}) = \CStr(\component{a}{\fillcon{U}{P_1}} \parallel P_2).$$   The thesis then follows by Definition \ref{def:cstr}: $\CStr(P) = \CStr(P')$.\qed
\end{desCription}


\subsection{Proof of Theorem \ref{stdynequiv}}\label{ap:stdynequiv}

We divide the proof into two lemmas.
We need some auxiliary results.

\begin{prop}\label{p:subsp}
Let $P_{1}$ and $P_{2}$ be \evols{} processes.
Then 
$\CStrs(P_{1} \parallel P_{2}) = \CStrs(P_{1}) \cup  \CStrs(P_{2})$.
\end{prop}
\proof
Immediate from the definition of $\CStrs(\cdot)$ (cf. Definition \ref{d:cstrs}).\qed

\begin{lem}\label{l:lemthenc1}
Let $P$  be an \evols{} process.
Also, let $S$ be 
a set of containment structure denotations, such that  $\CStrs(P) \subseteq S$.
Given the encoding $\dyn{\cdot}$ in Definition \ref{def:din}, 
if $P \arro{~~~}_s P'$ then $\dyn{P} \arro{~~~}_d \dyn{P'}$.
\end{lem}
\proof
By induction on the height of the derivation tree for $P \pired_{s} P'$, with a case analysis on the last applied rule. 
There are seven cases to check. 

\begin{desCription}
\item\noindent{\hskip-12 pt\bf Case \rulename{Act1}:}\
Then $P = P_{1} \parallel P_{2}$ and $P' = P'_{1} \parallel P_{2}$, with $P_{1} \pired_{s} P'_{1}$.
By inductive hypothesis, we have that 
$\dyns{P_{1}}{S'} \pired_{d} \dyns{P'_{1}}{S'}$ 
with $\CStrs(P_{1}) \subseteq S'$.
Now, since $\dyn{\cdot}$ is defined as an homomorphism with respect to parallel composition, 
and using Proposition \ref{p:subsp}, we can immediately infer 
that 
$\dyns{P_{1} \parallel P_{2}}{S} \pired_{d} \dyns{P'_{1} \parallel P_{2}}{S}$,
with $S' \cup \CStrs(P_{2}) \subseteq S$, as wanted.

 \item\noindent{\hskip-12 pt\bf Case \rulename{Act2}:}\ Analogous to the case for \rulename{Act1} and omitted. 

\item\noindent{\hskip-12 pt\bf Case \rulename{Loc}:}\
Then $P = \component{a}{Q}$ and $P' = \component{a}{Q'}$, with $Q \pired Q'$.
By inductive hypothesis, we have that 
$\dyns{Q}{S'} \pired_{w} \dyns{Q'}{S'}$, with $\CStrs(Q) \subseteq S'$.
From Definitions~\ref{def:din} and~\ref{d:cstrs} we immediately infer
that 
$\dyns{\component{a}{Q}}{S} \pired_{d} \dyns{\component{a}{Q'}}{S}$, with $S' \cup \CStr(\component{a}{Q}) \subseteq S$.

\item\noindent{\hskip-12 pt\bf Cases \rulename{Tau1}-\rulename{Tau2}:}\
Then $P \equiv  \fillcont{C_1}{A}\parallel \fillcont{C_2}{B}$, where 
\begin{enumerate}[$\bullet$]
\item $C_{1}, C_{2}$ are monadic contexts as in Definition \ref{d:mc}; 
\item $A$ is either $!b.Q$ or  $\sum_{i \in I} \pi_i.Q_i$ with $\pi_{l}=b$, for some $l\in I$;
\item $B$ is either  $!\outC{b}.R$
or $\sum_{i \in I} \pi_i.R_i$ with $\pi_{l}=\outC{b}$, for some $l\in I$.
\end{enumerate}

We consider only the case in which $A = \sum_{i \in I} \pi_i.Q_i$ and $B = !\outC{b}.R$;  the other cases are similar. 
Then, $P' \equiv \fillcont{C_1}{Q_l}\parallel \fillcont{C_2}{R \parallel!\outC{b}.R }$.
Using Definitions \ref{def:din} and \ref{d:mc} we verify that the reduction 
from $P$ is preserved in $\dyn{P}$:
\begin{align*}
 \dyn{P} & = \dyn{ \fillcontBig{C_1}{\sum_{i \in I} \pi_i.Q_i}\parallel \fillcontbig{C_2}{!\outC{b}.R}}\quad\text{with $\CStrs(P) \subseteq S$} \\
& = \fillcontBig{\dyn{C_1}}{\sum_{i \in I} \dyn{ \pi_i.Q_i}}\parallel \fillcontBig{\dyn{C_2}}{\dyn{!\outC{b}.R}} \\
& = \fillcontBig{\dyn{C_1}}{\sum_{i \in I} \pi_i.\dyn{Q_i}}\parallel \fillcontBig{\dyn{C_2}}{!\outC{b}.\dyn{R}}
 \end{align*}
 At this point, it is immediate to infer a reduction $\pired_{d}$ on $b$:
  \begin{align*}
  \dyn{P}  \pired_{d} ~& \fillcontBig{\dyn{C_1}}{\dyn{Q_l}}\parallel \fillcontBig{\dyn{C_2}}{\dyn{R} \parallel !\outC{b}.\dyn{R}}
 \end{align*}
 which is easily seen to correspond to $\dyn{P'}$, as wanted.

\item\noindent{\hskip-12 pt\bf Cases \rulename{Tau3}-\rulename{Tau4}:}\
Then $P \equiv \fillcont{C_1}{A} \parallel \fillcont{C_2}{B}$ where:
\begin{enumerate}[$\bullet$]
\item $C_{1},C_{2}$ are monadic contexts, as in Definition \ref{d:mc}; 
\item $A = \component{b}{P_1}$, for some $P_{1}$;  
\item $B  =  \sum_{i \in I} \pi_i.R_i$ with $\pi_{l}=\update{b}{\component{b}{U} \parallel A_2}$ for $l\in I$, or $ B = !\update{b}{\component{b}{U} \parallel P_2}.R$, for some $A_{2}, R$.
\item $\mathsf{cond}(U, P_{1})$ holds
\end{enumerate}
We consider only the case in which $B = !\update{b}{\component{b}{U} \parallel A_2}.R$; the other case is similar. 
Then, $P' \equiv \fillcontbig{C_1}{\component{a}{\fillcon{U}{P_1}} \parallel A_2}\parallel \fillcontbig{C_2}{ B \parallel R }$.
Since $\mathsf{cond}(U, P_{1})$ holds, we rely on Lemma \ref{lem:statvsdyn} 
to determine the possible cases for $U$ and $P_{1}$:
each of them entails a different encoding of $\dyn{P}$.
Consequently, we verify that in each case the actions that lead to reduction in $P$ are preserved in $\dyn{P}$.
\begin{enumerate}
  \setcounter{enumi}{-1}
\item  $\numholes{U}=0 \wedge \CStr(P_{1}) = \CStr(U)$. Then, using the definition of $\dyn{\cdot}$,  we have 
\begin{align*}
 \dyn{P} & = \dyn{ \fillcontbig{C_1}{\component{b}{P_1}}\parallel  \fillcontbig{C_2}{!\update{b}{\component{b}{U} \parallel A_2}.R}} \\
 & =  \fillcontbig{\dyn{C_1}}{\dyn{\component{b}{P_1}}}\parallel  \fillcontbig{\dyn{C_2}}{\dyn{!\update{b}{\component{b}{U} \parallel A_2}.R}} \\
  & =  \fillcontbig{\dyn{C_1}}{\component{\kappa}{\dyn{P_1}}}\parallel  \fillcontBig{\dyn{C_2}}{!\update{\kappa}{\component{\kappa}{\dyn{U}} \parallel \dyn{A_2}}.\dyn{R}}
 \end{align*}

  At this point, it is immediate to infer a reduction $\pired_{d}$ on $\kappa$:
  \begin{align*}
  \dyn{P}  \pired_{d} ~& \fillcontbig{\dyn{C_1}}{\star}\sub{ \fillcon{(\component{\kappa}{\dyn{U}} \parallel \dyn{A_2})}{P_{1}} }{\star} \parallel  \fillcontBig{\dyn{C_2}}{\dyn{B} \parallel \dyn{R}} \\
= ~& \fillcontBig{\dyn{C_1}}{\fillcon{(\component{\kappa}{\dyn{U}} \parallel \dyn{A_2})}{P_{1}}} \parallel  \fillcontBig{\dyn{C_2}}{\dyn{B} \parallel \dyn{R}} \\
     = ~& \fillcontBig{\dyn{C_1}}{\component{\kappa}{\dyn{P_{1}}} \parallel \dyn{A_2} } \parallel  \fillcontBig{\dyn{C_2}}{\dyn{B} \parallel \dyn{R}} = P''
  \end{align*}
  which is easily seen to correspond to $\dyn{P'}$, as desired. 

\item $\numholes{U}=1  \wedge \numap{U} = 0 \wedge (\numph{U} > 0 \Rightarrow \numap{Q} = 0)$. There are two subcases:
\begin{enumerate}[(1)]
\item\noindent{\hskip-12 pt\bf Case $\numph{U} > 0$:}\  Then, similarly as in the previous case, using the definition of $\dyn{\cdot}$ we can infer a reduction $\pired_{d}$ on name $\kappa_{b}$. 

\item Case $\numph{U} = 0$. Then, using the definition of $\dyn{\cdot}$,  we have 
\begin{align*}
 \dyn{P} & = \dyn{ \fillcontbig{C_1}{\component{b}{P_1}}\parallel  \fillcontbig{C_2}{!\update{b}{\component{b}{U} \parallel A_2}.R}} \\
 & =  \fillcontbig{\dyn{C_1}}{\dyn{\component{b}{P_1}}}\parallel  \fillcontbig{\dyn{C_2}}{\dyn{!\update{b}{\component{b}{U} \parallel A_2}.R}} \\
  & =  \fillcontbig{\dyn{C_1}}{\component{\kappa_{j}}{\dyn{P_1}}}\parallel  \fillcontBig{\dyn{C_2}}{\prod_{\kappa_{i} \in \ecs{\proj{S}{b}}} ! \, \updatebig{\kappa_{i}}{\componentbbig{\kappa_{i}}{\dyn{U}} \parallel \dyn{A_{2}}}.\dyn{R}}
 \end{align*}
with $\kappa_{j} = \ecs{\component{b}{P_1}}$.
At this point, it is immediate to infer a reduction $\pired_{d}$ on $\kappa_{j}$:
  \begin{align*}
  \dyn{P}  \pired_{d} ~& \fillcontbig{\dyn{C_1}}{\star}\sub{ \fillcon{(\component{\kappa_{j}}{\dyn{U}} \parallel \dyn{A_2})}{P_{1}} }{\star} \parallel  \fillcontBig{\dyn{C_2}}{\dyn{B} \parallel \dyn{R}} \\
= ~& \fillcontBig{\dyn{C_1}}{\fillcon{(\component{\kappa_{j}}{\dyn{U}} \parallel \dyn{A_2})}{P_{1}}} \parallel  \fillcontBig{\dyn{C_2}}{\dyn{B} \parallel \dyn{R}} \\
     = ~& \fillcontBig{\dyn{C_1}}{\component{\kappa_{j}}{\dyn{P_{1}}} \parallel \dyn{A_2} } \parallel  \fillcontBig{\dyn{C_2}}{\dyn{B} \parallel \dyn{R}} = P''
  \end{align*}
  which is easily seen to correspond to $\dyn{P'}$, as desired. 

\end{enumerate}

\item $\numholes{U}>1 \wedge \numap{U} = 0\wedge \numap{Q} = 0$.
Then, similarly as in case 1(a), using the definition of $\dyn{\cdot}$ we can infer a reduction $\pired_{d}$ on name $\kappa_{b}$. \qed
\end{enumerate}
\end{desCription}

\begin{lem}\label{l:lemthenc2}
Let $P$  be an \evols{} process.
Also, let $S$ be 
a set of containment structure denotations, such that  $\CStrs(P) \subseteq S$.
Given the encoding $\dyn{\cdot}$ in Definition \ref{def:din}, if $\dyn{P} \arro{~~~}_d \dyn{P'}$ then $P \arro{~~~}_s P'$.
\end{lem}
\proof
By induction on the height of the derivation tree for $P \pired_{d} P'$, with a case analysis on the last applied rule. 
There are seven cases to check. The analysis of all cases mirrors the one detailed in the proof of Lemma \ref{l:lemthenc1}, and we omit it.
The crucial point is the fact that the encoding uses the special 
name $\mathrm{err}$ to rename those update prefixes that may lead to incorrect reductions in \evols{}. 
Hence, adaptable processes included in the \evold{} process $\dyn{P}$ will be unable to interact with those ``error'' update prefixes. 
This ensures that for every reduction $\pired_{d}$ there is also a reduction $\pired_{s}$.\qed

We repeat the statement in Page \pageref{stdynequiv}:
\begin{thm}[\ref{stdynequiv}]
Let $P$  be an \evols{} process.
Also, let $S$ be 
a set of containment structure denotations, such that  $\CStrs(P) \subseteq S$.
 Then we have:
$$P \pired_s P'\text{ if and only if }\dyn{P} \pired_d \dyns{P'}{S}$$
\end{thm}

\proof
Immediate from Lemmas \ref{l:lemthenc1} and \ref{l:lemthenc2}.\qed

\section{Proofs from Section \ref{s:ev1}}

\subsection{Proof of Lemma \ref{th:corrE1}}\label{app:e1}
The proof relies on two  results:
completeness (Lemma~\ref{l:compl-pas}) and soundness (Lemma~\ref{l:sound-pas}).
We begin by defining the encoding of \mm configuration into~\evol{1}. 
\begin{defi}\label{def:confe1}
Let $N$ be a \mm 
with 
registers $r_j ~(j \in \{0,1\})$ and 
instructions $(1:I_1), \ldots, (n:I_n)$.
The encoding of a configuration $(i,m_0, m_1)$ of $N$, denoted
$\encp{(i, m_{0}, m_{1})}{\mmn{1}}$, is defined as:
\[
 \outC{p_i} \parallel \encp{r_0 = m_0}{\mmn{1}} \parallel \encp{r_1 = m_1}{\mmn{1}} \parallel \prod^{n}_{i=1} \encp{(i:I_i)}{\mmn{1}}
\]
where the encodings 
$\encp{r_j = m_j}{\mmn{1}}$ and  
$\encp{(i:I_i)}{\mmn{1}}, \ldots, \encp{(n:I_n)}{\mmn{1}}$ are as in Table~\ref{t:encod-pas}.
\end{defi}

\begin{lem}[Completeness]\label{l:compl-pas}
Let $(i, m_0, m_1)$ be a configuration of a \mm $N$. 
\begin{enumerate}[\em(1)]
\item  If $(i, m_0, m_1)\minskred (i', m_0', m_1')$ then, for some process $P$, 
it holds that\\ $\encp{(i, m_0, m_1) }{\mmn{1}} \pired^* P \equiv \encp{(i', m_0', m_1') }{\mmn{1}}$.

\item If  $(i, m_0, m_1)\notminskred $ then $\encp{(i, m_0, m_1) }{\mmn{1}} \barbk{e}$
\end{enumerate}

\end{lem}

\proof \hfill \\
\begin{enumerate}[(1)]
\item
We proceed by a case analysis on the instruction  performed by the \mma.
Hence, we distinguish three cases corresponding to the behaviors associated to rules
\textsc{M-Inc}, \textsc{M-Dec}, and \textsc{M-Jmp}.
Without loss of generality, we restrict our analysis to operations on register $r_0$.

\begin{desCription}
 \item\noindent{\hskip-12 pt\bf Case \textsc{M-Inc}:}\ We have a Minsky configuration $(i, m_0, m_1)$ with 
$(i: \mathtt{INC}(r_0))$. By Definition \ref{def:confe1}, its encoding into \evol{1} is as follows:
\begin{align*}
\encp{(i, m_0, m_1)}{\mmn{1}}  =   & ~\overline{p_i} \parallel \encp{r_0 = m_0}{\mmn{1}} \parallel \encp{r_1 = m_1}{\mmn{1}} \parallel  \\
&  \encp{(i: \mathtt{INC}(r_0))}{\mmn{1}} \parallel   \prod_{l=1..n,l\not = i} \encp{(l:I_l)}{\mmn{1}} 
\end{align*}

After consuming the program counter $p_i$  we have the following 
\[
\encp{(i, m_0, m_1)}{\mmn{1}}  \pired 
\component{r_0}{\encn{m_{0}}{0}} \parallel \update{r_0}{\component{r_0}{\overline{u_{0}}.\bullet}}.\overline{p_{i+1}}
\parallel S = P_1
\]
where $S = \encp{r_1 = m_1}{\mmn{1}} \parallel \prod_{i=1}^n \encp{(i:I_i)}{\mmn{1}}$ stands for the rest of the system.
The only reduction possible at this point is the synchronization on $r_0$, which allows 
the update of  the adaptable process at  $r_0$:
\[
 P_1 \pired \component{r_0}{\overline{u_{0}}.\encn{m_{0}}{0}} \parallel \overline{p_{i+1}} \parallel S = P_2 \, .
\]
By the encoding of numbers, it 
$P_{2}$ can be equivalently written as
\[
\component{r_0}{\encn{m_{0}+1}{0}} \parallel \overline{p_{i+1}} \parallel S
\]
and so it is easy to see that $P_2 \equiv \encp{(i+1, m_0+1, m_1)}{\mmn{1}}$, as desired.


\item\noindent{\hskip-12 pt\bf Case \textsc{M-Dec}:}\ We have a Minsky configuration $(i, c, m_1)$ such that 
$(i: \mathtt{DEC}(r_0,s))$ and $c > 0$. By Definition \ref{def:confe1}, its encoding into \evol{1}  is as follows:
\begin{eqnarray*}
\encp{(i, c, m_1)}{\mmn{1}} & = &   \overline{p_i} \parallel \encp{r_0 = c}{\mmn{1}} \parallel \encp{r_1 = m_1}{\mmn{1}} \parallel  \\
& & \encp{(i: \mathtt{DEC}(r_0,s))}{\mmn{1}} \parallel   \prod_{l=1..n,l\not = i} \encp{(l:I_l)}{\mmn{1}} 
\end{eqnarray*}
We begin by consuming the program counter $p_i$, which leaves the content of $\encp{(i: \mathtt{DEC}(r_0,s))}{\mmn{1}}$ exposed.
Using the encoding of numbers we have the following:
\[
 \encp{(i, c, m_1)}{\mmn{1}}  \pired   \component{r_0}{\overline{u_0}.\encn{c-1}{0}}
 \parallel (u_0.\overline{p_{i+1}} + z_0.\update{r_0}{\component{r_0}{\overline{z_0}}}.\overline{p_{s}}) \parallel S = P_1
\]
where $S = \encp{r_1 = m_1}{\mmn{1}} \parallel \prod_{i=1}^n \encp{(i:I_i)}{\mmn{1}}$ stands for the rest of the system.
Notice that only reduction possible at this point is the synchronization on $u_0$, 
which signals the fact we are performing a decrement instruction.
After this synchronization we have 
\begin{eqnarray*}
 P_1 & \pired  & \component{r_0}{\encn{c-1}{0}} \parallel \overline{p_{i+1}}  \parallel S \\
 & \equiv &  \encp{(i+1, c-1, m_1)}{\mmn{1}} 
\end{eqnarray*}
as desired.


\item\noindent{\hskip-12 pt\bf Case \textsc{M-Jmp}:}\ We have a Minsky configuration $(i, 0, m_1)$ and 
$(i: \mathtt{DEC}(r_0,s))$. By Definition \ref{def:confe1}, 
its encoding into \evol{1} is as follows:
\begin{eqnarray*}
\encp{(i, 0, m_1)}{\mmn{1}} & = &   \overline{p_i} \parallel \encp{r_0 = 0}{\mmn{1}} \parallel \encp{r_1 = m_1}{\mmn{1}} \parallel  \\
& & \encp{(i: \mathtt{DEC}(r_0,s))}{\mmn{1}} \parallel   \prod_{l=1..n,l\not = i} \encp{(l:I_l)}{\mmn{1}} \, .
\end{eqnarray*}
We begin by consuming the program counter $p_i$, which leaves the content of $\encp{(i: \mathtt{DEC}(r_0,s))}{\mmn{1}}$ exposed.
Using the encoding of numbers we have the following:
\[
 \encp{(i, 0, m_1)}{\mmn{1}}  \pired   \component{r_0}{\overline{z_0}}
 \parallel (u_0.\overline{p_{i+1}} + z_0.\update{r_0}{\component{r_0}{\overline{z_0}}}.\overline{p_{s}}) \parallel S = P_1
\]
where $S = \encp{r_1 = m_1}{\mmn{1}} \parallel \prod_{i=1}^n \encp{(i:I_i)}{\mmn{1}}$ stands for the rest of the system.
In $P_1$, the only reduction possible is through a synchronization on $z_0$, 
which signals the fact we are performing a jump.
Such a synchronization, in turn, enables an update action on $r_0$.
We then have:
\begin{eqnarray*}
P_1 & \pired &  \component{r_0}{\nil} \parallel \update{r_0}{\component{r_0}{\overline{z_0}}}.\overline{p_{s}} \parallel S \\
 & \pired &   \component{r_0}{\overline{z_0}} \parallel \overline{p_{s}} \parallel S \\
 & \equiv & \encp{(s, 0, m_1)}{\mmn{1}}
 \end{eqnarray*}
as desired.
\end{desCription}

\item 
We have a Minsky configuration $(i, m_{0}, m_1)$ with
$(i: \mathtt{HALT})$. By Definition \ref{def:confe1}, 
its encoding into \evol{1} is as follows:
\begin{eqnarray*}
\encp{(i, m_{0}, m_1)}{\mmn{1}} \! \! \!  & = &    
p_{i} \parallel \encp{r_0 = m_{0}}{\mmn{1}} \parallel \encp{r_1 = m_1}{\mmn{1}} \\&& \parallel  
  \encp{(i: \mathtt{HALT})}{\mmn{1}} \parallel   \prod_{l=1..n,\, l\not = i} \encp{(l:I_l)}{\mmn{1}} \\
   & \equiv &  \overline{p_i} \parallel !p_i.(e + \outC{p_i}) \parallel  S = P_{0}
\end{eqnarray*}
where $S = \encp{r_0 = m_{0}}{\mmn{1}} \parallel \encp{r_1 = m_1}{\mmn{1}} \parallel  \prod_{l=1..n,\, l\not = i} \encp{(l:I_l)}{\mmn{1}}$
stands for the part of the system that is not able to interact. 
It is easy to see that $P_{0} \Downarrow_{e}^1$. 
In fact, by synchronizing on $p_{i}$ and 
choosing the left-hand side process in the binary sum, we have  
$P_{0} \pired \arro{~e~}$.
The thesis is easily seen to hold by 
observing that by releasing new copies of the encoding of $(i: \mathtt{HALT})$, one always reaches a derivative $P_{j}$ of $P_{0}$
such that $P_{j} \Downarrow_{e}$.\qed
\end{enumerate}

\begin{lem}[Soundness]\label{l:sound-pas} 
Let $(i,m_0,m_1)$ be a configuration of a \mm $N$.  \\ 
If $\encp{(i,m_0,m_1)}{\mmn{1}}  \pired P_1$ then either:
\begin{enumerate}[\em(1)]
\item 
For every computation of $P_1$ there  exists a $P_j$ such that 
$$P_{1} \pired^{*} P_j = \encp{(i',m'_0,m'_1)}{\mmn{1}}$$ and $(i,m_0,m_1)\minskred (i',m'_0,m'_1)$; or  
\item 
$P_{1} \barbk{e}$ and  $(i,m_0,m_1) \notminskred$.
\end{enumerate}
\end{lem}
 
\proof
Consider the reduction  $\encp{(i,m_0,m_1)}{\mmn{1}}  \pired P_1$. An analysis of the structure of process
 $\encp{(i,m_0,m_1)}{\mmn{1}}$ reveals that, in all cases, the only possibility for the first step corresponds to the
consumption of the program counter $p_i$. This implies that there exists an instruction
labeled with $i$, that can be executed from the configuration $(i,m_0,m_1)$.
We proceed by a case analysis on the possible instruction, considering also the 
fact that the register on which the instruction acts can hold a value equal or greater than zero.

In the cases in which $(i:\mathtt{INC}(r_{j}))$ or $(i:\mathtt{DEC}(r_{j},s))$, it can be shown that computation evolves 
deterministically
until reaching a process in which a new program counter (that is, some $\overline{p_{i'}}$) appears.
The program counter $\overline{p_{i'}}$ is always inside a process that corresponds to $\encp{(i',m'_0,m'_1)}{\mmn{1}}$,
where $(i,m_0,m_1)\minskred (i',m'_0,m'_1)$.  
That is, for the cases $(i:\mathtt{INC}(r_{j}))$ and $(i:\mathtt{DEC}(r_{j},s))$, we have that Item (1) above holds.
The detailed analysis follows the same lines as the one reported for the proof of Lemma \ref{l:compl-pas}, and we omit it.

In the case in which $(i:\mathtt{HALT})$, 
we have that Item (2) holds. In order to see this, 
it suffices to observe that  
if $N$ does not terminate
(more precisely: if $N$ does not reach a program counter associated to a $\mathtt{HALT}$ instruction) 
then $\encp{N}{\mmn{1}}$ does not have a barb on $e$. In fact, 
by a simple inspection on the encodings in Table~\ref{t:encod-pas}
we can deduce that $e$ 
only appears in the encoding of halt instructions, and 
does not occur in the encodings of increment and decrement-and-jump instructions.
Hence, a barb on $e$ can only be observed when $P_{1}$ is the result of triggering a halt instruction.\qed

We are now ready to repeat the statement of Lemma \ref{th:corrE1}, in Page \pageref{th:corrE1}:

\begin{lem}[\ref{th:corrE1}]
Let $N$ be a \mm and $k \geq 1$. $N$ terminates iff $\encp{N}{1} \barbk{e}$.
\end{lem}

\proof
It follows directly from Lemmas \ref{l:compl-pas} and \ref{l:sound-pas}.\qed

\section{Proofs from Section \ref{s:ev2}}

\subsection{Proof of Lemma \ref{th:corrE2}}\label{app:e2}
The proof relies on two auxiliary results:
completeness (Lemma \ref{l:compl-pase2}) and soundness (Lemma~\ref{l:sound-pase2}).
Completeness relies on the auxiliary Lemma~\ref{lem:correctstep}.

%
We first introduce the notion of 
encoding of a MM
configuration 
into \evols{2}.
Notice that it in addition to the encodings of registers and instructions, it includes a number of resources 
$\outC{f}$ and $\outC{b}$ 
which are always available during the execution of the machine:


\begin{defi}\label{def:confe2}
Let $N$ be a \mm 
with 
registers $r_j ~(j \in \{0,1\})$ and 
instructions $(1:I_1), \ldots, (n:I_n)$.
The encoding of a configuration $(i,m_0, m_1)$ of $N$, denoted
$\encp{(i, m_{0}, m_{1})}{\mmn{2}}$, is defined as:
\[
 \outC{p_i} \parallel e \parallel \encp{r_0 = m_0}{\mmn{2}} \parallel \encp{r_1 = m_1}{\mmn{2}} \parallel \prod^{n}_{i=1} \encp{(i:I_i)}{\mmn{2}} \parallel \reso{\alpha,\beta,\gamma}
\]
where 
\begin{enumerate}[$\bullet$]
\item $\reso{\alpha,\beta,\gamma} \stackrel{\textrm{def}}{=} \prod^{\alpha} \outC{f} \parallel \prod^{\beta} \outC{b} \parallel \prod^{\gamma} \outC{g} \parallel !a.(\outC{f} \parallel \outC{b} \parallel \outC{a}) \parallel !h.(g.\outC{f} \parallel \outC{h})$, with  $\alpha,\beta,\gamma \geq 0$
\item the encodings 
$\encp{r_j = m_j}{\mmn{2}}$ and  
$\encp{(i:I_i)}{\mmn{2}}, \ldots, \encp{(n:I_n)}{\mmn{2}}$ are as in Table~\ref{tab:minskyccsbs}.
\end{enumerate}
\end{defi}
Notice that $\reso{\alpha,\beta,\gamma}$ abstracts the evolution of 
process \textsc{Control} in Table~\ref{tab:minskyccsbs}, and the resources
that it produces and maintains (namely, $\alpha$ copies of $\outC{f}$, $\beta$ copies of $\outC{b}$, and $\gamma$ copies of $\outC{g}$).

\begin{rem}\label{rem:enough}
As we have discussed, the presence of copies of $\outC{f}$ is required for the execution of increment and
decrement-and-jump instructions. 
In their absence, the encoding of the \mm would reach a deadlocked state.
Such outputs are produced at the beginning of the execution of the encoding of a \mm, by means of a replicated process.
In the proofs below, we assume that the initialization of the encoding always produces enough copies of $\outC{f}$
so as to ensure the existence of a correct simulation of the machine. 
That is to say, we assume that the absence of copies of $\outC{f}$ is not a possible source of deadlocks.
\end{rem}

We prove that given a \mm $N$ there exists a 
computation of process $\encp{N}{\mmn{2}}$ which correctly mimics its behavior.

\begin{lem}\label{lem:correctstep}
Let $(i, m_0, m_1)$ be a configuration of a \mm $N$. 
\begin{enumerate}[\em(1)]
\item If $(i, m_0, m_1)\minskred (i', m_0', m_1')$ then, for some process $P$, 
it holds that $$\encp{(i, m_0, m_1) }{\mmn{2}} \pired^* P \equiv \encp{(i', m_0', m_1') }{\mmn{2}}$$
\item If $(i, m_0, m_1)\notminskred$ then $\encp{(i, m_0, m_1) }{\mmn{2}}\! \Downarrow_{\outC{p_1}}^1$.
\end{enumerate}
\end{lem}
\proof \hfill \\
\begin{enumerate}[(1)]
\item
We proceed by a case analysis on the instruction  performed by the \mma.
Hence, we distinguish three cases corresponding to the behaviors associated to rules
\textsc{M-Inc}, \textsc{M-Dec}, and \textsc{M-Jmp}.
Without loss of generality, we restrict our analysis to operations on register $r_0$. 

\begin{desCription}
\item\noindent{\hskip-12 pt\bf Case \textsc{M-Inc}:}\ We have a Minsky configuration $(i, m_0, m_1)$ with 
$(i: \mathtt{INC}(r_0))$. By Definition \ref{def:confe2}, its encoding into \evols{2} is as follows:
\begin{eqnarray*}
\encp{(i, m_0, m_1) }{\mmn{2}}  & = & 
 \outC{p_i} \parallel e  \parallel
\encp{r_0 = m_0}{\mmn{2}} \parallel \encp{r_1 = m_1}{\mmn{2}} \parallel \\
& & !p_i.f.(\outC{g} \parallel b.\outC{inc_0}.\outC{p_{i+1}}) \parallel \prod_{l=1..n,l\neq i} \encp{(l:I_l)}{\mmn{2}} \parallel \reso{\alpha,\beta,\gamma}
\end{eqnarray*}

We then have:
$$\encp{(i, m_0, m_1) }{\mmn{2}}   \pired e  \parallel
\encp{r_0 = m_0}{\mmn{2}} \parallel f.(\outC{g} \parallel b.\outC{inc_0}.\outC{p_{i+1}}) \parallel \reso{\alpha,\beta,\gamma} \parallel S \! =\! P$$
where $S = \encp{r_1 = m_1}{\mmn{2}} \parallel \prod_{l=1}^{n} \encp{(l:I_l)}{\mmn{2}}$
stands for the rest of the system.
Starting from $P$, a possible sequence of reductions is the following:
\begin{eqnarray*}
P \! \! \! & \pired & e  \parallel \encp{r_0 = m_0}{\mmn{2}} \parallel b.\outC{inc_0}.\outC{p_{i+1}} \parallel \reso{\alpha-1,\beta,\gamma+1} \parallel S\\
& = & e  \parallel \component{r_0}{!inc_0.\outC{u_0} \parallel \prod^{m_{0}}\overline{u_0} \parallel \outC{z_0} }  \parallel b.\outC{inc_0}.\outC{p_{i+1}} \parallel \reso{\alpha-1,\beta,\gamma+1} \parallel S\\
& \pired & e  \parallel \component{r_0}{!inc_0.\outC{u_0} \parallel \prod^{m_{0}}\overline{u_0} \parallel \outC{z_0} }  \parallel \outC{inc_0}.\outC{p_{i+1}} \parallel \reso{\alpha-1,\beta-1,\gamma+1}\parallel S\\
& \pired \equiv \! \! & e  \parallel \component{r_0}{!inc_0.\outC{u_0} \parallel \prod^{m_{0}+1}\overline{u_0} \parallel \outC{z_0} } \parallel \outC{p_{i+1}} \parallel \reso{\alpha-1,\beta-1,\gamma+1} \parallel S = P'
\end{eqnarray*}
It is easy to see that $P' \equiv \encp{(i+1, m_0+1, m_1)}{\mmn{2}}$, as desired. 
Observe how the number of resources changes: in the first reduction, a copy of $\outC{f}$ is consumed, and  a copy of $\outC{g}$ is released in its place.
Notice that we are assuming that $\beta > 0$, that is, that there is at least one copy of $\outC{b}$.
In fact, since the instruction only takes place after a synchronization on $b$ (i.e., the second reduction above)
the presence of at least one copy of $\outC{b}$ in $\reso{\alpha-1,\beta,\gamma+1}$ is essential to avoid  deadlocks.


\item\noindent{\hskip-12 pt\bf Case \textsc{M-Dec}:}\ We have a Minsky configuration $(i, m_0, m_1)$ with 
$m_0 > 0$ and 
$(i: \mathtt{DEC}(r_0,s))$. By Definition \ref{def:confe2}, its encoding into \evols{2}  is as follows:
\begin{eqnarray*}
\encp{(i, m_0, m_1) }{\mmn{2}} \! \! \!  & = \!\!\!& 
 \outC{p_i} \parallel e  \parallel
\encp{r_0 = m_0}{\mmn{2}} \parallel 
\encp{r_1 = m_1}{\mmn{2}} \parallel \\
& & !p_i.f.\big(\outC{g} \parallel (u_0.(\outC{b} \parallel
           \outC{p_{i+1}}) +   z_0.\update{r_0}{\component{r_0}{!inc_0.\outC{u_0} \parallel \outC{z_0}}}. \outC{p_s})\big)\\
           & & \parallel  \prod_{l=1..n,l\neq i} \encp{(l:I_l)}{\mmn{2}} \parallel \reso{\alpha,\beta,\gamma}
\end{eqnarray*}

We then have:
\begin{eqnarray*}
\encp{(i, m_0, m_1) }{\mmn{2}} \!\!\!&  \pired \equiv\!\!\! &  
f.\big(\outC{g} \parallel (u_0.(\outC{b} \parallel
           \outC{p_{i+1}}) +   z_0.\update{r_0}{\component{r_0}{!inc_0.\outC{u_0} \parallel \outC{z_0}}}. \outC{p_s})\big)       \\
           & &\parallel e  \parallel  \encp{r_0 = m_0}{\mmn{2}} \parallel           \reso{\alpha,\beta,\gamma} \parallel S = P
\end{eqnarray*}
where $S =  \encp{r_1 = m_1}{\mmn{2}} \parallel \prod_{l=1}^{n} \encp{(l:I_l)}{\mmn{2}}$ stands for the rest of the system.
Starting from $P$, a possible sequence of reductions is the following:
\begin{eqnarray*} 
P \!\!\!& \pired \!\!\! &  
u_0.(\outC{b} \parallel
           \outC{p_{i+1}}) +   z_0.\update{r_0}{\component{r_0}{!inc_0.\outC{u_0} \parallel \outC{z_0}}}. \outC{p_s} \parallel      \\
           & & e  \parallel  \encp{r_0 = m_0}{\mmn{2}} \parallel           \reso{\alpha-1,\beta,\gamma+1} \parallel S \\
& = \!\!\!&  
u_0.(\outC{b} \parallel
           \outC{p_{i+1}}) +   z_0.\update{r_0}{\component{r_0}{!inc_0.\outC{u_0} \parallel \outC{z_0}}}. \outC{p_s} \parallel      \\
           & & e  \parallel  \component{r_0}{!inc_0.\outC{u_0} \parallel \prod^{m_{0}}\overline{u_0} \parallel \outC{z_0} }  \parallel           \reso{\alpha-1,\beta,\gamma+1} \parallel S ~= P'\\
& \pired \!\!\!&  \outC{p_{i+1}} \parallel  e \parallel \component{r_0}{!inc_0.\outC{u_0} \parallel \prod^{m_{0}-1}\overline{u_0} \parallel \outC{z_0} }  \parallel           \reso{\alpha-1,\beta+1,\gamma+1} \parallel S ~= P'' 
\end{eqnarray*}
It is easy to see that $P' \equiv \encp{(i+1, m_0-1, m_1)}{\mmn{2}}$, as desired. 
Observe how 
in the last reduction 
the presence of at least  a copy of $\outC{u_{0}}$ in $r_{0}$ is fundamental for releasing both an extra copy of $\outC{b}$
and the trigger for the next instruction.



\item\noindent{\hskip-12 pt\bf Case \textsc{M-Jmp}:}\ We have a Minsky configuration $(i, 0, m_1)$ and 
$(i: \mathtt{DEC}(r_0,s))$. By Definition \ref{def:confe2}, 
its encoding into \evols{2} is as follows:

\begin{eqnarray*}
\encp{(i, 0, m_1) }{\mmn{2}} \!\!\!  & = \!\!\!& 
 \outC{p_i} \parallel e  \parallel
 \encp{r_0 = 0}{\mmn{2}} \parallel \encp{r_1 = m_1}{\mmn{2}} \parallel \\
& & !p_i.f.\big(\outC{g} \parallel (u_0.(\outC{b} \parallel
           \outC{p_{i+1}}) +   z_0.\update{r_0}{\component{r_0}{!inc_0.\outC{u_0} \parallel \outC{z_0}}}. \outC{p_s})\big)  \\
           & & \parallel \prod_{l=1..n,l\neq i} \encp{(l:I_l)}{\mmn{2}} \parallel \reso{\alpha,\beta,\gamma}
\end{eqnarray*}

We then have:
\begin{eqnarray*}
\encp{(i, 0, m_1) }{\mmn{2}} \!\!\! &  \pired \equiv \!\!\! &  
f.\big(\outC{g} \parallel (u_0.(\outC{b} \parallel
           \outC{p_{i+1}}) +   z_0.\update{r_0}{\component{r_0}{!inc_0.\outC{u_0} \parallel \outC{z_0}}}. \outC{p_s})\big)       \\
           & & \parallel e  \parallel  \encp{r_0 = m_0}{\mmn{2}} \parallel           \reso{\alpha,\beta,\gamma} \parallel S = P
\end{eqnarray*}
where $S =  \encp{r_1 = m_1}{\mmn{2}} \parallel \prod_{l=1}^{n} \encp{(l:I_l)}{\mmn{2}}$ stands for the rest of the system.
Starting from $P$, a possible sequence of reductions is the following:
\begin{eqnarray*} 
P & \pired &  
u_0.(\outC{b} \parallel
           \outC{p_{i+1}}) +   z_0.\update{r_0}{\component{r_0}{!inc_0.\outC{u_0} \parallel \outC{z_0}}}. \outC{p_s} \parallel      \\
           & & e  \parallel  \encp{r_0 = 0}{\mmn{2}} \parallel           \reso{\alpha-1,\beta,\gamma+1} \parallel S \\
& = &  
u_0.(\outC{b} \parallel
           \outC{p_{i+1}}) +   z_0.\update{r_0}{\component{r_0}{!inc_0.\outC{u_0} \parallel \outC{z_0}}}. \outC{p_s} \parallel      \\
           & & e  \parallel  \component{r_0}{!inc_0.\outC{u_0} \parallel  \outC{z_0} }  \parallel           \reso{\alpha-1,\beta,\gamma+1} \parallel S \\
& \pired &  \update{r_0}{\component{r_0}{!inc_0.\outC{u_0} \parallel \outC{z_0}}}. \outC{p_s} \parallel e \parallel \component{r_0}{!inc_0.\outC{u_0}}  
 \parallel            \reso{\alpha-1,\beta,\gamma+1} \parallel S \\
 & \pired &  \component{r_0}{!inc_0.\outC{u_0} \parallel \outC{z_0}} \parallel  \outC{p_s} \parallel e 
 \parallel            \reso{\alpha-1,\beta,\gamma+1} \parallel S  = P'
\end{eqnarray*}
It is easy to see that $P' \equiv \encp{(s, 0, m_1)}{\mmn{2}}$, as desired. 
Observe how the number of copies of $\outC{b}$ remains invariant when the \mm is correctly simulated.

\end{desCription}

\item
If $(i, m_0, m_1)\notminskred$ then $i$ corresponds to the $\mathtt{HALT}$ instruction.
Then, by Definition~\ref{def:confe2}, 
its encoding into \evols{2} is as follows:
\begin{eqnarray*}
\encp{(i, m_0, m_1) }{\mmn{2}}  & = & 
 \outC{p_i} \parallel e  \parallel
 \encp{r_0 = m_0}{\mmn{2}} \parallel \encp{r_1 = m_1}{\mmn{2}} \parallel \\
& & !p_i.\outC{h}.h.\update{r_0}{\component{r_0}{!inc_0.\outC{u_0} \parallel \outC{z_0}}}.\update{r_1}{\component{r_1}{!inc_1.\outC{u_1} \parallel \outC{z_1}}}.\outC{p_1} \parallel \\
           & & \prod_{l=1..n,l\neq i} \encp{(l:I_l)}{\mmn{2}} \parallel \reso{\alpha,\beta,\gamma}
\end{eqnarray*}

We then have: 
\begin{eqnarray*}
\encp{(i, m_0, m_1) }{\mmn{2}}  & \pired  \equiv & 
\outC{h}.h.\update{r_0}{\component{r_0}{!inc_0.\outC{u_0} \parallel \outC{z_0}}}.\update{r_1}{\component{r_1}{!inc_1.\outC{u_1} \parallel \outC{z_1}}}.\outC{p_1} \parallel \\
& & e  \parallel  \encp{r_0 = m_0}{\mmn{2}} \parallel \reso{\alpha,\beta,\gamma} \parallel S = P
\end{eqnarray*}
where $S =  \encp{r_1 = m_1}{\mmn{2}} \parallel \prod_{l=1}^{n}\encp{(l:I_l)}{\mmn{2}}$ stands for the rest of the system.
Starting from $P$, a possible sequence of reductions is the following:
\begin{eqnarray*} 
P & \pired^{*} &  \update{r_0}{\component{r_0}{!inc_0.\outC{u_0} \parallel \outC{z_0}}}.\update{r_1}{\component{r_1}{!inc_1.\outC{u_1} \parallel \outC{z_1}}}.\outC{p_1} \parallel \\
& & e  \parallel  \encp{r_0 = m_0}{\mmn{2}} \parallel \reso{\alpha+c,\beta,\gamma-c} \parallel S= P_{1}
\end{eqnarray*}
where the output on $h$ in $P$ interacted with  process $\reso{\alpha,\beta,\gamma}$
so as to replace $c$ outputs on $g$ with $c$ outputs on $f$.
After that,  a synchronization on $h$ took place between 
the evolutions of $\reso{\alpha,\beta,\gamma}$ and of $P$.
We now have:
 \begin{eqnarray*} 
P_{1} & \pired \equiv  & e  \parallel  \encp{r_0 = 0}{\mmn{2}} \parallel \encp{r_1 = m_1}{\mmn{2}} \\
& & 
\update{r_1}{\component{r_1}{!inc_1.\outC{u_1} \parallel \outC{z_1}}}.\outC{p_1} \parallel \prod_{l=1}^{n} \encp{(l:I_l)}{\mmn{2}} \parallel \reso{\alpha+c,\beta,\gamma-c} \\
& \pired  & e  \parallel  \encp{r_0 = 0}{\mmn{2}} \parallel \encp{r_1 = 0}{\mmn{2}} \parallel \outC{p_1} \parallel \prod_{l=1}^{n} \encp{(l:I_l)}{\mmn{2}} \parallel \reso{\alpha+c,\beta,\gamma-c} 
\end{eqnarray*}
which 
corresponds to $\encp{(1, 0, 0) }{\mmn{2}}$. 
In turn, it can be seen  that $\encp{(i, m_0, m_1) }{\mmn{2}}\!
\Downarrow_{\outC{p_1}}^1$.\qed
\end{enumerate}

\begin{rem}\label{rem:wrong}
It is instructive to identify the exact point in which an erroneous computation can be made when mimicking the behavior
of a decrement-and-jump instruction. Consider again the process $P'$, as analyzed in the case \textsc{M-Dec} above:
\begin{eqnarray*}
P' & = & u_0.(\outC{b} \parallel
           \outC{p_{i+1}}) +   z_0.\update{r_0}{\component{r_0}{!inc_0.\outC{u_0} \parallel \outC{z_0}}}. \outC{p_s} \parallel      \\
           & & e  \parallel  \component{r_0}{!inc_0.\outC{u_0} \parallel \prod^{m_{0}}\overline{u_0} \parallel \outC{z_0} }  \parallel           \reso{\alpha-1,\beta,\gamma+1} \parallel S
\end{eqnarray*}
where $S =  \encp{r_1 = m_1}{\mmn{2}} \parallel \prod_{l=1}^{n} \encp{(l:I_l)}{\mmn{2}}$ stands for the rest of the system.
Above, we analyzed the correct computation from $P'$, namely a synchronization on $u_{0}$:
\begin{eqnarray*}
P' & \pired &  \outC{p_{i+1}} \parallel  e \parallel \component{r_0}{!inc_0.\outC{u_0} \parallel \prod^{m_{0}-1}\overline{u_0} \parallel \outC{z_0} }  \parallel           \reso{\alpha-1,\beta+1,\gamma+1} \parallel S ~= P'' 
\end{eqnarray*}
with $P'' \equiv \encp{(i+1,m_{0}-1,m_{1})}{\mmn{2}}$.
The erroneous computation takes place when there is a synchronization on $z_{0}$, rather than on $u_{0}$. We then have:
\begin{eqnarray*}
P' & \pired &  \update{r_0}{\component{r_0}{!inc_0.\outC{u_0} \parallel \outC{z_0}}}. \outC{p_s} \parallel  e \parallel \\
& & \component{r_0}{!inc_0.\outC{u_0} \parallel \prod^{m_{0}}\overline{u_0} }  \parallel           \reso{\alpha-1,\beta,\gamma+1} \parallel S \\
& \pired \equiv &   \outC{p_s} \parallel  e \parallel \component{r_0}{!inc_0.\outC{u_0} \parallel \outC{z_0}} \parallel   \reso{\alpha-1,\beta,\gamma+1} \parallel S~= P'''
\end{eqnarray*}
with $P''' \equiv \encp{(s,0,m_{1})}{\mmn{2}}$.
The side effect of the above erroneous computation can be seen on the number of copies of $\outC{b}$ that remain after
the (erroneous) synchronization on $z_{0}$.
In fact, while a correct computation (as $P''$ above) increases in one the number of such copies, 
in an incorrect computation (as $P'''$ above) the number of copies of $\outC{b}$ remains invariant. 
Notice also that copies of $\outC{b}$ can be only produced at the beginning of the execution of the encoding of the \mm.
This is significant since, as discussed at the end of the case \textsc{M-Inc}, the number of copies of $\outC{b}$ has a direct influence on potential deadlocks  of the encoding of a \mm.
\end{rem}

\begin{lem}[Completeness]\label{l:compl-pase2}
Let $N$ be a \mm, if $N$ terminates then $\encp{N}{\mmn{2}} \barbw{e}$.
\end{lem}
\proof
Recall that $N$ is said to terminate if there exists a computation 
$$(1,0,0) \minskred ^* (h, 0, 0)$$
such that $(h: \mathtt{HALT})$.
Lemma~\ref{lem:correctstep} guarantees the existence of a process $P$ such that 
$\encp{(1, 0, 0)}{\mmn{2}} \pired ^* P \equiv \encp{(h, 0, 0)}{\mmn{2}}$, with 
$P \Downarrow_{\outC{p_{1}}}^1$.
This  ensures that every time that the encoding of $N$ reaches $\mathtt{HALT}$ the simulation is restarted.
Therefore,  termination of $N$ ensures that $\encp{N}{\mmn{2}}$
has an infinite computation:
since the encoding always 
exhibits barb $e$, we can conclude that $\encp{N}{\mmn{2}} \barbw{e}$.\qed

\begin{lem}[Soundness]\label{l:sound-pase2} 
Let $N$ be a \mm. If $N$ does not terminate then $\encp{N}{\mmn{2}} \negbarbw{e}$.
\end{lem}
\proof
It is enough to prove that if $N$ does not terminate 
(that is, if $N$ does not reach a  $\mathtt{HALT}$ instruction)
then all the computations of $\encp{N}{\mmn{2}}$ are finite.
Since the encoding can mimic the behavior of $N$ both correctly and incorrectly, we have two possible cases:
\begin{enumerate}[(1)]
 \item
 In the first case, 
  the simulation of $\encp{N}{\mmn{2}}$ is correct and no erroneous steps are introduced.
Notice that at every instruction an output on $f$ is consumed permanently: these
copies of $\outC{f}$ are only recreated when invoking a $\mathtt{HALT}$ instruction, which converts every $\outC{g}$ into a $\outC{f}$.
Since a $\mathtt{HALT}$ instruction is never reached, 
new copies  of $\outC{f}$ are never recreated, and 
the computation of process $\encp{N}{\mmn{2}}$ has necessarily to be  finite.

 \item 
 In the second case, 
 the simulation is not correct and one or more wrong guesses occurred
 in the simulation of a decrement-and-jump instruction.
 Here, in addition to the possibility of deadlocks described in Item (1) above, 
 erroneous computations constitute another source of deadlocks.
 In fact, as detailed in Remark \ref{rem:wrong}, 
 for each one of such wrong guesses a copy of $\outC{b}$ is permanently lost. 
%
An arbitrary number of wrong guesses may thus lead to a state in which there are no outputs on $b$.
As discussed at the end of the case of the \textsc{M-Inc} in the proof of Lemma \ref{lem:correctstep}, 
the encoding of an increment instruction
reaches a deadlock if a copy of $\outC{b}$ is not available.
This means that wrong guesses in simulating a decrement-and-jump instruction may induce deadlocks when simulating an increment instruction.
%
\end{enumerate}

\noindent Hence, as all the computations of $\encp{N}{\mmn{2}}$ are finite, therefore $\encp{N}{\mmn{2}}$  barb $e$
cannot be exposed an infinite number of times.\qed

We are now ready to repeat the statement of Lemma \ref{th:corrE2}, in Page~\pageref{th:corrE2}:

\begin{lem}[\ref{th:corrE2}]
Let $N$ be a \mm. $N$ terminates iff $\encp{N}{\mmn{2}} \barbw{e}$.
\end{lem}

\proof
It follows directly from Lemmas \ref{l:compl-pase2} and \ref{l:sound-pase2}.\qed

\section{Proofs from Section \ref{s:ev3}}

\subsection{Proof of Lemma \ref{th:corrE3}}\label{app:e31}
The proof relies on two  results:
completeness (Lemma~\ref{l:compl-pase3}) and soundness (Lemma~\ref{l:sound-pase3}).
The proof is very similar to the one presented for the case of \evols{2}, and 
considerations concerning the handling of resources (i.e., process \textsc{Control})
are exactly the same. Hence, Remarks \ref{rem:enough} and \ref{rem:wrong} are valid also in this proof.

We first introduce the notion of  encoding of a MM
configuration into \evold{3}. 

\begin{defi}\label{def:confe3}
Let $N$ be a \mm 
with 
registers $r_j ~(j \in \{0,1\})$ and 
instructions $(1:I_1), \ldots, (n:I_n)$.
The encoding of a configuration $(i,m_0, m_1)$ of $N$, denoted
$\encp{(i, m_{0}, m_{1})}{\mmn{3}}$, is defined as:
\[
 \outC{p_i} \parallel e \parallel \encp{r_0 = m_0}{\mmn{3}} \parallel \encp{r_1 = m_1}{\mmn{3}} \parallel \prod^{n}_{i=1} \encp{(i:I_i)}{\mmn{3}} \parallel \reso{\alpha,\beta,\gamma}
\]
where 
\begin{enumerate}[$\bullet$]
\item $\reso{\alpha,\beta,\gamma} \stackrel{\textrm{def}}{=} \prod^{\alpha} \outC{f} \parallel \prod^{\beta} \outC{b} \parallel \prod^{\gamma} \outC{g} \parallel !a.(\outC{f} \parallel \outC{b} \parallel \outC{a}) \parallel !h.(g.\outC{f} \parallel \outC{h})$, with  $\alpha,\beta,\gamma \geq 0$;
\item $\encp{(i:I_i)}{\mmn{3}}, \ldots, \encp{(n:I_n)}{\mmn{3}}$ are as in Table~\ref{t:encod-evold3d};
\item $\encp{r_j = m_j}{\mmn{3}}$ stands for $\componentbbig{r_j}{\prod^{m_j}U_j \parallel Reg_j \parallel \component{c_j}{\gar{\delta}}}$ with 
\begin{enumerate}[--]
\item $Reg_j = !inc_j.\update{c_j}{\component{c_j}{\bullet}}.\outC{ack}.u_j.\update{c_j}{\component{c_j}{\bullet}}.\outC{ack}$ (as in Table~\ref{t:encod-evold3d})
\item $U_j \stackrel{\textrm{def}}{=}  u_j.\update{c_j}{\component{c_j}{\bullet}}.\outC{ack}$
\item $ \garb{\delta}{j} \stackrel{\textrm{def}}{=}  Reg_{j} \parallel \prod^{\delta} U_j $
\end{enumerate}
\end{enumerate}\smallskip
\end{defi}

\noindent Similarly as before, in addition to the encodings of registers and instructions, 
the encoding of a \mm configuration 
includes a number of resources 
$\outC{f}$ and $\outC{b}$ which are always available during the execution of the machine.
These are represented by process 
$\reso{\alpha,\beta,\gamma}$, which  abstracts the evolution of 
process \textsc{Control} in Table~\ref{t:encod-evold3d}, and the resources
that it produces and maintains (namely, $\alpha$ copies of $\outC{f}$, $\beta$ copies of $\outC{b}$, and $\gamma$ copies of $\outC{g}$).
In addition, 
the encoding of register $j$ in \evold{3} includes a 
``garbage''
process  $ \garb{\delta}{j} $ 
representing residual processes 
which are accumulated during the execution of the encoding; 
as we will see, every interaction with  such a garbage process will result into a deadlocked process.

We prove that given a \mm $N$ there exists a 
computation of process $\encp{N}{\mmn{3}}$ which correctly mimics its behavior. 
We remind that Remark \ref{rem:enough} applies to this case too.

\begin{lem}\label{lem:correctstep3}
Let $(i, m_0, m_1)$ be a configuration of a \mm $N$. 
\begin{enumerate}[\em(1)]
\item If $(i, m_0, m_1)\minskred (i', m_0', m_1')$ then, for some process $P$, 
it holds that $$\encp{(i, m_0, m_1) }{\mmn{3}} \pired^* P \equiv \encp{(i', m_0', m_1') }{\mmn{3}}$$
\item If $(i, m_0, m_1)\notminskred$ then $\encp{(i, m_0, m_1) }{\mmn{3}}\! \Downarrow_{\outC{p_1}}^1$.
\end{enumerate}
\end{lem}
\proof \hfill \\
\begin{enumerate}[(1)]
\item
We proceed by a case analysis on the instruction  performed by the \mma.
Hence, we distinguish three cases corresponding to the behaviors associated to rules
\textsc{M-Inc}, \textsc{M-Dec}, and \textsc{M-Jmp}.
Without loss of generality, we restrict our analysis to operations on register $r_0$. 

\begin{desCription}
\item\noindent{\hskip-12 pt\bf Case \textsc{M-Inc}:}\ We have a Minsky configuration $(i, m_0, m_1)$ with 
$(i: \mathtt{INC}(r_0))$. By Definition \ref{def:confe3}, its encoding into \evold{3} is as follows:
\begin{eqnarray*}
\encp{(i, m_0, m_1) }{\mmn{3}}  & = & 
 \outC{p_i} \parallel e  \parallel
\encp{r_0 = m_0}{\mmn{3}} \parallel \encp{r_1 = m_1}{\mmn{3}} \parallel  \reso{\alpha,\beta,\gamma} \parallel\\
& & !p_i.f.(\outC{g} \parallel b.\outC{inc_0}.ack.\outC{p_{i+1}}) \parallel \prod_{l=1..n,l\neq i} \encp{(l:I_l)}{\mmn{3}} 
\end{eqnarray*}

We then have: $\encp{(i, m_0, m_1) }{\mmn{3}}   \pired R$
where 
$$R = \encp{r_0 = m_0}{\mmn{3}} \parallel f.(\outC{g} \parallel b.\outC{inc_0}.ack.\outC{p_{i+1}}) \parallel \reso{\alpha,\beta,\gamma} \parallel S ,$$
and $S = e \parallel \encp{r_1 = m_1}{\mmn{3}} \parallel \prod_{l=1}^{n} \encp{(l:I_l)}{\mmn{3}}$
stands for the rest of the system.
Starting from $R$, a possible sequence of reductions is the following:
\begin{eqnarray*}
R & \pired & \encp{r_0 = m_0}{\mmn{3}} \parallel b.\outC{inc_0}.ack.\outC{p_{i+1}} \parallel \reso{\alpha-1,\beta,\gamma+1} \parallel S\\
& = &  \componentbig{r_0}{\prod^{m_0}U_0 \parallel !inc_0.\update{c_0}{\component{c_0}{\bullet}}.\outC{ack}.u_0.\update{c_0}{\component{c_0}{\bullet}}.\outC{ack} \parallel \component{c_0}{\garb{\delta}{0}}} \parallel \\
&&  b.\outC{inc_0}.ack.\outC{p_{i+1}} \parallel \reso{\alpha-1,\beta,\gamma+1} \parallel S\\
& \pired &    \componentbig{r_0}{\prod^{m_0}U_0 \parallel !inc_0.\update{c_0}{\component{c_0}{\bullet}}.\outC{ack}.u_0.\update{c_0}{\component{c_0}{\bullet}}.\outC{ack} \parallel \component{c_0}{\garb{\delta}{0}}} \parallel \\
&&  \outC{inc_0}.ack.\outC{p_{i+1}} \parallel \reso{\alpha-1,\beta,\gamma+1} \parallel S = R'\\
& \pired  &   \componentbig{r_0}{\prod^{m_0}U_0 \parallel \update{c_0}{\component{c_0}{\bullet}}.\outC{ack}.u_0.\update{c_0}{\component{c_0}{\bullet}}.\outC{ack} \parallel Reg_0 \parallel \component{c_0}{\garb{\delta}{0}}} \\
&&  \parallel ack.\outC{p_{i+1}} \parallel \reso{\alpha-1,\beta,\gamma+1} \parallel S\\
& \pired  &   \componentbig{r_0}{\prod^{m_0}U_0 \parallel \outC{ack}.u_0.\update{c_0}{\component{c_0}{\bullet}}.\outC{ack} \parallel Reg_0 \parallel \component{c_0}{\garb{\delta}{0}}} \parallel \\
&&  ack.\outC{p_{i+1}} \parallel \reso{\alpha-1,\beta,\gamma+1} \parallel S \\
& \pired  &   \componentbig{r_0}{\prod^{m_0}U_0 \parallel u_0.\update{c_0}{\component{c_0}{\bullet}}.\outC{ack} \parallel Reg_0 \parallel \component{c_0}{\garb{\delta}{0}}} \parallel \\
&&  \outC{p_{i+1}} \parallel \reso{\alpha-1,\beta,\gamma+1} \parallel S \\
& = &   \componentbig{r_0}{\prod^{m_0+1}U_0  \parallel Reg_0 \parallel \component{c_0}{\garb{\delta}{0}}} \parallel  \outC{p_{i+1}} \parallel \reso{\alpha-1,\beta,\gamma+1} \parallel S = P
\end{eqnarray*}
It is easy to see that $P \equiv \encp{(i+1, m_0+1, m_1)}{\mmn{3}}$, as desired. 
Observe how the number of resources changes: in the first reduction, a copy of $\outC{f}$ is consumed, and  a copy of $\outC{g}$ is released in its place.
Notice that we are assuming that $\beta > 0$, that is, that there is at least one copy of $\outC{b}$.
In fact, since the instruction only takes place after a synchronization on $b$ (i.e., the second reduction above)
the presence of at least one copy of $\outC{b}$ in $\reso{\alpha-1,\beta,\gamma+1}$ is essential to avoid  deadlocks.
For the same reason, it is interesting to observe that 
in $R'$ the computation can only evolve if $\outC{inc_0}$ synchronizes with the replicated input process $inc_0$ inside $r_0$. 
Had it synchronized with the input on $inc_0$ inside $c_0$, the simulation would have reached a deadlock state, 
as there are no other adaptable processes at $c_0$ inside it.


\item\noindent{\hskip-12 pt\bf Case \textsc{M-Dec}:}\ We have a Minsky configuration $(i, m_0, m_1)$ with 
$m_0 > 0$ and 
$(i: \mathtt{DEC}(r_0,s))$. By Definition \ref{def:confe3}, its encoding into \evold{3}  is as follows:
\begin{eqnarray*}
\encp{(i, m_0, m_1) }{\mmn{3}}  & = & 
 \outC{p_i} \parallel e  \parallel
\encp{r_0 = m_0}{\mmn{3}} \parallel 
\encp{r_1 = m_1}{\mmn{3}}  \parallel \reso{\alpha,\beta,\gamma} \parallel \\
& & !p_i.f.\big(\outC{g} \parallel(\outC{u_0}.ack.(\outC{b} \parallel \outC{p_{i+1}})  + \\
&& \update{c_0}{\bullet}.\update{r_0}{\component{r_0}{Reg_0 \parallel
\component{c_0}{\bullet}}}.\outC{p_s})\big) \parallel  \prod_{l=1..n,l\neq i} \encp{(l:I_l)}{\mmn{3}} 
\end{eqnarray*}

We then have:
\begin{eqnarray*}
\encp{(i, m_0, m_1) }{\mmn{3}} &  \pired &  \encp{r_0 = m_0}{\mmn{3}} \parallel           \reso{\alpha,\beta,\gamma} \parallel \\
& & f.\big(\outC{g} \parallel(\outC{u_0}.ack.(\outC{b} \parallel \outC{p_{i+1}})  + \\
& & \qquad \update{c_0}{\bullet}.\update{r_0}{\component{r_0}{Reg_0 \parallel
\component{c_0}{\bullet}}}.\outC{p_s})\big) \parallel      S = R
\end{eqnarray*}
where $S =  e  \parallel  \encp{r_1 = m_1}{\mmn{3}} \parallel \prod_{l=1}^{n} \encp{(l:I_l)}{\mmn{3}}$ stands for the rest of the system.
Starting from $R$, a possible sequence of reductions is the following:
\begin{eqnarray*} 
R \!\!\!& \pired \!\!\!&  
\encp{r_0 = m_0}{\mmn{3}} \parallel           \reso{\alpha-1,\beta,\gamma+1} \parallel \\
& & \big((\outC{u_0}.ack.(\outC{b} \parallel \outC{p_{i+1}})  + \update{c_0}{\bullet}.\update{r_0}{\component{r_0}{Reg_0 \parallel
\component{c_0}{\bullet}}}.\outC{p_s})\big) \parallel      S\\
& = \!\!\!&  
\componentbig{r_0}{\prod^{m_0-1}U_0 \parallel u_{0}.\update{c_0}{\component{c_0}{\bullet}}.\outC{ack} \parallel Reg_0 \parallel \component{c_0}{\garb{\delta}{0}}} \parallel 
          \reso{\alpha-1,\beta,\gamma+1}  \\
&& \parallel \big((\outC{u_0}.ack.(\outC{b} \parallel \outC{p_{i+1}})  + \update{c_0}{\bullet}.\update{r_0}{\component{r_0}{Reg_0 \parallel
\component{c_0}{\bullet}}}.\outC{p_s})\big) \parallel S ~ = R'\\ 
& \pired \!\!\!&  \componentbig{r_0}{\prod^{m_0-1}U_0 \parallel\update{c_0}{\component{c_0}{\bullet}}.\outC{ack} \parallel 
Reg_0 \parallel \component{c_0}{\garb{\delta}{0}}} \parallel \reso{\alpha-1,\beta,\gamma+1} \parallel \\
&& ack.(\outC{b} \parallel \outC{p_{i+1}})\parallel S\\  
& \pired \!\!\!&  \componentbig{r_0}{\prod^{m_0-1}U_0 \parallel \outC{ack} \parallel 
Reg_0 \parallel \component{c_0}{\garb{\delta}{0}}} \parallel \reso{\alpha-1,\beta,\gamma+1} \parallel \\
&& ack.(\outC{b} \parallel \outC{p_{i+1}})\parallel S\\  
& \pired \!\!\!&  \componentbig{r_0}{\prod^{m_0-1}U_0   \parallel 
Reg_0 \parallel \component{c_0}{\garb{\delta}{0}}} \parallel \reso{\alpha-1,\beta,\gamma+1} \parallel  \outC{b} \parallel \outC{p_{i+1}}\parallel S\\  
& \equiv \!\!\!&  \componentbig{r_0}{\prod^{m_0-1}U_0   \parallel 
Reg_0 \parallel \component{c_0}{\garb{\delta}{0}}} \parallel \reso{\alpha-1,\beta+1,\gamma+1} \parallel   \outC{p_{i+1}}\parallel S =P
\end{eqnarray*}
It is easy to see that $P \equiv \encp{(i+1, m_0-1, m_1)}{\mmn{3}}$, as desired. 
Observe how 
in the last reduction 
the presence of at least  a copy of $\outC{u_{0}}$ in $r_{0}$ is fundamental for releasing both an extra copy of $\outC{b}$
and the trigger for the next instruction. 
Notice also that if $\outC{u_{0}}$ in $R'$ synchronizes with $u_0$ inside adaptable process $c_0$ then,
 as in the case of the increment, the simulation would be deadlocked.


\item\noindent{\hskip-12 pt\bf Case \textsc{M-Jmp}:}\ We have a Minsky configuration $(i, 0, m_1)$ and 
$(i: \mathtt{DEC}(r_0,s))$. By Definition \ref{def:confe3}, 
its encoding into \evold{3} is as follows:

\begin{eqnarray*}
\encp{(i, m_0, m_1) }{\mmn{3}}  & = & 
 \outC{p_i} \parallel e  \parallel
\encp{r_0 = 0}{\mmn{3}} \parallel 
\encp{r_1 = m_1}{\mmn{3}}  \parallel \reso{\alpha,\beta,\gamma} \parallel \\
& & !p_i.f.\big(\outC{g} \parallel(\outC{u_0}.ack.(\outC{b} \parallel \outC{p_{i+1}})  + \\
&& \qquad \quad \qquad \update{c_0}{\bullet}.\update{r_0}{\component{r_0}{Reg_0 \parallel
\component{c_0}{\bullet}}}.\outC{p_s})\big) \parallel  \\
& & \prod_{l=1..n,l\neq i} \encp{(l:I_l)}{\mmn{3}} 
\end{eqnarray*}

We then have:
\begin{eqnarray*}
\encp{(i, m_0, m_1) }{\mmn{3}} &  \pired &  \encp{r_0 = 0}{\mmn{3}} \parallel           \reso{\alpha,\beta,\gamma} \parallel
f.\big(\outC{g} \parallel(\outC{u_0}.ack.(\outC{b} \parallel \outC{p_{i+1}})  + \\
&&\update{c_0}{\bullet}.\update{r_0}{\component{r_0}{Reg_0 \parallel
\component{c_0}{\bullet}}}.\outC{p_s})\big) \parallel      S = R
\end{eqnarray*}
where $S =  e  \parallel  \encp{r_1 = m_1}{\mmn{3}} \parallel \prod_{l=1}^{n} \encp{(l:I_l)}{\mmn{3}}$ stands for the rest of the system.
Starting from $R$, a possible sequence of reductions is the following:
\begin{eqnarray*} 
R & \pired &  
\encp{r_0 = 0}{\mmn{3}} \parallel           \reso{\alpha-1,\beta,\gamma+1} \parallel \\
& & \big(\outC{u_0}.ack.(\outC{b} \parallel \outC{p_{i+1}})  + 
\update{c_0}{\bullet}.\update{r_0}{\component{r_0}{Reg_0 \parallel
\component{c_0}{\bullet}}}.\outC{p_s}\big) \parallel      S\\
& = &  
\componentbbig{r_0}{ Reg_0 \parallel \component{c_0}{\garb{\delta}{0}}} \parallel 
          \reso{\alpha-1,\beta,\gamma+1} \parallel \\
&& \big(\outC{u_0}.ack.(\outC{b} \parallel \outC{p_{i+1}})  + \update{c_0}{\bullet}.\update{r_0}{\component{r_0}{Reg_0 \parallel
\component{c_0}{\bullet}}}.\outC{p_s}\big) \parallel S \\ 
& \pired &  \componentbbig{r_0}{Reg_0 \parallel \garb{\delta}{0}} \parallel \reso{\alpha-1,\beta,\gamma+1} \parallel  \update{r_0}{\component{r_0}{Reg_0 \parallel
\component{c_0}{\bullet}}}.\outC{p_s} \parallel S\\  
& \pired \equiv&  \componentbbig{r_0}{Reg_0 \parallel \component{c_0}{Reg_0 \parallel \garb{\delta}{0}}} \parallel \reso{\alpha-1,\beta,\gamma+1} \parallel   \outC{p_s}\parallel S \\
& \simeq &  \componentbbig{r_0}{Reg_0 \parallel \component{c_0}{\garb{\delta}{0}}} \parallel \reso{\alpha-1,\beta,\gamma+1} \parallel   \outC{p_s}\parallel S=P
\end{eqnarray*}
It is easy to see that $P \equiv \encp{(s, 0, m_1)}{\mmn{3}}$, as desired. 
Notice that the first reduction results from a synchronization on $f$. 
The second reduction arises from an update action on $c_{0}$, which removes that ``boundary'' for $\garb{\delta}{0}$.
Finally, the third reduction is an update action on $r_{0}$.
We use $\simeq$ to denote the extension of  structural congruence with the axiom $!\pi.P \parallel !\pi.P = !\pi.P$.
\end{desCription}

\item If $(i, m_0, m_1)\notminskred$ then $i$ corresponds to the $\mathtt{HALT}$ instruction.
Then, by Definition~\ref{def:confe3}, 
its encoding into \evold{3} is as follows:
\begin{eqnarray*}
\encp{(i, m_0, m_1) }{\mmn{3}}  & = & 
 \outC{p_i} \parallel e  \parallel
 \encp{r_0 = m_0}{\mmn{3}} \parallel \encp{r_1 = m_1}{\mmn{3}} \parallel \reso{\alpha,\beta,\gamma} \parallel \\
& & !p_i.\outC{h}.h.\update{c_0}{\bullet}.\update{r_0}{\component{r_0}{Reg_0 \parallel \component{c_0}{\bullet}}}. \\
&& \update{c_1}{\bullet}.\update{r_1}{\component{r_1}{Reg_1 \parallel
\component{c_1}{\bullet}}}.\outC{p_1} \parallel \prod_{l=1..n,l\neq i} \encp{(l:I_l)}{\mmn{3}} 
\end{eqnarray*}
We then have: $\encp{(i, m_0, m_1) }{\mmn{3}}   \pired  R$ where
\begin{eqnarray*}
 R &=&
 \encp{r_0 = m_0}{\mmn{3}}  \parallel \encp{r_1 = m_1}{\mmn{3}} \parallel \reso{\alpha,\beta,\gamma} \parallel\\
 && \outC{h}.h.\update{c_0}{\bullet}.\update{r_0}{\component{r_0}{Reg_0 \parallel \component{c_0}{\bullet}}}. \update{c_1}{\bullet}.\update{r_1}{\component{r_1}{Reg_1 \parallel \component{c_1}{\bullet}}}.\outC{p_1}  \parallel S
\end{eqnarray*}
where $S =  e  \parallel  \prod_{l=1}^{n}\encp{(l:I_l)}{\mmn{3}}$ stands for the rest of the system.
Starting from $R$, a possible sequence of reductions is the following:
\begin{eqnarray*} 
R & \pired^{*} &   \encp{r_0 = m_0}{\mmn{3}}  \parallel \encp{r_1 = m_1}{\mmn{3}} \parallel \reso{\alpha+c,\beta,\gamma-c} \parallel S \parallel\\
 && \update{c_0}{\bullet}.\update{r_0}{\component{r_0}{Reg_0 \parallel \component{c_0}{\bullet}}}. \update{c_1}{\bullet}.\update{r_1}{\component{r_1}{Reg_1 \parallel \component{c_1}{\bullet}}}.\outC{p_1} =R_1
\end{eqnarray*}
where the output on $h$ in $R$ interacted with  process $\reso{\alpha,\beta,\gamma}$
so as to replace $c$ outputs on $g$ with $c$ outputs on $f$.
After that,  a synchronization on $h$ took place between 
the evolutions of $\reso{\alpha,\beta,\gamma}$ and of $R$.
We now have:
 \begin{eqnarray*} 
R_1 \!\!\!& \pired\!\!\! &  \componentbig{r_0}{\prod^{m_0}U_0 \parallel Reg_0 \parallel \garb{\delta_0}{0}}  \parallel \componentbig{r_1}{\prod^{m_1}U_1 \parallel Reg_1 \parallel \component{c_1}{\garb{\delta_1}{1}}} \parallel S  \parallel\\
 && \update{r_0}{\componentbbig{r_0}{Reg_0 \parallel \component{c_0}{\bullet}}}. \update{c_1}{\bullet}.\update{r_1}{\componentbbig{r_1}{Reg_1 \parallel \component{c_1}{\bullet}}}.\outC{p_1} \parallel \reso{\alpha+c,\beta,\gamma-c}\\
& \pired  \simeq \!\!\!& \componentbig{r_0}{Reg_0 \parallel \component{c_0}{\garb{\delta_0+m_0}{0}}}  \parallel \componentbig{r_1}{\prod^{m_1}U_1 \parallel Reg_1 \parallel \component{c_1}{\garb{\delta_1}{1}}} \parallel S  \parallel\\
 && \update{c_1}{\bullet}.\update{r_1}{\componentbbig{r_1}{Reg_1 \parallel \component{c_1}{\bullet}}}.\outC{p_1} \parallel \reso{\alpha+c,\beta,\gamma-c}\\
& \pired \!\!\! & \component{r_0}{Reg_0 \parallel \component{c_0}{\gar{\delta_0+m_0}}}  \parallel \component{r_1}{\prod^{m_1}U_1 \parallel Reg_1 \parallel \garb{\delta_1}{1}} \parallel S  \parallel\\
 && \update{r_1}{\componentbbig{r_1}{Reg_1 \parallel \component{c_1}{\bullet}}}.\outC{p_1} \parallel \reso{\alpha+c,\beta,\gamma-c}\\
& \pired  \simeq \!\!\!& \componentbig{r_0}{Reg_0 \parallel \component{c_0}{\garb{\delta_0+m_0}{0}}}  \parallel \componentbig{r_1}{Reg_1 \parallel \component{c_1}{\garb{\delta_1+m_1}{1}}} \parallel S \parallel\\
&&  \outC{p_1} \parallel \reso{\alpha+c,\beta,\gamma-c}  
\end{eqnarray*}
which is easily seen to correspond to $\encp{(1, 0, 0) }{\mmn{3}}$,
and thus $\encp{(i, m_0, m_1) }{\mmn{3}}\!
\Downarrow_{\outC{p_1}}^1$.\qed
\end{enumerate}\smallskip

\noindent It is straightforward to see that Remark \ref{rem:wrong} is valid for $\encp{N}{\mmn{3}}$ too.
Unsurprisingly, the proof concludes following the same lines of the proof of Lemma \ref{th:corrE2}.

\begin{lem}[Completeness]\label{l:compl-pase3}
Let $N$ be a \mm. If $N$ terminates then $\encp{N}{\mmn{3}} \barbw{e}$.
\end{lem}
\proof
Recall that $N$ is said to terminate if there exists a computation 
$$(1,0,0) \minskred ^* (h, 0, 0)$$
such that $(h: \mathtt{HALT})$.
Lemma~\ref{lem:correctstep3} guarantees the existence of a process $P$ such that 
$\encp{(1, 0, 0)}{\mmn{3}} \pired ^* P \equiv \encp{(h, 0, 0)}{\mmn{3}}$, with 
$P \Downarrow_{\outC{p_{1}}}^1$.
This  ensures that every time that the encoding of $N$ reaches $\mathtt{HALT}$ the simulation is restarted.
Therefore,  termination of $N$ ensures that $\encp{N}{\mmn{3}}$
has an infinite computation:
since the encoding always 
exhibits barb $e$, we can conclude that $\encp{N}{\mmn{3}} \barbw{e}$.\qed

\begin{lem}[Soundness]\label{l:sound-pase3} 
Let $N$ be a \mm. If $N$ does not terminate then $\encp{N}{\mmn{3}} \negbarbw{e}$.
\end{lem}
\proof
It is enough to prove that if $N$ does not terminate 
(that is, if $N$ does not reach a  $\mathtt{HALT}$ instruction)
then all the computations of $\encp{N}{\mmn{3}}$ are finite.
Since the encoding can mimic the behavior of $N$ both correctly and incorrectly, we have two possible cases:
\begin{enumerate}[(1)]
 \item
 In the first case, 
  the simulation of $\encp{N}{\mmn{3}}$ is correct and no erroneous steps are introduced.
Notice that at every instruction an output on $f$ is consumed permanently: these
copies of $\outC{f}$ are only recreated when invoking a $\mathtt{HALT}$ instruction, which converts every $\outC{g}$ into a $\outC{f}$.
Since a $\mathtt{HALT}$ instruction is never reached, 
new copies  of $\outC{f}$ are never recreated, and 
the computation of process $\encp{N}{\mmn{3}}$ has necessarily to be  finite.

 \item 
 In the second case, 
 the simulation is not correct and one or more wrong guesses occurred
 in the simulation of a decrement-and-jump instruction.
 Here, in addition to the possibility of deadlocks described in Item (1) above, 
 erroneous computations constitute another source of deadlocks.
 In fact, as detailed in Remark \ref{rem:wrong}, 
 for each one of such wrong guesses a copy of $\outC{b}$ is permanently lost. 
 Finally, the last source of error is represented by a wrong synchronization with $inc_j$ (in case of an increment) or $u_j$ (in case of a decrement) inside the adaptable process $c_j$. As 
 described above, those wrong synchronizations lead to a deadlock. 
An arbitrary number of wrong guesses may thus lead to a state in which there are no outputs on $b$.
As discussed at the end of the case of the \textsc{M-Inc} in the proof of Lemma \ref{lem:correctstep3}, 
the encoding of an increment instruction
reaches a deadlock if a copy of $\outC{b}$ is not available.
This means that wrong guesses in simulating a decrement-and-jump instruction may induce deadlocks when simulating an increment instruction.

\end{enumerate}

\noindent Hence, as all the computations of $\encp{N}{\mmn{3}}$ are finite, therefore $\encp{N}{\mmn{3}}$  barb $e$
cannot be exposed an infinite number of times.\qed

We are now ready to repeat the statement of Lemma \ref{th:corrE3}, in Page~\pageref{th:corrE3}:

\begin{lem}[\ref{th:corrE3}]
Let $N$ be a \mm. $N$ terminates iff $\encp{N}{\mmn{3}} \barbw{e}$.
\end{lem}

\proof
It follows directly from Lemmas \ref{l:compl-pase3} and \ref{l:sound-pase3}.\qed

\subsection{Proof of Lemma \ref{l:pn}}\label{app:e32}
Here we prove that given an \evols{3} process $P$ 
its associated Petri net representation \pnr{P}{\emptyset}
faithfully preserves its behavior.
We need some auxiliary propositions and definitions.
The following proposition states how to build a Petri net for the parallel composition of two processes starting from the Petri nets of the two processes. 

\begin{prop}\label{p:pn-pc}
Let  $P_1$ and $P_2$ 
be
two  \evols{3} processes with 
associated Petri nets $\pnr{P_1}{\emptyset}$ and $\pnr{P_2}{\emptyset}$, as in Definition \ref{d:pn}.
Then, the Petri net \pnr{P_{1} \parallel P_{2}}{\emptyset} is defined as:
$$\pnr{P_{1} \parallel P_{2}}{\emptyset} = (\places{P_{1} \parallel P_{2}}, \transit{P_{1} \parallel P_{2}}, \initMark{P_{1} \parallel P_{2}})$$
where 
\begin{align*}
\places{P_{1} \parallel P_{2},\emptyset}&=\places{P_1, \emptyset} \cup \places{P_2, \emptyset},\\
\transit{P_{1} \parallel P_{2},\emptyset}& =  \transit{P_1, \emptyset} \cup \transit{P_2, \emptyset} \cup T,\\
\initMark{P_{1} \parallel P_{2}}&=\initMark{P_1} \uplus \initMark{P_2}
\end{align*}
with $T$ representing the set of instances of transition schemata in Table \ref{tab:pn} 
that become possible 
due to the interplay of places in $\places{P_1, \emptyset}$ and in $\places{P_2, \emptyset}$. 
\end{prop}

\proof
Immediate from the definitions.\qed

Similarly, the next proposition shows how to obtain the Petri net associated to $\component{a}{P}$ starting from the one of $P$.

\begin{prop}\label{p:pn-ap}
Let  $P$ 
be
an \evols{3} process with 
associated Petri net $\pnr{P}{\emptyset}$ as in Definition~\ref{d:pn}.
Then, the Petri net \pnr{\component{a}{P}}{\emptyset}
is defined as:
$$ \pnr{\component{a}{P}}{\emptyset} = (\places{\component{a}{P}}, \transit{\component{a}{P}}, \initMark{\component{a}{P}})$$
where:
\begin{enumerate}[$\bullet$]
\item $\places{\component{a}{P},\emptyset}=\{ \coppia{Q}{a\sigma} \mid  \coppia{Q}{\sigma} \in \places{P, \emptyset}\} \cup 
\{a \sigma \mid \sigma \in  \places{P, \emptyset} \}$ $\cup
\{ a \}$.
\item $\transit{\component{a}{P}}$ is obtained from $\transit{P,\emptyset}$ by replacing places in $\places{P, \emptyset}$ with places in $\places{\component{a}{P},\emptyset}$, as defined above.
\item $\initMark{\component{a}{P}}$ is obtained from $\initMark{P}$ by 
(i) replacing places in $\places{P, \emptyset}$ with places in $\places{\component{a}{P},\emptyset}$, as defined above, 
and (ii)  adding a token in  the place for the adaptable process $a$.
\end{enumerate}
\end{prop}
\proof
Immediate from the definitions.\qed

\begin{lem}\label{l:auxpn}
 Let $P$ and $(\places{P,\emptyset}, \transit{P,\emptyset}, \initMark{P})$
 be an \evols{3} process and its associated Petri net,  as in Definition \ref{d:pn}.
 Then we have:
\begin{enumerate}[\em(1)]
\item If $P \pired P'$ then $\decc{\varepsilon}{P} \uplus \{go\} \rightarrow  \decc{\varepsilon}{P'} \uplus \{go\}$.

\item If $\decc{\varepsilon}{P} \uplus \{go\} \rightarrow  m\uplus \{go\}$ then, for some $P'$, 
$P \pired P' \text{ and } \decc{\varepsilon}{P'}=m$
\end{enumerate}

\end{lem}
\proof
The proof of  (1) proceeds by induction on the derivation tree of $P \pired P'$, with a case analysis on the last applied rule. 
There are seven cases to check. 

\begin{desCription}
 \item\noindent{\hskip-12 pt\bf Case \rulename{Act1}:}\ Then we have $P = P_1 \parallel P_2$ and:
$$  
\inferrule{P_1 \arro{~~} P_1'}{P_1 \parallel P_2 \arro{\ } P'_1 \parallel P_2}			
$$
By inductive hypothesis,  we have $\decc{\varepsilon}{P_1} \uplus \{go\} \rightarrow  \decc{\varepsilon}{P_1'} \uplus \{go\}.$
By Proposition~\ref{p:pn-pc},
$\transit{P_{1},\emptyset} \subseteq \transit{P_{1} \parallel P_{2},\emptyset}$.
Since by Definition \ref{d:pn} $\decc{\varepsilon}{P_1 \parallel P_2}  =  \decc{\varepsilon}{P_1} \uplus \decc{\varepsilon}{P_2}$,
then we can conclude that $\decc{\varepsilon}{P} \uplus \{go\} \rightarrow \decc{\varepsilon}{P'} \uplus \{go\} $ with $\decc{\varepsilon}{P'}=\decc{\varepsilon}{P_1'} \uplus \decc{\varepsilon}{P_2}$.
This concludes the proof for this case.

 \item\noindent{\hskip-12 pt\bf Case \rulename{Act2}:}\ Analogous to the case for \rulename{Act1} and omitted. 

 \item\noindent{\hskip-12 pt\bf Case \rulename{Loc}:}\ Then we have $P=\component{a}{P_1}$ and
$$  
\inferrule{P_1 \arro{~~} P_1'}{\component{a}{P_1} \arro{~~}  \component{a}{P_1'}}
$$
By inductive hypothesis, we have  $\decc{\varepsilon}{P_1} \uplus \{go\} \rightarrow  \decc{\varepsilon}{P_1'} \uplus \{go\}$.
Proposition \ref{p:pn-ap} states that 
$\transit{P,\emptyset}$ is obtained by extending the addresses of places
in $\transit{P_{1},\emptyset}$ with name $a$.
Since by Definition \ref{d:pn} $\decc{\varepsilon}{\component{a}{P_1}}  =  \decc{a}{P_1} \uplus \{a\}$ 
then we can conclude that $\decc{\varepsilon}{P} \uplus \{go\} \rightarrow \decc{\varepsilon}{P'} \uplus \{go\} $ with $\decc{\varepsilon}{P'}=\decc{a}{P_1'} \uplus \{a\}$. This concludes the proof for this case.

\item\noindent{\hskip-12 pt\bf Cases \rulename{Tau1}-\rulename{Tau2}:}\
Then $P \equiv  \fillcont{C_1}{A}\parallel \fillcont{C_2}{B}$, where $C_{1}, C_{2}$ are monadic contexts as in Definition \ref{d:mc}.
Moreover, 
$A$ is either 
$!b.Q$ or 
$\sum_{i \in I} \pi_i.Q_i$ with $\pi_{l}=b$, for some $l\in I$, and 
$B$ is either 
$!\outC{b}.R$
or $\sum_{i \in I} \pi_i.R_i$ with $\pi_{l}=\outC{b}$, for some $l\in I$.

We consider only the case in which $A = \sum_{i \in I} \pi_i.Q_i$ and $B = !\outC{b}.R$;  the other cases are similar. 
Let us denote with $\sigma$ and $\theta$ 
the address (with respect to the hole) induced by adaptable processes in $C_{1}$ and $C_{2}$, respectively.
That is, the address of $A$ in $C_{1}$ is $\sigma$ and the address of $B$ in $C_{2}$ is $\theta$.
Then, by construction of the Petri net, there is a token in the places $\coppia{\sum_{i \in I} \pi_i.Q_i}{\sigma}$ and $\coppia{!\outC{b}.R}{\theta}$.
Therefore,  transition 
$$\big\{go, \coppia{\sum_{i \in I} \pi_i.Q_i}{\sigma},
        \coppia{!\outC{b}.R}{\theta}\big\}
         \derriv
\big\{go, \coppia{!\outC{b}.R}{\theta}\big\} \uplus
\decc{\theta}{R} \uplus \decc{\sigma}{Q_{l}}
$$
(denoted (4) in Table~\ref{tab:pn}) can fire.
By definition of $\mathsf{dec}$ (cf. Definition \ref{d:pn})
it is easy to see that this corresponds to $\decc{\varepsilon}{P'} \uplus \{go\}$. This concludes the proof for this case.
 
\item\noindent{\hskip-12 pt\bf Cases \rulename{Tau3}-\rulename{Tau4}:}\
Then $P \equiv \fillcont{C_1}{A} \parallel \fillcont{C_2}{B}$ where:
\begin{enumerate}[$\bullet$]
\item $C_{1},C_{2}$ are monadic contexts, as in Definition \ref{d:mc}; 
\item $A = \component{b}{P_1}$, for some $P_{1}$;  
\item $B  =  \sum_{i \in I} \pi_i.R_i$ with $\pi_{l}=\update{b}{\component{b}{U} \parallel P_2}$ for $l\in I$, or $ B = !\update{b}{\component{b}{U} \parallel P_2}.R$, for some $P_{2}, R$.
\end{enumerate}

\noindent We consider the case in which $B = !\update{b}{\component{b}{U} \parallel P_2}.R$; the other case is similar. 
Let us denote with $\sigma$ and $\theta$ 
the address (with respect to the hole) induced by adaptable processes in $C_{1}$ and $C_{2}$, respectively.
That is, the address of $A$ in $C_{1}$ is $\sigma$ and the address of $B$ in $C_{2}$ is $\theta$.
Then, by construction of the Petri net, 
we have a token in the places $\{\sigma b\}$ and $\coppia{!\update{b}{\component{b}{U} \parallel P_2}.R}{\theta}$. 
At this point, we should distinguish two cases, depending on whether $\sigma b$ is contained in $\theta$ or not.
Suppose $\sigma b$ is not contained in $\theta$. That is, there is no process with the nesting structure of $C_{1}$ inside $C_{2}$.
Then   transition 
$$
\begin{array}{ll}
\big\{go, \sigma b, \coppia{!\update{b}{\component{b}{U} \parallel P_2}.R}{\theta} \big\}
 \derriv&
 \big\{go, \coppia{!\update{b}{\component{b}{U} \parallel P_2}.R}{\theta} ,\sigma b\big\} \uplus \\
 &\decc{\theta}{R} \uplus 
 \decc{\sigma }{P_2} 
 \uplus \decc{\sigma b}{U} 
 \end{array}
              $$
(denoted (8) in Table~\ref{tab:pn}) can fire.
It easy to see that this corresponds to $\decc{\varepsilon}{P'} \uplus \{go\}$ and we are done.

Similarly,  if $\sigma b$ is contained in $\theta$ then it means that there exists a process with the same structure of $C_1$ inside $C_2$.
Therefore, the place $\sigma b$ is duplicated and a token is present in both places. Then transition 
$$
\begin{array}{ll}
\big\{go, \sigma b, \sigma b, \coppia{!\update{b}{\component{b}{U} \parallel P_2}.R}{\theta}, \big\}
 \derriv \!\!\!&
 \big\{go, \sigma b, \sigma b,  \coppia{!\update{b}{\component{b}{U} \parallel P_2}.R}{\theta} \big\} \uplus \\
 &\decc{\theta}{R} \uplus 
 \decc{\sigma }{P_2} 
 \uplus \decc{\sigma b}{U} 
  \end{array}
              $$
(denoted (9) in Table~\ref{tab:pn}) can fire.
It easy to see that this corresponds to $\decc{\varepsilon}{P'} \uplus \{go\}$ and this concludes the proof.\smallskip
\end{desCription}

\noindent We now move on the proof of (2), which proceeds by a case analysis on the transition fired by the Petri net. 
The transition schemata in 
Table~\ref{tab:pn}
can be divided into two groups: (1) transitions mimicking a synchronization (i.e., an interaction between  
an input and an output prefix)
and (2) transitions mimicking an update action
(i.e., an interaction between  
an update prefix and an adaptable process). We consider these two groups separately:

\begin{enumerate}[(1)]
 
 \item  This group comprises transition schemata (3)--(5) in Table~\ref{tab:pn}.
 For simplicity we concentrate only on transitions of kind (3), as the others are similar.
If a transition of this kind can fire, then we have tokens in  $$\big\{go, \sum_{i \in I} \coppia{\pi_i.A_i}{\alpha},\sum_{j \in J} \coppia{\rho_j.B_j}{\beta}\big\}$$
which, by construction of the Petri net, 
implies that 
$$P \equiv \fillcont{D}{\sum_{i \in I}\pi_i.A_i,  \sum_{j \in J}\rho_j.B_j}$$
where $D$ is a biadic context, as in Definition \ref{d:mc}.
After the fire of the transition, tokens move to 
$\{go\}\uplus \decc{\alpha}{A_{l}} \uplus \decc{\beta}{B_{m}}$; by construction, 
this  corresponds to a process $P' \equiv \fillcont{D}{A_l, A_m}$, and we are done.

 \item  This group comprises transition schemata (6)--(9) in Table~\ref{tab:pn}.
For simplicity, we concentrate only on transitions of kind (6), as the others are similar.  
If a transition of this kind can fire, then we have tokens in   $$\big\{go, \coppia{\sum_{i \in I} \pi_i.A_i}{\alpha},\beta \big\}$$ 
which, by construction of the Petri net, implies 
$$P\equiv \fillcont{D}{\sum_{i \in I}\pi_i.A_i, \component{a}{Q} }$$ where $D$ is a biadic context, as in Definition \ref{d:mc}.
After the fire of the transition the tokens move to $\{go\} \uplus \decc{\alpha}{A_{l}} \uplus \decc{\beta}{A}  \uplus \decc{\beta a}{U} \uplus \{\beta a\}$;
by construction, this corresponds to a process $P' \equiv \fillcont{D}{A_l, \component{a}{\fillcon{U}{Q}} \parallel A} $, and we are done.\qed
\end{enumerate}

We can now restate Lemma \ref{l:pn}, as in Page~\pageref{l:pn}:
\begin{lem}[\ref{l:pn}]
 Let $P$ and $(\places{P,\emptyset}, \transit{P,\emptyset}, \initMark{P})$
 be an \evols{3} process and its associated Petri net,  as in Definition \ref{d:pn}.
 Then we have:
$$P \pired P' \textrm{ iff } \decc{\varepsilon}{P} \uplus \{go\} \rightarrow  \decc{\varepsilon}{P'} \uplus \{go\}.$$
\end{lem}
\proof
Immediate from Lemma \ref{l:auxpn}.\qed


 \end{document}